\documentclass{aa}  
\newcommand{\micron}{$\mu$m}
\newcommand{\Ks}{$K_{\rm{s}}$}
\newcommand{\msun}{M$_{\sun}$}

\newcommand{\HII}{\ion{H}{ii}}
\newcommand{\hei}{He{\sc i}}
\newcommand{\paa}{Pa$\alpha$}

\newcommand{\brg}{Br$\gamma$}
\newcommand{\kms}{km~s$^{-1}$}
\newcommand{\halpha}{H$\alpha$}
\newcommand{\hbeta}{H$\beta$}

\usepackage{graphicx}
\usepackage{subcaption}
\usepackage{txfonts}
\usepackage{color}

\begin{document}

   \title{Spatially resolved gas and stellar kinematics in compact starburst galaxies\thanks{Based on observations collected at the European Southern Observatory at Paranal, Chile (ESO programme 097.B-1061).} \thanks{The data cubes, emission line, velocity and dispersion maps are available in electronic form at the CDS via anonymous ftp to cdsarc.u-strasbg.fr (130.79.128.5) or via http://cdsweb.u-strasbg.fr/cgi-bin/qcat?J/A+A/} }

        \author{A. Bik
        \inst{1}
        \and
          G. \"Ostlin
          \inst{1}
 	\and
	M. Hayes
	\inst{1}
	J. Melinder
	\inst{1}
	\and
	V. Menacho
	\inst{1}
	}
	
   \institute{Department of Astronomy, Stockholm University, Oscar Klein Centre, AlbaNova University Centre, 106 91 Stockholm, Sweden
             \email{arjan.bik@astro.su.se}          
}
   \date{Received ; accepted }

  \abstract
   {The kinematics of galaxies provide valuable insights into their physics and assembly history. Kinematics are governed not only by the gravitational potential, but also by merger events and stellar feedback processes such as stellar winds and supernova explosions.}
   {We aim to identify what governs the kinematics in a sample of  SDSS-selected nearby starburst galaxies, by obtaining spatially resolved measurements of the gas and stellar kinematics.}
   {We obtained near-infrared integral-field $K$-band spectroscopy with VLT/SINFONI for 15 compact starburst galaxies. We derived the integrated as well as spatially resolved stellar and gas kinematics.   The stellar kinematics were derived from the CO absorption bands, and \paa\ and \brg\ emission lines were used for the gas kinematics.}
   {Based on the integrated spectra, we find that the majority of galaxies have gas and stellar velocity dispersion that are comparable.
    A spatially resolved comparison shows that the six  galaxies that deviate show evidence for a bulge or stellar feedback. Two galaxies are identified as mergers based on their double-peaked emission lines. In our sample, we find a negative correlation between the ratio of the rotational velocity over the velocity dispersion (v$_{rot}$/$\sigma$) and the star formation rate surface density.}
   {We propose a scenario where the global kinematics of the galaxies are determined by gravitational instabilities that affect both the stars and gas. This process could be driven by mergers or accretion events. Effects of stellar feedback on the ionised gas are more localised and detected only in the spatially resolved analysis. The mass derived from the velocity dispersion provides  a reliable mass even if the galaxy cannot be spatially resolved.  The technique used in this paper is applicable to galaxies at low and high redshift with the next generation of infrared-focussed telescopes (JWST and ELT).}


   \maketitle
%

\section{Introduction}

Observing the internal motions of stars and gas in galaxies provides invaluable insights into their physics and assembly history. This is a powerful way to trace the gravitational potential of the galaxy in which the stars and gas reside. Kinematics of local spiral galaxies have revealed that galaxies are surrounded by large dark matter halos \citep{Rubin80,Bosma89}. However, the movement of gas in galaxies  not only traces the gravitational potential, but is also influenced by energy and momentum input from stellar feedback processes such as stellar winds and supernova explosions. Stellar feedback can, for example, increase the turbulence in the galaxy as well as drive powerful galactic-scale outflows \citep[e.g.][]{Chevalier85,Strickland09}. 

Over the last 20 years, large galaxy surveys with integral field spectrographs at low (e.g. SAURON: \citet{Bacon01}, GHASP: \citet{Epinat10}, CALIFA: \citet{Sanchez12}, DYNAMO: \citet{Green14}, MaNGA: \citet{Bundy15}, SAMI: \citet{Scott18,Oh22}) and high redshift  (e.g. \citet{Law09}, SINS: \citet{ForsterSchreiber06}, \citet{Newman13}, KMOS$^{\mathrm{3D}}$: \citet{Wisnioski15}, KMOS-KROSS: \citet{Johnson18} have provided enormous insights into the spatially resolved as well as global kinematics of galaxies and its evolution as a function of  redshift \citep[see also][]{Glazebrook13}. 

\begin{table*}[!t]
\caption{Galaxy sample}
\centering
\begin{tabular}{l|rrcccc}
\hline\hline 
 
Name & $\alpha$ (J2000)&$\delta$ (J2000) & redshift &petroR50$\_$r & EW & K$_\mathrm{s}$ \\
	&   				&			&		&  & (H$\alpha$) & \\
          &(h m s)&($^\circ$\ \arcmin\  \arcsec)	   &   ($z$)           & (\arcsec) &  (\AA) & (mag) \\
          \hline
SDSS J0018$-$0903	&		00:18:09.16&	-09:03:35.83 	&	0.02282963	&	1.60 $\pm$ 0.01		&	120.1	$\pm$	0.8 	& 14.43 \\
SDSS J0230$-$0720	&		02:30:05.86 &	-07:20:32.13	&	0.05879955	&	2.10 $\pm$ 0.01		&	120.4	$\pm$	0.7	& 14.35\\
SDSS J1140$-$0024		&		11:40:13.23 &	-00:24:42.17	&	0.0220188\phantom{0}&	1.46 $\pm$ 0.01	&	213.0 $\pm$		1.5	& 14.35\\
SDSS J1155$-$0100		&		11:55:59.21&	-01:00:00.97	&	0.03648656	&	1.83 $\pm$ 0.01		&	152.3	$\pm$	0.9	& 14.47\\
SDSS J1211$+$1424	&		12:11:14.37&	+14:24:35.20	&	0.04529009	&	2.97 $\pm$ 0.04		&	106.7	$\pm$	0.9	& 13.57\\	
SDSS J1336$+$0748	&		13:36:27.75&	+07:48:58.49	&	0.02333842	&	1.97 $\pm$ 0.04 		&	104.7	$\pm$	0.8	& 14.03\\
SDSS J1339$-$0045	&		13:39:51.46&	-00:45:11.08	&	0.02227835	&	1.69 $\pm$ 0.02		&	132.6	$\pm$	0.9	& 14.93\\
SDSS J1340$+$0616	&		13:40:49.02&	+06:16:54.80	&	0.02902982	&	2.06 $\pm$ 0.10		&	173.9	$\pm$	1.0	& 14.43	\\
SDSS J1353$-$0231	&		13:53:56.15&	-02:31:47.62	&	 0.05145098	&	2.85 $\pm$ 0.02		&	119.5	$\pm$	0.7	& 13.48\\
SDSS J1400$+$0404	&		14:00:09.00 &	+04:04:50.82	&	0.04046174	&	3.04 $\pm$ 0.01		&	134.1	$\pm$	1.0	& 13.57\\
SDSS J1412$-$0049	&		14:12:24.85 &	-00:49:55.21	&	 0.02600801	&	2.59 $\pm$ 0.02		&	116.7	$\pm$	1.0	& 14.39\\
SDSS J1513$+$0431	&		15:13:36.39&	+04:31:45.38	&	 0.02482945	&	2.82 $\pm$ 0.03		&	116.7	$\pm$ 	0.8	& 14.08\\
SDSS J1525$+$0500	&		15:25:27.48 &	+05:00:29.99	&	0.03574942	&	3.46 $\pm$ 0.02		&	106.5	$\pm$	0.8	& 12.66\\
SDSS J1545$+$0814	&		15:45:56.30&	+08:14:59.33	&	 0.04175024	&	2.10 $\pm$ 0.02		&	147.0	$\pm$	1.1	& 14.03\\
SDSS J2256$+$1305	&		22:56:09.41&	+13:05:51.50	&	 0.03743555&	3.00 $\pm$ 0.04			&	145.3	$\pm$	1.2	& 13.97\\
\hline\label{tab:sample}
\end{tabular}
\end{table*}

\begin{table*}[!t]
\caption{Star formation rates and masses}
\centering
\begin{tabular}{l|ccccc}
\hline\hline 
 
Name &  E(B-V)$_\mathrm{optical}$ & E(B-V)$_\mathrm{IR}$  & SFR &SFR & log(Mass)\\
	& \halpha/\hbeta & \brg/\halpha&   (\halpha) & (\brg)\tablefootmark{a} & \\
          &(mag) &  (mag) & (\msun/yr) & (\msun/yr)\\
          \hline
 J0018$-$0903	&	 0.39	$\pm$ 0.02&0.30 $\pm$ 0.05	&0.8	&0.6		&9.12 $\pm$	     0.15 \\
 J0230$-$0720	&	0.31	$\pm$ 0.01&\emph{0.35 $\pm$ 0.05}	&6.6	&\emph{7.2}		&9.79 $\pm$     0.28\\
 J1140$-$0024	&	0.29	$\pm$ 0.02&0.00 $\pm$ 0.05	&1.3	&0.6		&---\\
 J1155$-$0100	&	0.21	$\pm$ 0.03&\emph{0.70 $\pm$ 0.05}	&2.8	&\emph{8.7}		&9.21 $\pm$  0.16\\
 J1211$+$1424	&	0.83	$\pm$ 0.05&\emph{1.05 $\pm$ 0.05}	&5.9	&\emph{9.9}		&10.25 $\pm$ 0.12\\	
 J1336$+$0748	&	0.28	$\pm$ 0.03&0.50 $\pm$ 0.05	&1.0	&1.8		&9.31	$\pm$ 0.24\\
 J1339$-$0045	&	0.21	$\pm$ 0.02&0.00 $\pm$ 0.05	&0.5 	&0.2		&8.91$\pm$ 0.16\\
 J1340$+$0616	&	0.47	$\pm$ 0.02&0.39 $\pm$ 0.05	&1.7	&1.4		&---	\\
 J1353$-$0231	&	0.65	$\pm$ 0.02&\emph{1.22 $\pm$ 0.05}	&7.3 &\emph{27.5}		&10.55 $\pm$ 0.01\\
 J1400$+$0404	&	0.53	$\pm$ 0.02&\emph{0.66 $\pm$ 0.05}	&10.1 &\emph{13.6}		& 9.93 $\pm$  0.01\\
 J1412$-$0049	&	0.36	$\pm$ 0.03&0.00 $\pm$ 0.05	&1.5	&0.5		& 9.39 $\pm$ 0.19\\
 J1513$+$0431	&	0.27	$\pm$ 0.02&0.24 $\pm$ 0.05	& 1.0	&0.9		&9.47	$\pm$	0.25\\
 J1525$+$0500	&	1.00	$\pm$ 0.05&0.70 $\pm$ 0.05	&17.5&9.0		&10.48 $\pm$ 0.11\\
 J1545$+$0814	&	0.53	$\pm$0.05&\emph{0.29 $\pm$ 0.05}	&4.7 	&\emph{2.7}	& 9.97 $\pm$ 0.17\\
 J2256$+$1305	&	0.40	$\pm$0.01&0.23 $\pm$ 0.05	&3.4 	&2.3		& 9.85 $\pm$ 0.35 \\
\hline\label{tab:sfr_mass}
\end{tabular}
\tablefoot{
\tablefoottext{a}{The values printed in italics are derived using \paa\ instead of \brg.}
}

\end{table*}

 Most late-type galaxies in the local Universe form the so-called star formation main sequence \citep{Chang15}. These galaxies are typically disk-like galaxies with ordered kinematics and form stars on a moderate rate at a fixed mass. A small fraction of galaxies ($\sim$1\%), however, do not fall on this star formation sequence, but have much higher star formation rates (SFRs) compared to other galaxies with the same mass \citep{Bergvall16}. These so-called starburst galaxies typically have much more complicated kinematics and show perturbed disks, or dispersion-dominated kinematics, driven by merger events and/or starbursts  \citep[e.g.][]{Ostlin01,Goncalves10,Green14,Herenz16,Cresci17, Bik18,Menacho19}.

These starburst galaxies are rare at low redshift, but become much more common at higher redshift. Therefore galaxy surveys at higher redshift reveal a much higher fraction of galaxies with complex dynamics due to mergers and extreme star formation. Additionally, even the  galaxies that do show disk-like kinematics behave differently than the local disk galaxies \citep[e.g.][]{ForsterSchreiber06,Cresci09,Wisnioski18}. They show much larger velocity dispersions than the local disk galaxies \citep[e.g.][]{Swinbank12,Johnson18}.

Observations show a  relation between the  SFR and the luminosity-weighted velocity dispersion ($\sigma_{m}$).  \citet{Green10,Green14}  found a tight SFR-$\sigma_{m}$ relation, but a much weaker relation between $\sigma_{m}$ and the stellar mass.  The apparent correlation of $\sigma_{m}$with the \halpha\ luminosity, that is to say the SFR, led Green et al. to propose that the SFR itself is the prime source of the line width, that is star formation causes turbulence. Similar relations between $\sigma_{m}$ and SFR were found at high redshift  \citep{Cresci09,Lehnert09,Lehnert13}, making these authors also conclude that the turbulence is star formation induced.

By compiling a large set of literature data covering both low- and high-redshift galaxies, \citet{Krumholz16} found that gravity is the ultimate source of the turbulence, where turbulence is created by gravitational instabilities in a marginally stable galaxy disk. In a follow-up paper, \citet{Krumholz18} propose a new unified model for the structure and evolution of gas in galactic disks. In this model both star formation feedback and radial transport are included as a source of turbulence. 
This results in a much better explanation for the observed $\sigma_{m}$ versus SFR relation, where gravity-driven turbulence is the dominant source of turbulence for the high $\sigma_{m}$, high SFR galaxies, while the velocity dispersions of main sequence galaxies with lower SFR are predominantly caused by feedback-induced turbulence.  

The measurements discussed above  all concern properties of the gas (mostly ionised gas (\halpha) or neutral gas (\ion{H}{I})). However, by just the velocities of the gas alone, it is hard to derive the cause of the observed kinematics for an individual galaxy. We cannot discriminate between velocity dispersion caused by gravitational instabilities or stellar feedback through outflows. By including  stellar kinematics in the analysis we can put stronger constraints on the origin of the observed kinematics. Stellar feedback mostly affects the kinematics of the gas by increasing the turbulence in the ISM, creating outflows or expanding bubbles in the ISM. Gravitational instabilities in the disk  have an effect on both the stellar and gas kinematics.  The kinematics of the older stellar populations can be decoupled from the gas due to for example the presence of a gas-free bulge, where random stellar motions dominate the kinematics \citep{Kormendy04}. In merger systems on the other hand, the gas relaxes earlier than the stars due to the dissipative nature of the gas.  Therefore, by a careful comparison of the spatially-resolved kinematics on both the gas and the stars, we gain insight into the structure and history of the galaxy.

This comparison is done the most easily in late-type spiral galaxies,  where both the gas and stellar kinematics can be measured.  A comparison between the HI, [OIII] emission and the H \& K stellar absorption in a sample of spiral galaxies by \citet{Kobulnicky00} showed that the kinematics agree within 20 \%. Similar results were found by other studies  \citep{CatalanTorrecilla20, Ganda06, FalconBarroso06}. A common feature observed is that the stellar rotation speed is slower than the rotation speed of the gas. This asymmetric drift is caused by the random motions of the stars \citep{Martinsson13,Oh22}. Galaxy bulges stand out in the stellar kinematics as they typically have a larger velocity dispersion than the stellar disk \citep[e.g.][]{Oh20}. In massive early type, elliptical galaxies, containing typically very little gas, only the stellar kinematics can be derived \citep[e.g.][]{Emsellem04,FalconBarroso17}.

\begin{figure*}[!t]
   \includegraphics[width=\hsize]{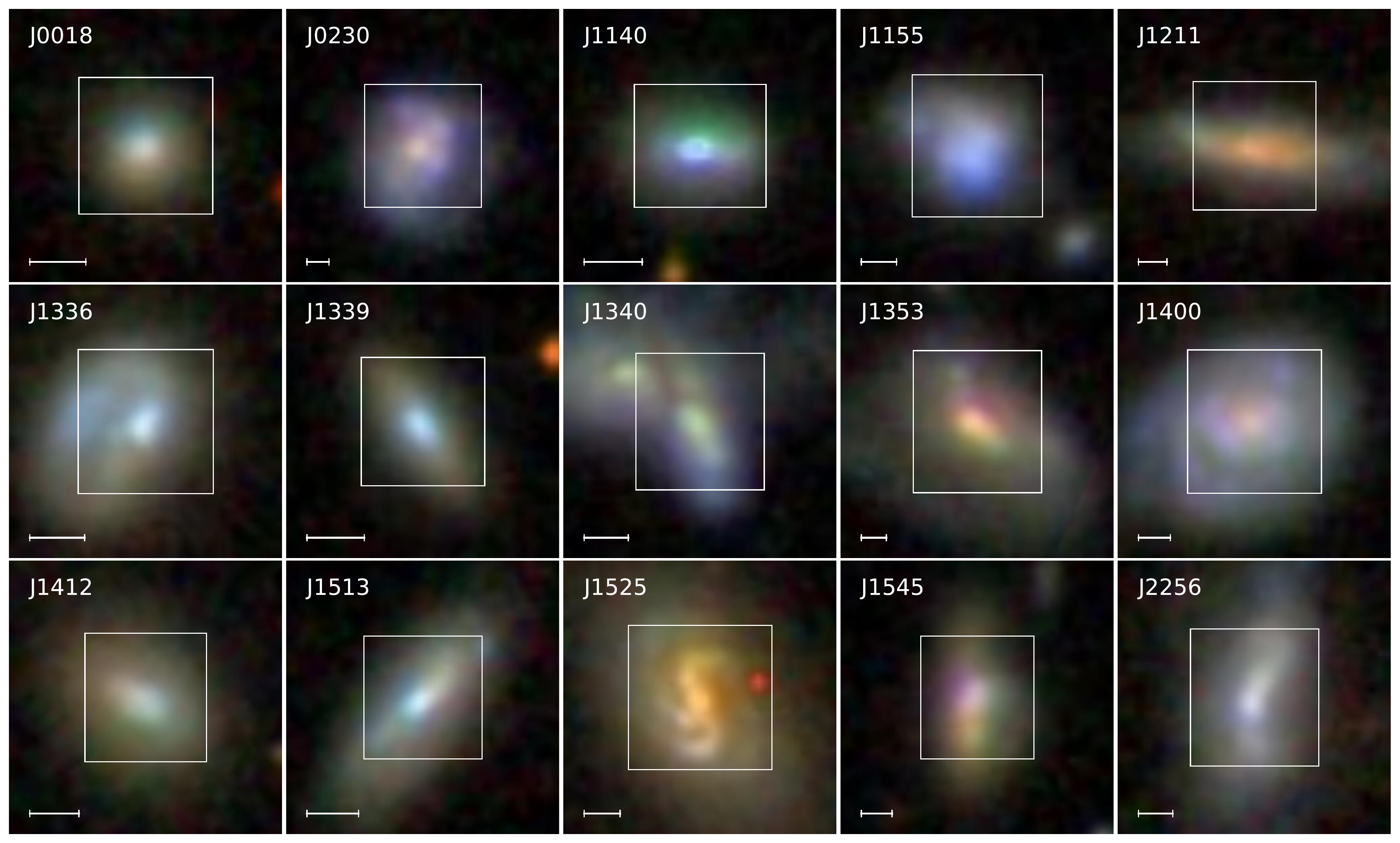}
      \caption{Galaxy sample as seen by SDSS. The images are the SDSS postage stamp images 20\arcsec $\times$ 20\arcsec\ in size. The squared boxes on the images show the field of view of our SINFONI observations to show what part of the galaxies is covered by the SINFONI observations. The horizontal line in the  bottom of each panel shows a  2 kpc scaling bar.}
    \label{fig:SDSS_images}
\end{figure*}

Due to their high star formation rates, starburst galaxies are much more promising targets to measure the effect of stellar feedback. However, starburst galaxies have very strong line  and nebular continuum emission, making the detection of the underlying stellar absorption  much harder. For stellar absorption the calcium-triplet lines are among the most used lines as they are typically strong, even in a young star-bursting galaxy.  Observations of several starburst galaxies as well as ULIRGs show that the spatially-averaged velocity dispersion measured from the nebular emission lines and (mostly optical) stellar absorptions are similar \citep[e.g.][]{Ostlin04, Colina05, Marquart07, Ostlin15}. The spatially-resolved data shows more discrepancies, mostly linked to features related to stellar feedback. \citet{Ostlin04} found differences in the spatially-resolved velocity field of the blue compact galaxy  ESO 400-G43, which they attribute to the presence of an outflow.  Similar results are found for the blue compact galaxies ESO 338-IG04 and Haro 11 \citep{Cumming08,Ostlin15}. In the latter galaxy, the measured irregularities in the velocity maps support the fact that this galaxy is undergoing a merger. In the case of ESO 338-IG04 an outflow is observed in the spatially resolved emission line kinematics \citep{Bik15,Bik18}.

On the other hand in the star-bursting galaxy He 2-10, a significant difference between the stellar and gaseous kinematics is found. \citet{Marquart07} found differences in both the velocity field, suggesting the presence of an outflow, and the velocity dispersion in the central areas of the galaxy. The authors interpreted the kinematic decoupling between the gas and the stars in the core as a sign of the transformation of He 2-10 to a dwarf elliptical galaxy. More recent observations with MUSE by \citet{Cresci17} confirmedt this result and show that the stellar and gaseous kinematics are largely decoupled. 

In this paper we explore a method to derive the  kinematics from the near-infrared $K$-band emission of a sample of starburst galaxies. With future space based near- and mid-infrared integral field instruments such as NIRSPEC and MIRI at the James Web Space Telescope (JWST), this wavelength regime becomes much more accessible then currently from the ground and this technique can be applied to fainter and more distant galaxies. The technique uses the \brg\ and \paa\ emission lines for probing the gas kinematics and the CO bandheads at 2.3\micron\ for the stellar kinematics. The CO-bandhead absorption originates in young populations due to red super giants and  in old stellar populations it comes from red-giant branch stars. Especially red super giants result in very deep stellar CO absorption \citep{Leitherer99}, making it easier to detect this absorption even in the presence of strong nebulosity. This technique has been used on the central regions of spiral galaxies \citep{Boker08}, ULIRG galaxies \citep{Colina05} as well as hosts of Active Galactic Nuclei  \citep[AGN, e.g.][]{Riffel15}.  In this paper we apply this technique to starburst galaxies.

This paper is organised as follows; section 2 presents the galaxy sample and the reference samples from the literature. In section 3 we describe the observations, data reduction and extraction of the spectra and emission line properties. Section 4 presents the  results from the comparison between the stellar and gas kinematics in the integrated as well as spatially resolved spectra. Section 5 focuses on the spatially resolved analysis of the emission line maps as they extend significantly further than the continuum emission.  In section 6 we discuss the results and the paper ends with conclusions in section 7.  In this paper we adopt the cosmological parameters from \citet{Planck15}. 

\begin{table*}[!t]
\caption{Reference samples from the literature}
\centering
\begin{tabular}{lrrrl}
\hline\hline 
Comparison  & redshift & log(Mass) & SFR & reference\\
	& &   (\msun) & (\msun\ $yr^{-1})$ & \\
	\hline
This sample & 0.012 < z < 0.06 &  9.1 - 10.55 & 0.3 - 3.2 & --- \\
\hline
LARS	& 0.03 < z < 0.2& 9.24 - 10.97 & 0.6 - 20.6&  \citet{Herenz16,Guaita15,Hayes14}  \\
DYNAMO	&	z $\sim 0.1$ &  9 - 11  & 0.2 - 160& \citet{Green14}\\
LBA 	& z $\sim$ 0.2& 9.1 - 10.9 & 2.5 - 108 & \citet{Goncalves10} \\
MASSIV	& 0.9 < z < 1.6& 9.7  - 11.2 & 6 - 300 & \citet{Epinat12,Contini12}\\
High-z SFR		& 2 < z < 3& 9 - 11.2 & 6 - 190 & \citet{Law09}  \\
SINS		& 1.3 < z < 2.6& 9.3 - 11.0 & 5 - 210&  \citet{Forster09}  \\
\hline
\label{tab:refsamples}
\end{tabular}
\end{table*}

\section{Galaxy sample}

We select the galaxies  from the Sloan Digital Sky Survey \citep[SDSS]{Eisenstein11} data release 12 \citep{Alam15}. They are selected to be star-forming with \halpha\ equivalent widths (EW) of 100 \AA\ or higher and required to be detected by \emph{GALEX}. Galaxies classified in SDSS DR12 as  AGN have been removed from the sample.  Additionally, we applied a constraint on the apparent size of the galaxy in order to fit in the SINFONI field of view. The galaxies are selected to have a \emph{petroR50$\_$r} between 1.5\arcsec\ and 4\arcsec, with \emph{petroR50$\_$r} being the radius containing 50\% of the Petrosian flux of the galaxy. The galaxies are selected in few small redshift ranges  (0.012 $< z <$ 0.032,  0.0355 $< z <$  0.052, 0.056  $< z <$ 0.06), to make sure the CO absorption is not contaminated by strong telluric absorption lines. The brightest galaxies with a 2MASS $K_{\mathrm{s}}$ magnitude above $K_{\mathrm{s}}$  = 15 mag (Vega) and observable from Paranal  between May and September were included in the target list. A total of  15 galaxies are observed, fulfilling these criteria.

Figure \ref{fig:SDSS_images} shows the 20\arcsec\ $\times$ 20\arcsec\ SDSS cutout images of our targets with the observed SINFONI field of view  ($\sim$8\arcsec $\times$ $\sim$8\arcsec) overlaid on top of it.  With our SINFONI observations we recover only the emission from the brighter central regions.  Their morphologies range from disk-like galaxies, to galaxies in state of a merger. Table \ref{tab:sample} and \ref{tab:sfr_mass} present the detailed properties of the observed galaxies.  In order to derive the star formation rate (SFR) and the nebular extinction, E(B-V), of the galaxies, we fitted the [\ion{N}{II}] lines and \halpha\  in the SDSS spectra with three Gaussians simultaneously. The \hbeta\ emission line is fitted with a single Gaussian. We corrected for stellar absorption by simultaneously fitting a broad absorption line in those galaxies where \hbeta\ absorption is present. The \hbeta\ stellar absorption is very weak, and in some galaxies not even detected and does not affect the  extinction derivation significantly. The E(B-V) is derived from the \halpha\ over \hbeta\ ratio, assuming case B with an intrinsic ratio of 2.86 \citep{Osterbrockbook} and the extinction law from \citet{Cardelli89}. The derived E(B-V) values are presented in Table \ref{tab:sfr_mass}. 
 
 The SFR is  derived from the integrated flux of the \halpha\ line using the calibration of \citet{Kennicutt12}.  Figure \ref{fig:SDSS_images}  shows that several galaxies are larger than the  aperture of the SDSS spectroscopic fibres (3\arcsec diameter). The emission-line maps derived in Sect. \ref{sec:cont_snr} show that several galaxies have more extended ionised emission. We calculated an aperture correction for the SDSS spectra to correct for the missing H$\alpha$ flux by calculating the ratio of the \paa\ or \brg\ flux inside the 3\arcsec\ aperture and the total line flux measured in the SINFONI observations. The derived  aperture corrections (between 1 and 1.9) are applied to the H$\alpha$ line fluxes derived from the Gaussian fits to the SDSS spectra. Additionally, we corrected the SFRs for the measured extinction from  the \halpha\ over \hbeta\ ratio. The derive SFRs range from 0.5 to 18 \msun\,yr$^{-1}$.  

The stellar masses of 13 out of 15 galaxies are taken from the SDSS database (J1140 and J1340 did not have these quantities derived). We  used the masses derived using  Principle Component Analysis \citep{Chen12} with the \citet{Maraston11} stellar population models.  The masses are in the range of log(\msun) = 9.1 to 10.55.  Figure \ref{fig:sfr_mass} shows the relation between the SFR and the galaxy mass of our sample. Comparing the location of the galaxies in this diagram with that of the location of the star-formation main-sequence \citep{Chang15} shows that  our galaxies are located above the main sequence and are starbursting, consistent with their selection criteria.

We compare our galaxies with low- and high- redshift samples presented in the literature. We selected studies performed with integral field spectrographs, having the dataset as comparable as possible. From the low redshift studies we compare our sample to the DYNAMO survey \citep{Green14}, the LARS galaxies  \citep{Ostlin14,Herenz16} as well as a sample of Lyman Break Analogues \citep[LBA,][]{Goncalves10}. The high redshift samples we are using for comparison come from the MASSIV survey \citep{Epinat12}, the SINS survey \citep{Forster09} and a sample of z=2-3 star forming galaxies \citep{Law09}. The basic properties of the reference samples as well as our galaxy sample are listed in Tab. \ref{tab:refsamples}. We note that different methods have been used to derive the galaxy properties. 

We plot the galaxies of the reference samples in Fig. \ref{fig:sfr_mass} to make a more direct comparison.  All galaxy samples overlap in mass with our sample, however the high-redshift galaxies have typically more galaxies at  higher mass than our sample. In terms of measured SFR, the high-redshift galaxies show much higher values. This is driven by the fact that the fainter lines from galaxies with lower SFRs are hard to detect with the current generation of telescopes and instruments.   The galaxies in our sample show most similarities to the galaxies studied in the DYNAMO survey \citep{Green14} and the LARS sample \citep{Ostlin14,Herenz16}. The galaxies in this study overlap with the lower-end of the SFR distribution of the two surveys.

\section{Observations and data reduction}

In this section we describe the observations and the data reduction of the SINFONI observations in order to obtain the final data cubes. Additionally, the procedures to measure the stellar and gas kinematics are described.

\subsection{Observations}\label{sec:obs}
The observations of the compact starburst galaxies (Tab. \ref{tab:sample}) were performed using the SINFONI \citep{Eisenhauer03,Bonnet04}  integral field instrument, mounted at UT4 (Yepun) of the VLT at Paranal, Chile. The observations were carried out in service mode between 2016-05-19 and  2016-08-21. The non-AO 0.25\arcsec camera was used in combination with the $K$-band grating, providing a spectral resolution of $R$ = 4490 and an instrument field of view of 8\arcsec $\times$ 8\arcsec.
The detector integration time (DIT) in each of the observations was set to 300 sec and each galaxy was observed with a total exposure time varying from 2400 s to 5400 s (Tab. \ref{tab:obslog}).
The seeing during the observations varied strongly from observations to observation and is listed in Tab. \ref{tab:obslog}.

For each observation we applied a dither pattern with a jitter box between 2\arcsec\ and 4\arcsec\ and eight or nine offsets per OB. No separate sky frames were taken. The dither pattern resulted in a larger field of view than the 8\arcsec\ of one SINFONI pointing. Telluric standards of B spectral type were observed immediately after the science OBs as part of the standard calibration plan. 

\begin{figure}[!t]
   \includegraphics[width=\hsize]{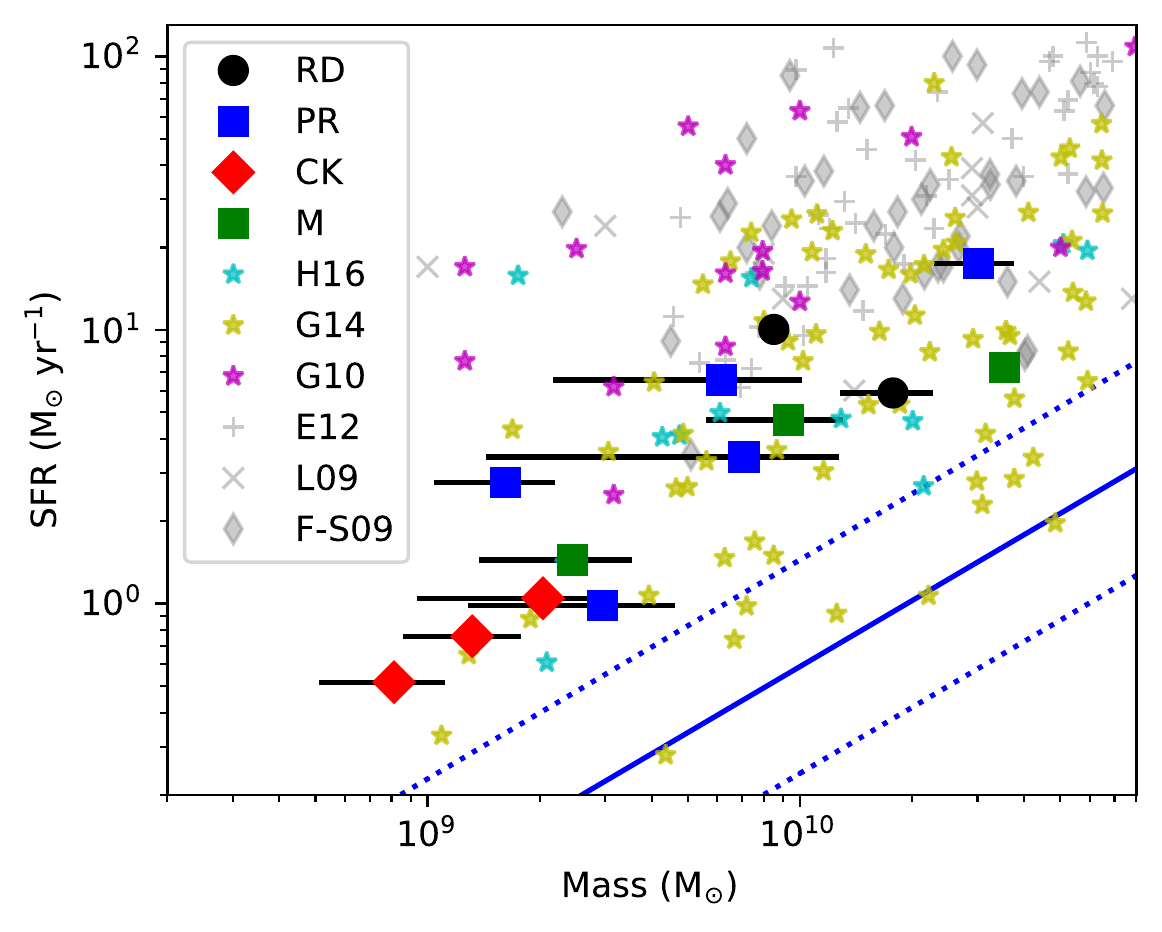}
      \caption{Star formation rate versus galaxy mass diagram. The star formation rates are corrected for extinction derived from \halpha/\hbeta\ and are corrected for the missing flux outside the 3\arcsec SDSS fibre, derived from the emission line maps in our SINFONI data. The symbols represent  their morphological classification introduced in section \ref{subsec:resolved}: (RD, rotating disks: black circles, PR, perturbed rotators: blue squares, CK, complex kinematics: red diamonds, M, mergers: green squares). Over-plotted are sample from the literature (Tab. \ref{tab:refsamples}). The low redshift samples shown are LARS, cyan asterisk \citep[H16]{Herenz16,Hayes14,Guaita15}, LBA analogues; magenta asterisk \citep[G10]{Goncalves10} and the DYNAMO sample, yellow asterisk \citep[G14]{Green14}. The high redshift sample are plotted as grey plus signs for MASSIV \citep[E12]{Epinat12,Contini12}, grey crosses for  \citep[L09]{Law09} and grey diamonds for SINS \citep[F-S09]{Forster09}.
         The blue solid line is the star formation main-sequence derived by \citet{Chang15} with their 1$\sigma$ scatter (0.39 dex) plotted as blue dashed lines.}
    \label{fig:sfr_mass}
\end{figure}

\begin{table}[!h]
\caption{Log of observations. \label{tab:obslog}}
\centering
\begin{tabular}{l|cccc}
\hline\hline 
 
Name &  Observing  & Int. time & Seeing & IQ(Ks)  \\ 
          &  date& (s)	&           (\arcsec)	&  (\arcsec) \\ 
          \hline
 J0018$-$0903	&	2016-07-22	& 2400 & 0.76	&0.69 \\ 
			&	2016-07-22	& 2400 & 0.46	&0.56 \\
 J0230$-$0720	&	2016-08-21	& 2700 & 0.56	& 0.58 \\
 J1140$-$0024	&	2016-05-23	& 2400 & 0.96	& 0.82 \\
 J1155$-$0100	&	2016-05-24	& 2400 & 0.47	&  0.57 \\
 J1211$+$1424	&	2016-07-13	& 2400 & 1.24	& 1.10 \\
 J1336$+$0748	&	2016-05-23	& 2400 & 0.95 & 0.81 \\
 			&	2016-07-06	& 2400 & 1.20	& 0.99 \\
 J1339$-$0045	&	2016-07-06	& 2400 & 	1.44 &  1.04 \\
 J1340$+$0616	&	2016-07-05	& 2400 & 1.58	&  1.14 \\
 J1353$-$0231	&	2016-05-23	& 2400 &	0.55 &  0.61 \\
 			&	2016-07-06	& 2400 & 	1.49& 1.21 \\
 J1400$+$0404	&	2016-07-16	& 2700 & 	1.33& 1.02 \\
 			& 	2016-08-11	& 2700 & 	1.63& 1.30 \\
 J1412$-$0049	&	2016-07-16	& 2700 &	0.84&  0.77\\
 			& 	2016-07-29	& 2700 &	0.85& 0.73\\
 J1513$+$0431	&	2016-07-02	& 2700 & 	0.92& 0.79\\
 			&	2016-07-02	& 2700 & 0.69& 0.70\\
 J1525$+$0500	&	2016-06-08	& 2400 & 1.35 	&1.03\\
 J1545$+$0814	&	2016-05-19	& 2400 & 1.83	& 1.33\\
 J2256$+$1305	&	2016-05-21	& 2700 & 1.30	&  1.16\\
\hline
\end{tabular}
\end{table}

\subsection{Data reduction}\label{sec:reduction}

The SINFONI data are reduced with the ESO pipeline version version 3.1.1 using ESOREX version 3.13.1 together with custom written IDL routines. The SINFONI pipeline is used for applying all the basic steps of the data reduction, such as dark removal, flat field and distortion correction,  cosmic ray removal, and wavelength calibration. As no individual sky frames were taken, we developed a custom procedure to construct master sky frames. Each science frame is reduced and individually reconstructed to a 3D cube with the ESO pipeline. The reconstruction is done \emph{without} subtracting the sky and combining the individual exposures.

\begin{figure*}
\begin{center}
   \includegraphics[width=\hsize]{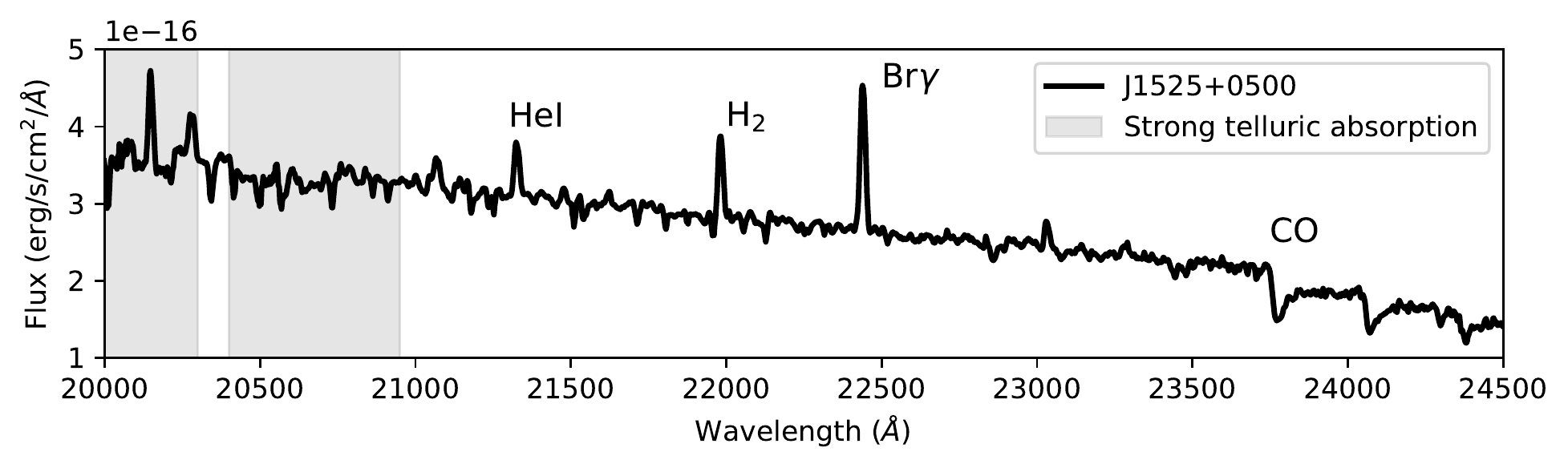}
      \caption{Integrated K-band spectrum of J1525+0500 smooth with a Gaussian kernel of 3 pixels. The most important emission and absorption lines  are highlighted. The grey shaded areas mark the wavelength range with very strong telluric absorption.
    \label{fig:spectraJ1525}}
\end{center}
\end{figure*}

\begin{table*}[!ht]
\caption{Kinematic measurements integrated spectra. \label{tab:fitresults}}
\centering
\begin{tabular}{l|rr|rr|rr}
\hline\hline 
 
Name &  v$_{\mathrm{stars}}$ &$\sigma_{\mathrm{tot,stars}}$ & v$_{Pa\alpha}$ &$\sigma_{tot,Pa\alpha}$  & v$_{Br\gamma}$ &$\sigma_{tot,Br\gamma}$  \\
	&  (\kms) & (\kms)    &  (\kms) & (\kms) &  (\kms) & (\kms)\\
          \hline
 J0018$-$0903 & $-7.7^{3.2}_{3.4}$ &$ 50.4^{7.4}_{7.7} $& ---	 &---	  & -2.2  $\pm$ 1.7 & 43.6  $\pm$ 1.7 \\
 J0230$-$0720 & $5.2^{10.0}_{10.9} $& $27.6^{24.5}_{27.3}$ & -17.5 $\pm$ 0.5 & 53.0 $\pm$ 0.5 & -30.8  $\pm$ 2.4 & 28.0  $\pm$ 2.2 \\
 J1140$-$0024 & $-14.8^{17.2}_{9.4} $& $52.9^{22.4}_{32.4} $& ---	  &--- & 20.6  $\pm$1.6 & 42.6  $\pm$ 1.5 \\
 J1155$-$0100 & $-30.2^{30.2}_{22.7} $& $74.4^{47.8}_{69.9} $& -3.6 $\pm$ 0.6 & 50.8 $\pm$ 0.6 & -13.5  $\pm$ 2.3 & 48.2  $\pm$ 2.3\\
 J1211$+$1424 &$ 61.6^{30.2}_{23.2} $& $176.5^{18.7}_{26.7}$ & 70.5 $\pm$ 1.7 & 116.1 $\pm$ 1.7 & 82.8  $\pm$ 4.3 & 117.2  $\pm$ 4.5 \\
 J1336$+$0748 &$ 3.3^{4.2}_{5.3}$ & $43.4^{24.8}_{12.7}$ & --- &  --- & 11.8  $\pm$ 1.6 & 43.6  $\pm$ 1.6 \\
 J1339$-$0045 &$ -12.8^{29.2}_{27.0}$ & $59.3^{86.5}_{19.3}$ & --- &--- & 3.2  $\pm$ 2.4 & 34.5  $\pm$ 2.3 \\
 J1340$+$0616 &$ -23.1^{39.6}_{20.3}$ & $68.2^{27.9}_{37.8}$ & --- & --- & -27.7  $\pm$ 1.5 & 60.7  $\pm$ 1.5 \\
 J1353$-$0231 & $-82.8^{16.1}_{5.1} $&$ 91.7^{8.6}_{6.5}$ & -7.1 $\pm$ 4.2 & 149.5 $\pm$ 4.5 & -8.8  $\pm$ 6.1 & 150.5  $\pm$ 6.7 \\
J1353 blue & --- & ---           & -128.0 $\pm$ 3.7  & 79.9 $\pm$ 2.2& -116.7  $\pm$ 8.7 &88.1  $\pm$   5.4          \\
J1353 red  & --- & ---          &  79.8 $\pm$ 2.6 & 66.8 $\pm$ 2.3  & 83.4 $\pm$  5.2& 59.4  $\pm$ 4.6\\
 J1400$+$0404 & $-82.6^{8.1}_{8.0} $& $224.8^{7.9}_{9.1} $& -16.5 $\pm$ 8.7 & 68.8 $\pm$ 8.7 & -16.1  $\pm$ 2.3 & 70.5  $\pm$ 2.3 \\
 J1412$-$0049 & $-22.7^{18.7}_{30.1}$ &$ 69.7^{38.7}_{35.5}$ & --- & --- & -0.3  $\pm$ 1.9 & 40.9  $\pm$ 1.8 \\
 J1513$+$0431 & $42.6^{5.7}_{5.3}$ &$ 62.1^{9.9}_{6.0} $& --- & --- & 0.7  $\pm$1.1 & 50.3  $\pm$ 1.1 \\
 J1525$+$0500 & $38.5^{4.3}_{3.5}$ & $91.6^{5.2}_{4.7}$ & --- & ---         & 6.5  $\pm$ 2.1 & 102.0  $\pm$ 2.2 \\
 J1545$+$0814 & $5.7^{24.6}_{6.5}$ & $20.8^{45.8}_{17.2} $& 12.3 $\pm$ 1.3 & 91.6 $\pm$ 1.3 & 7.8  $\pm$ 5.4 & 93.2 $\pm$ 5.6 \\
J1545 blue & --- & ---           & -49.2 $\pm$ 0.5 & 51.7 $\pm$  0.9 & --- & ---           \\
J1545 red  & --- & ---          &  70.5 $\pm$ 0.7 &  50.9 $\pm$ 0.8 & --- & --- \\
 J2256$+$1305 & $4.2^{21.9}_{18.3}$ & $88.4^{16.2}_{20.8}$ & --- & ---	 & -12.4  $\pm$ 2.0 & 48.0  $\pm$ 1.9 \\
\hline\hline
\end{tabular}
\end{table*}

The individual cubes in each OB are  median averaged (without correcting for the dither offset) to construct an average sky frame. Before averaging the cubes, we corrected for differences in wavelength solution due to flexure of the instrument by cross-correlating the OH emission lines from each frame. When applying the median averaging of all the frames in to a single sky cube we reject the four brightest values for each pixel to make sure the emission of the galaxy is removed when averaging. This works well in most cases, only if the galaxy is extended some residuals are remaining in the final cube.

Six galaxies are at a  redshift high enough to  have the \paa\ emission line shifted in the $K$-band. As this emission line is very bright, it is typically also much more spatially extended than the continuum or \brg\ emission of the galaxy, which results in residual emission in the master sky cubes. To remove this residual emission, we cut out the wavelength range effected by the \paa\ line and replaced that with the sky emission taken from the standard star observation which does not have any \paa\ emission.  In order to account for atmospheric variations of the sky emission, the sky spectrum of the standard star frame is fitted to the master sky spectrum allowing for variations in intensity, background level and wavelength. This recovered some of the \paa\ emission, however faint over subtraction of  \paa\ emission remains in the science frames, affecting especially the faint outskirts of the extended emission.

After these operations, the master sky frame was subtracted from the individual science cubes allowing for variations in the OH emission line spectra using the procedure of \citet{Davies07}. Finally, the individual sky corrected cubes are combined to the final data cube using the \emph{sinfo\_utl\_cube\_combine} procedure of the SINFONI pipeline. Before combining the cubes, the spatial mean of each cube plane was subtracted to remove sky background residuals. 

The standard-star observations were fully reduced  with the SINFONI pipeline. Their spectra, extracted from the data cube with a radius of 10 pixels, are used for the removal of the telluric absorption lines and flux calibration. For telluric absorption line correction, the \brg\ and sometimes \hei\ absorption lines have to be removed first. Identical to the procedure in \citet{Bik10}, this is done in two steps. First the standard star spectrum is corrected for telluric absorption by a high signal-to-noise atmospheric spectrum\footnote{Obtained from the SINFONI website and obtained at NSO/Kitt Peak Observatory}. This is not taken at the same atmospheric conditions, but  provides a first rudimentary correction in order to properly remove the \brg\ and \hei\ absorption lines. The \brg\ and \hei\ lines are then removed by fitting a Lorentzian profile to the cleaned spectrum. The standard-star spectrum with the stellar absorption lines removed is then used to correct for the telluric absorption in the science data. The flux calibration is done by  using the 2MASS \citep{Cohen03,Skrutskie06} magnitudes of the standard star.

Some galaxies are observed twice (Tab. \ref{tab:obslog}) and a combined cube is created out of the two cubes form the individual OBs.  The cubes are corrected for the difference in their heliocentric velocity and spatially aligned by fitting a two dimensional Sers\'ic function to the galaxy 
image in Qfitsview, after which their offset is corrected and they are combined to a final data cube.

Finally, the observations are converted  to the heliocentric rest frame by using the heliocentric velocities calculated by the SINFONI pipeline. In order to check the wavelength solution of our final data cubes, we processed a wavelength-calibration frame in the same way as the science frames and fitted the profile of five bright, isolated lines in the arc spectrum. This revealed a velocity shift of $\sim$30 \kms\ for each of the observations, and  $\sigma$ = 34 \kms, which is adopted as the $\sigma_{instr}$ of the instrumental profile. The derived velocities are corrected for the 30 \kms\ shift measured in the arc spectra. 

The cosmic-ray removal in the pipeline is not perfect and as a final step in the data reduction we removed the cosmic rays in the final reduced cubes by hand in the wavelength planes used to construct the line and continuum maps in this paper.
 We used the iraf task \emph{imedit} to mark the cosmic rays in each plane and replaced the value of the affected pixel with the mean of the neighbouring pixels.

\subsection{Extraction of the integrated spectra}\label{sec:spec_extract}

In order to extract the integrated $K$-band spectra of the galaxies, we first isolated the area of the cube where continuum flux from the galaxy is present.
The cubes were convolved  with the 2MASS \Ks\ transmission curve and  a pseudo \Ks\ image is constructed for each galaxy. We defined a background region in the \Ks\  image and the pixels which have flux  3$\sigma$ above the background  are selected to contain emission of the galaxy. After this we constructed a mask to remove the background pixels from the image and apply this mask to the data cube. This procedure could remove line emission when it is more extended than the Ks continuum. In Sect. \ref{sec:integrated_kin} we show that this does not affect the integrated gas kinematics. 

The total flux of the galaxy  is extracted with each spaxel weighted by its relative contribution to the galaxies emission using the pseudo \Ks\ image. This gives less weight to the low signal-to-noise ratio (S/N) regions of the galaxy, increasing the S/N of the final spectrum. We constructed this 3D weighting function as follows, following partly the optimal extraction recipe of \citet{Horne86}. For each spatial pixel containing galaxy flux, an initial linear fit is made to the spectrum. We subtracted this linear fit so the outliers and emission lines can be easier removed. We removed them by five times iteratively clipping away 3$\sigma$ deviations, resulting in the effective removal of the emission lines and other outliers.  After that the initial fit is added again resulting in a cleaned spectrum with only continuum (and absorption lines). 

We then performed a linear fit to the cleaned spectrum in order to get the weighting function. Applying this procedure to each pixel provides us with a 'noise-free' image of the galaxy at each wavelength bin. Normalisation of these image at each wavelength results in the weighting function for the extraction of the integrated spectrum.  Such a 3D analysis naturally also takes into account possible displacements as function of wavelength due to flexure or other distortions. All the spectra are extracted using this scheme. This weighting results in significant reduction of sky line residuals and increases the S/N of the spectrum.  The observed S/N per spectral pixel varies from S/N = 8 for J1339 to S/N = 50 for J1525  (Fig. \ref{fig:spectraJ1525}). All galaxies show emission lines from Hydrogen (\brg\ and \paa) and most galaxies also  H$_{2}$ and \ion{He}{I}. All galaxies show at least 1 CO absorption band. For the higher redshift galaxies, one or more bands are shifted beyond 24600 \AA, at which the SINFONI observations stop. For the high S/N spectra, also some  stellar atomic lines (\ion{Na}{i}, \ion{Ca}{i}) are seen in absorption. Of the entire sample, six galaxies show \paa\ emission.

The absolute flux calibration of data  is derived from the observations of the standard star. This assumes that the atmospheric conditions were identical between the two observations. Calibrating one standard star observation with another standard star observation shows deviations of a factor of two in flux. This could be due to bad atmospheric conditions. Additionally, calibrating the observed flux levels  to the 2MASS magnitude of the galaxies (Tab. \ref{tab:sample}) is difficult due to the small field of view of SINFONI compared to the aperture used in 2MASS (radius of 4\arcsec). This does not affect the main analysis in the paper; since it is focussed on the kinematics, it only affects the derivation of the infrared extinction.

\subsection{Spatially resolved emission and absorption maps.}\label{sec:cont_snr}

For the spatial resolved analysis we created emission-line maps  by numerically integrating under the emission line. The continuum is subtracted by interpolating  the continuum on the blue and red side of the emission line.  After that, we applied a weighted Voronoi-tessellation  binning algorithm by \citet{Diehl06}, which is a generalisation of the \citet{Cappellari03} algorithm, to obtain the required S/N = 30  per cell needed for the analysis of the resolved emission lines. The Voronoi binning pattern is applied to the data cube for further analysis.

For the spatially-resolved maps of the absorption lines we constructed a continuum S/N map by measuring the S/N for each pixel between $\lambda_{rest} \sim $2.18 and 2.2 \micron, over 100 wavelength elements. This S/N map is used as input for the weighted Voronoi tessellation algorithm to create a pattern to achieve a S/N of 25 per cell for the continuum analysis. 

\section{Integrated infrared spectra}

\subsection{Infrared star formation rates}
By comparing the total flux in \brg\ and \paa\ to that of \halpha\ we can derive an additional measure of the extinction. As \brg\ and \paa\ are at longer wavelengths and less sensitive to extinction, there lines can probe gas at much higher extinction than probed by \hbeta\ and \halpha.  Especially in star-forming galaxies containing large amounts of dust, this can result in  much higher extinction \citep[e.g.][]{Lopez16,Cleri22}. 

We measured the total \brg\  and \paa\ emission line flux in the spatially resolved emission maps. For the galaxies where \paa\ is detected we derived the E(B-V) and SFR using the \paa\ line, for the other galaxies we used the \brg\ line. Using an intrinsic \brg\ over \halpha\ ratio of 0.00973 and \paa\ over \halpha = 0.1182  \citep{Osterbrockbook}, the extinction law of \citet{Cardelli89} we derived the E(B-V) values from \brg\ and \paa\ using the extinction coefficients calculated with \citet{Barbary16}. Additionally, we derived the extinction-corrected SFRs from the \brg\ (\paa) emission using the calibration of  \citet{Kennicutt12}, taking into account the intrinsic \brg\ (\paa) over \halpha\ ratio. The results are shown in Table. \ref{tab:sfr_mass}, together with the results derived from the optical emission lines.
 
 Comparison between the optical SFR and the infrared SFR shows indeed six galaxies have an infrared SFR which is larger than the optically derived SFR, suggesting the presence of optically-obscured gas. Especially the merging galaxy J1353 is an extreme example, where the infrared SFR is more than three times higher than the optical SFR. The merger in J1353 could have initiated an starburst which is still partially embedded. Additionally, a clumpy geometry of the dust could result in a higher infrared E(B-V) as well \citep{Natta84,Calzetti94}
 
Four galaxies show similar SFR values (within 0.5 \msun/yr) and five galaxies have optical SFRs higher than the infrared SFRs.
For those galaxies where both the \paa\ and \brg\ lines are present, typically the \paa\ lines results in a higher extinction. As described in Sect. \ref{sec:spec_extract}, the uncertainties in the flux calibration make the derivation of the infrared extinction much less reliable than the optical extinction. These low infrared values could therefore be the result of this uncertain calibration. Even though some galaxies show a higher infrared SFR, revealing optically-obscured star formation, the flux calibration of the SINFONI spectra are not very reliable. Therefore, for the remaining of the analysis we use the optical SFRs. This makes it also easier to compare with other samples in the literature where the star formation is typically derived from \halpha.

\begin{figure*}[!t]

\begin{center}
   \includegraphics[width=0.9\hsize]{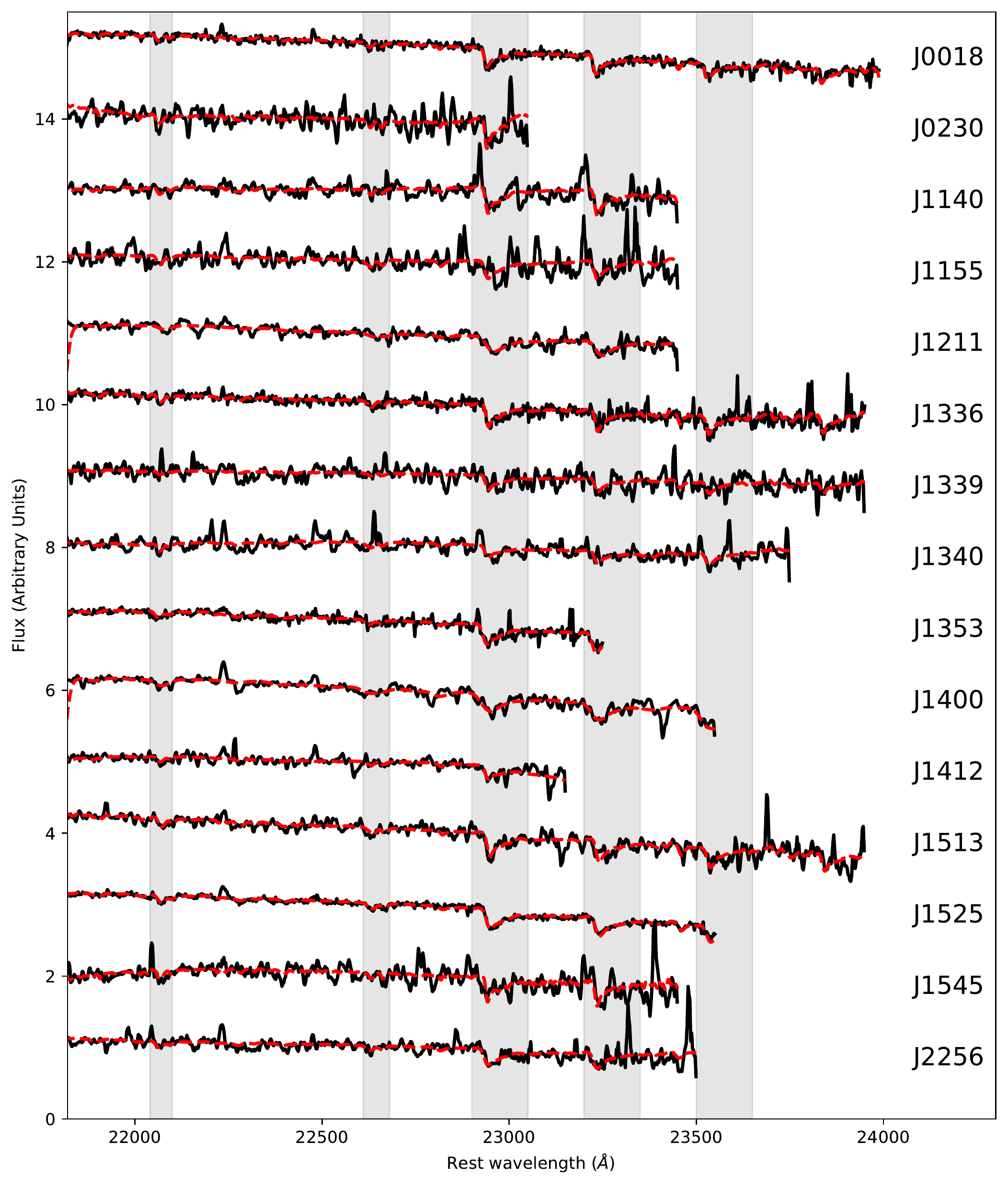}
      \caption{Integrated $K$-band spectra, corrected for their redshift between 2.18 and 2.4 \micron, the wavelength range over which the pPXF fit was performed. The spectra are divided by the flux scaling factor determined by pPXF to match the template spectra to the observed spectra. This approximately scales the spectra by the median flux. For display purposes the spectra are shifted after that by adding 1 between each spectrum.
        No strong emission lines are present in this wavelength range.  The spectra are smoothed with a 3 pixel boxcar filter. Only the spectra of J0018, J1336 and J1525 are not smoothed. The red-dashed lines are the best pPXF fits. Highlighted in grey are the strongest absorption bands in this part of the spectrum: \ion{Na}{I}, $\lambda_{\mathrm{rest}}$ = 2.206 \micron, \ion{Ca}{I}, $\lambda_{\mathrm{rest}}$ = 2.26 \micron, and the CO overtone absorption bands red wards of 2.29 \micron\ \citep{Rayner09}.}
    \label{fig:spectrawithfits}
\end{center}
\end{figure*}

\subsection{Comparing gas and stellar kinematics}\label{sec:integrated_kin}

To derive the stellar kinematics from the absorption lines, we made use of the Penalised Pixel-Fitting procedure \citep[pPXF,][]{Cappellari04,Cappellari17}. As templates in the pPXF fitting we used the  spectral library of \citet{Winge09}. This is a library of cool, mostly giant stars (spectral type F to M), showing CO bandheads in absorption observed with GNIRS \citep{Elias98} at Gemini South and NIFS \citep{McGregor02} at Gemini North with a spectra resolution of R = 5300 - 5900, higher than our SINFONI observations.  For this study we only used the GNIRS sample observed with both the blue and the red setting (2.15 - 2.43 \micron), rebinned to a common grid of 1\AA/pixel. We fitted the spectra redwards of $\lambda_{\mathrm{rest}}$=21800 \AA, covering absorption lines sensitive to cool stars (\ion{Na}{I}, $\lambda_{\mathrm{rest}}$ = 2.206 \micron, \ion{Ca}{I}, $\lambda_{\mathrm{rest}}$ = 2.26 \micron, \ion{Mg}{I}, $\lambda_{\mathrm{rest}}$ = 2.28 \micron) and the CO overtone absorption bands redwards of 2.29 \micron\ \citep{Rayner09}. This wavelength range is free from strong emission lines.

During the fitting process, pPXF constructs linear combinations of the template spectra in order to reproduce the observed spectra. The fit takes into account the instrumental broadening of both the reference spectra and the science spectra.
The errors on the derived velocity and velocity dispersion are derived by Monte-Carlo simulations by running the pPXF fitting 1000 times while varying the input spectra randomly based on the measured S/N of the spectrum. The derived errors correspond to the 16\% and 84\% values of the resulting distributions for the velocity and $\sigma$. 

 \citet{Winge09} extensively discussed the effect of the shape of the CO absorption in the template stars on the derived velocity dispersion. They performed a pPXF fit to a galaxy spectrum for each template star in their sample. The found a spread of several 10s of \kms, especially for small velocity dispersion, close to the resolution limit. They did not find a relation between the EW(CO) of the template star and the value of the velocity dispersion, as found by \citet{Wallace97}, but concluded that the deviations are caused by the fact that one template fits the science spectrum better than other templates. They concluded that the most reliable kinematics is derived from using the entire sample of reference stars. Some tests with our objects shows the same results where fits to small sub-samples of the reference stars result in differences of several 10s of \kms, but also a higher $\chi^2$ for the pPXF fit. Fig. \ref{fig:spectrawithfits} shows the integrated K-band spectra corrected for redshift, redwards of 21800 \AA\  of all the galaxies. Over-plotted is the best pPXF fit. 

To determine the integrated  kinematics of the ionised gas ($v$ and $\sigma_{tot}$), we fitted the \paa\ and \brg\ emission lines with a Gaussian profile and quadratically removed the instrumental broadening. For targets with a redshift below z=0.036 the \paa\ line is not observed and only the \brg\ is fitted. The errors on the $v$ and  $\sigma_{tot}$ are taken from the fitting routine. The derived velocities and velocity dispersions are listed in Tab. \ref{tab:fitresults}. 
We note that the values for the integrated velocity dispersions measured this way \citep[$\sigma_{\mathrm{tot}}$, e.g.][]{Herenz16} cannot directly be compared with the flux-weighted velocity dispersion from spatially-resolved emission lines  \citep[$\sigma_m$ or $\sigma_0$][]{Glazebrook13,Herenz16},  as in the latter one the systematic velocity patterns are being removed.  In Sect. \ref{sec:resolved} we derive the flux-weighted velocity dispersion for the galaxies.

\begin{figure}[!t]
	   \includegraphics[width=0.49\linewidth]{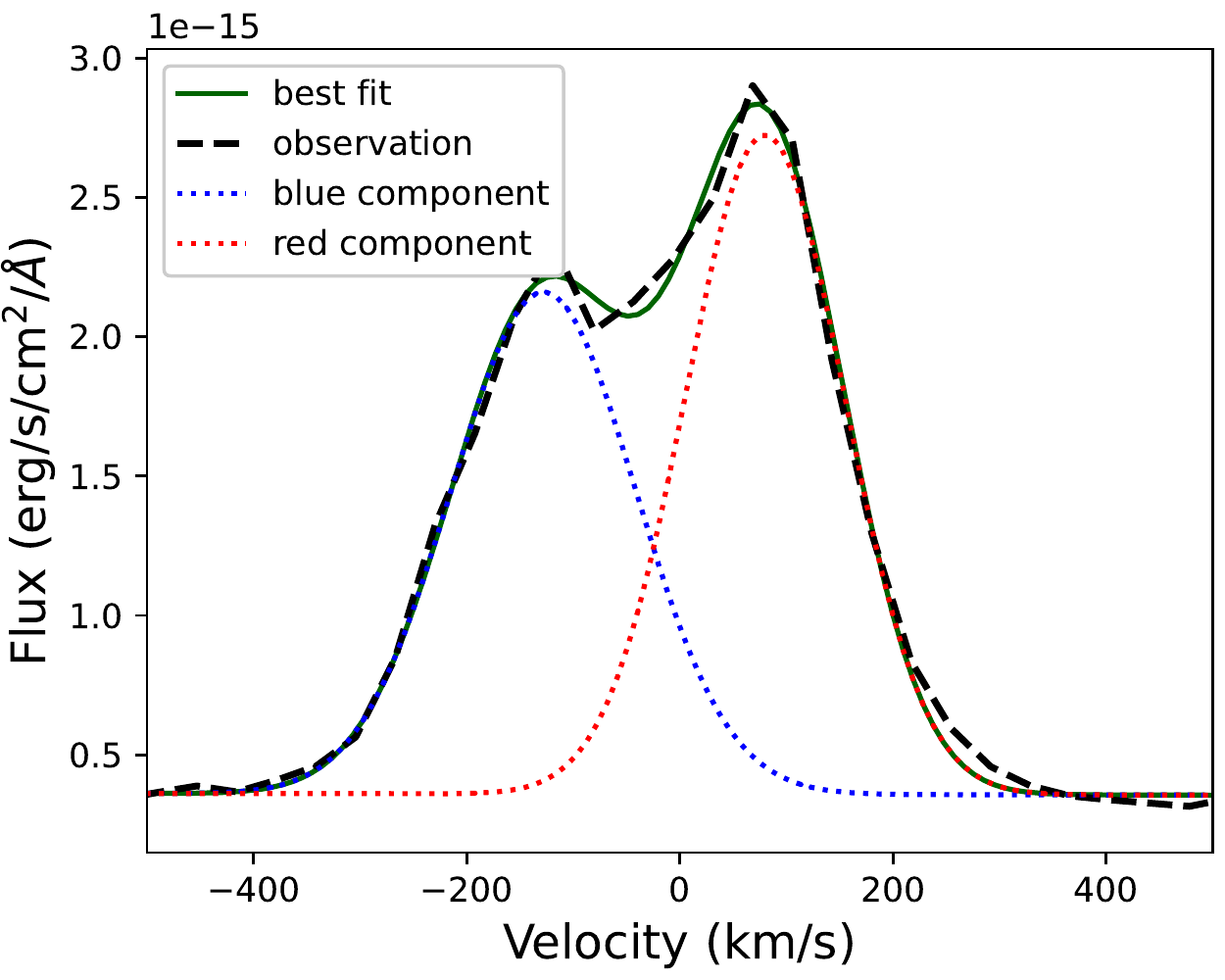}     
   \includegraphics[width=0.49\linewidth]{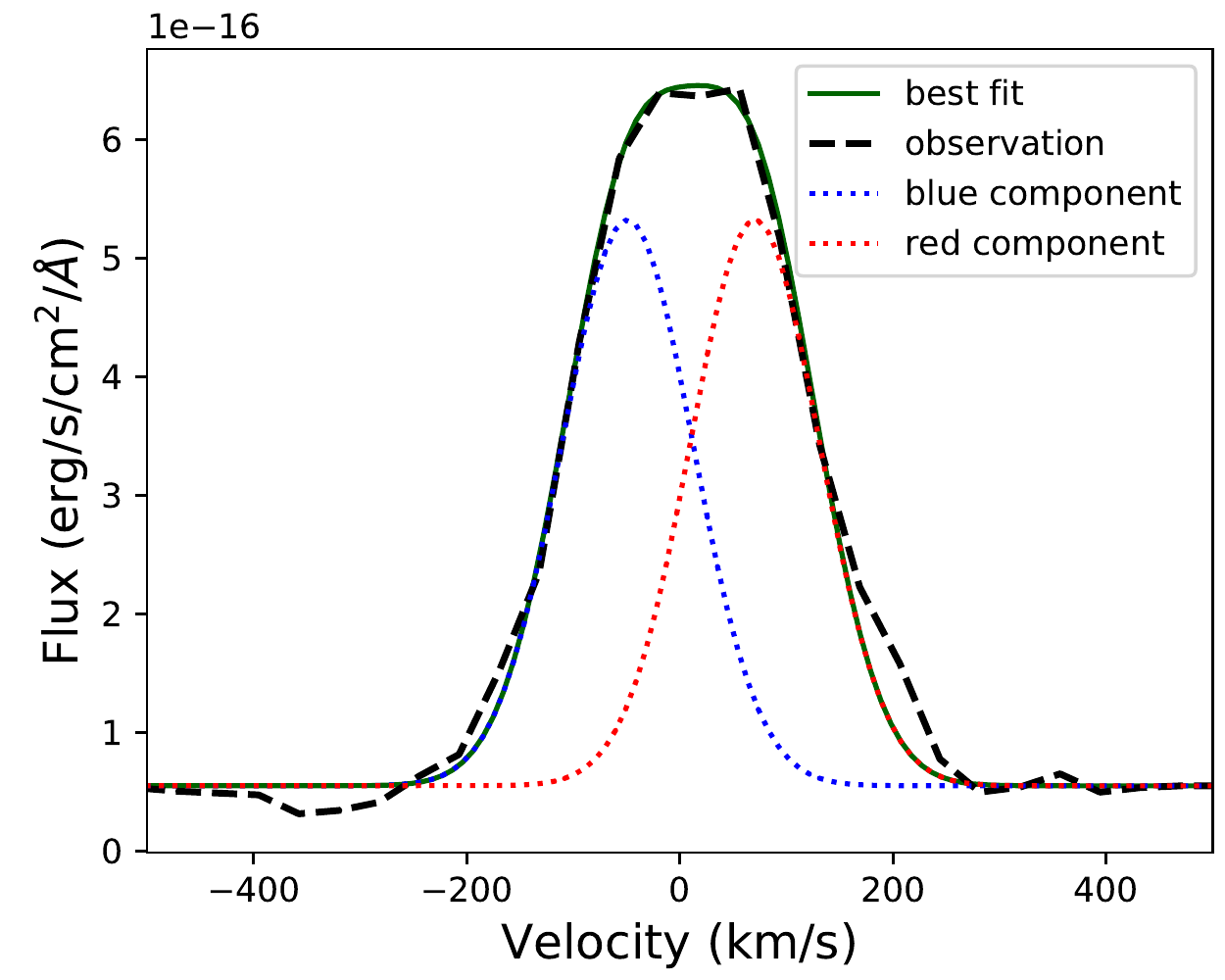}    
   \caption{	\paa\ emission line of the integrated spectra of J1353 (left) and J1545 (right). Over-plotted is the best fit with two Gaussian components.\label{fig:doublepeaked}}
\end{figure}

\begin{figure}[!t]
   \includegraphics[width=\hsize]{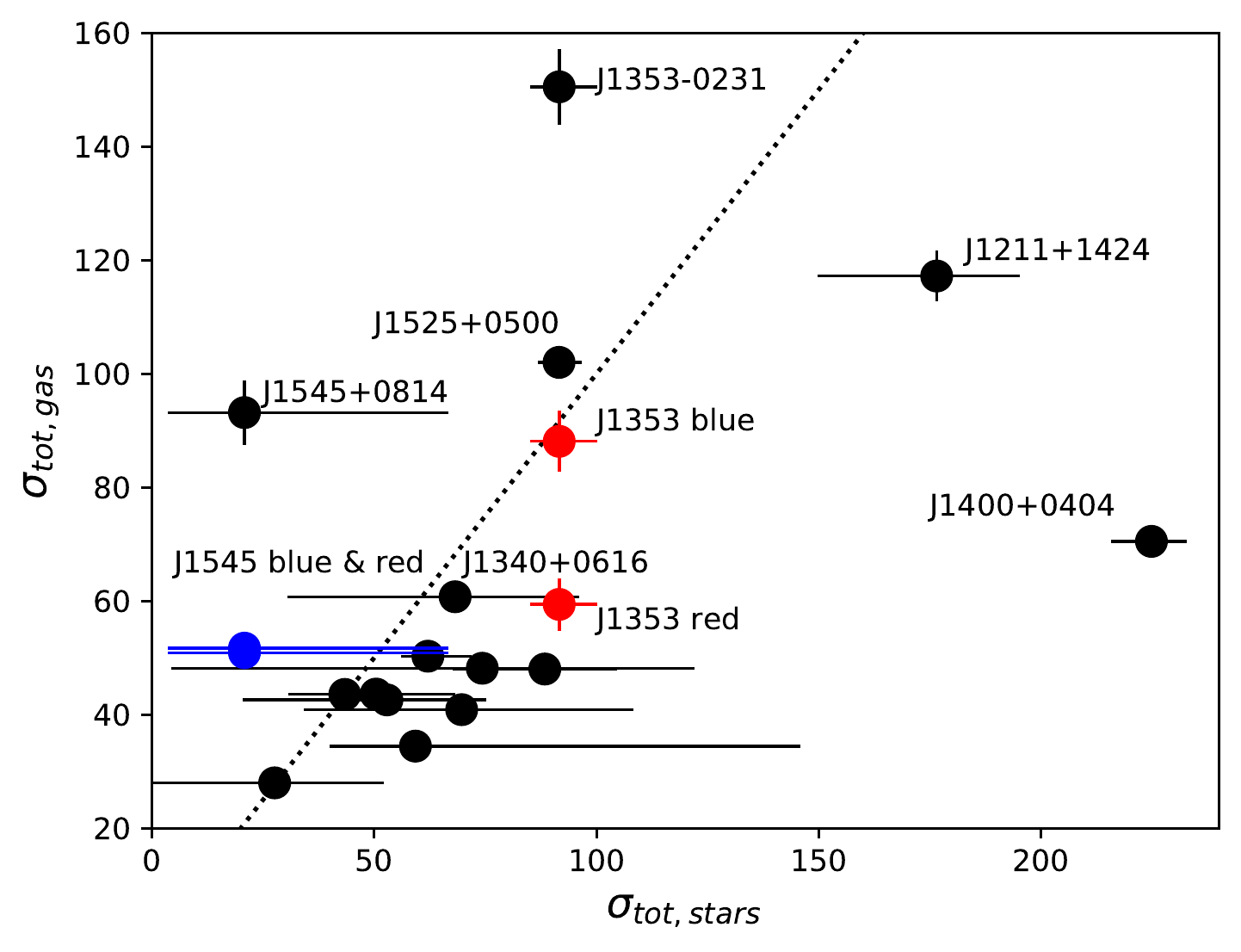}
      \caption{Relation between $\sigma_{tot,stellar}$ and  $\sigma_{tot,Br\gamma}$ (Table \ref{tab:fitresults}). The red points are the blue and redshifted components of J1353, the blue points those of J1545. The black dotted line shows the line where both $\sigma$ values are equal.}
    \label{fig:sigmaplot}
\end{figure}

Two galaxies show double-peaked emission lines: J1353 and J1545 (Fig. \ref{fig:doublepeaked}). For these galaxies we fitted the emission lines with a double Gaussian profile and  report the central velocity and dispersion for the two components in Table. \ref{tab:fitresults}. The two components of J1353 (Fig \ref{fig:doublepeaked}, left panel) are separated by 208 \kms, while the separation for J1545 (right panel) is only 120 \kms. This makes the fitting of the latter galaxy more challenging and we fix the peak flux ratio to 1 in order to obtained a reasonable fit to the line profile. The \brg\ line is  too low S/N to fit a double Gaussian profile reliably.

Comparison  between the line-of-sight velocities of \paa\ and \brg\ from the same objects, measured in the same apertures, shows differences larger than the fitting errors, with values up to 13.3 \kms\ for J0230 (Tab. \ref{tab:fitresults}).
 Inspecting the arc frames we found that there is slight increase in velocity of the fitted arc lines with decreasing wavelength. For J0230 we measured a difference of 6 \kms\ between the arc close to \paa\ and close to \brg\ and  this does not explain the measured differences in velocity between.  J0230 is the only galaxy which shows a difference between the $\sigma_{tot}$ measurements of \paa\ (53 \kms) and \brg\ (28 \kms). For all the other galaxies, the velocity dispersion of \paa\ and \brg\ are in good agreement with each other.  Inspection of \paa\ emission line map and spatially resolved kinematics (Fig. \ref{fig:2D_J0230}) suggests that the most extreme velocities (responsible for the broader emission wings) are at lower surface brightness. As \brg\ is intrinsically much fainter, this emission is not observed in \brg, resulting in a narrower line profile. This could also be the cause of the observed velocity shift, where more extreme velocities at low surface brightness would change the measured velocity of \paa\ with respect to \brg. In our 1D comparison between the gas and stellar kinematics we only use the \brg\ line, as that is observed in each galaxy, while the \paa\ line is only seen in 6 galaxies. 

As discussed in Sect. \ref{sec:spec_extract} the spectra are extracted using the continuum as weighting function. In order to check whether a different weighting would result in different measurements, we also extracted the spectra by using the \brg\ or \paa\ spatial distribution as weighting function. We repeated the fitting of the absorption and emission lines on these spectra. The emission line fits resulted in very similar values with differences less than 10 \kms\ in measured velocity and 4 \kms\ in velocity dispersion. This  demonstrates that our method to calculate the integrated spectrum by using the continuum shape as weighting function does not affect the emission line kinematics.

In Fig. \ref{fig:sigmaplot} we compare the stellar velocity dispersion ($\sigma_{tot,stars}$) and the ionised-zgas velocity dispersion  ($\sigma_{tot,gas}$) as measured from the integrated \brg\ emission line. Most galaxies (ten out of  15) have modest values of $\sigma_{tot}$ ($\sigma_{tot,gas} <$ 70 \kms), and their stellar and gas values are within the errors consistent with each other, they all lie within 1.5 sigma of the one-to-one line. The errors on the stellar measurement are larger than those of the gas emission as the S/N of the continuum is much lower than that of the emission lines and the lower $\sigma$ values are close or below the instrumental resolution.

The remaining five galaxies show a very large spread in $\sigma_{tot,star}$ with the J1400 to be the highest with  $\sim$200 \kms.  In these galaxies, also the difference between  $\sigma_{tot,star}$  and $\sigma_{tot,gas}$  is much larger than the other ten galaxies. As described above, J1353 and J1545 have double peaked profiles and their single Gaussian values are over estimating the $\sigma_{tot,gas}$. The stellar velocity dispersion of J1545 is poorly constrained due to the faint CO bandheads, but much smaller than  $\sigma_{tot,gas}$ of the single component fit. 

The result of the two component fits for J1353 (red points) and J1545 (blue points) show much lower values for  $\sigma_{tot,gas}$, making them more compatible with the measured $\sigma_{tot,star}$. For J1545 we find similar values for both components, while for J1353 the blue component is about twice as broad as the red component. The blue component shows a value similar to the stellar velocity dispersion.  This would suggest that the stellar kinematics is dominated by the emission of the blue component.  In section \ref{sec:doublepeaked} we discuss the nature of the two components in more detail.

J1525 is, within two sigma consistent with the one-to-one line.  Both J1211 and J1400 show a much larger $\sigma_{tot,star}$ compared to their gas measurements. This suggests that the gas and stellar kinematics are not coupled to each other. If the continuum light of the two galaxies is dominated by the presence of a bulge, a higher   $\sigma_{tot,stars}$  could be expected \citep[e.g.][]{Oh20}.

\begin{figure*}[!t]
   \includegraphics[width=\hsize]{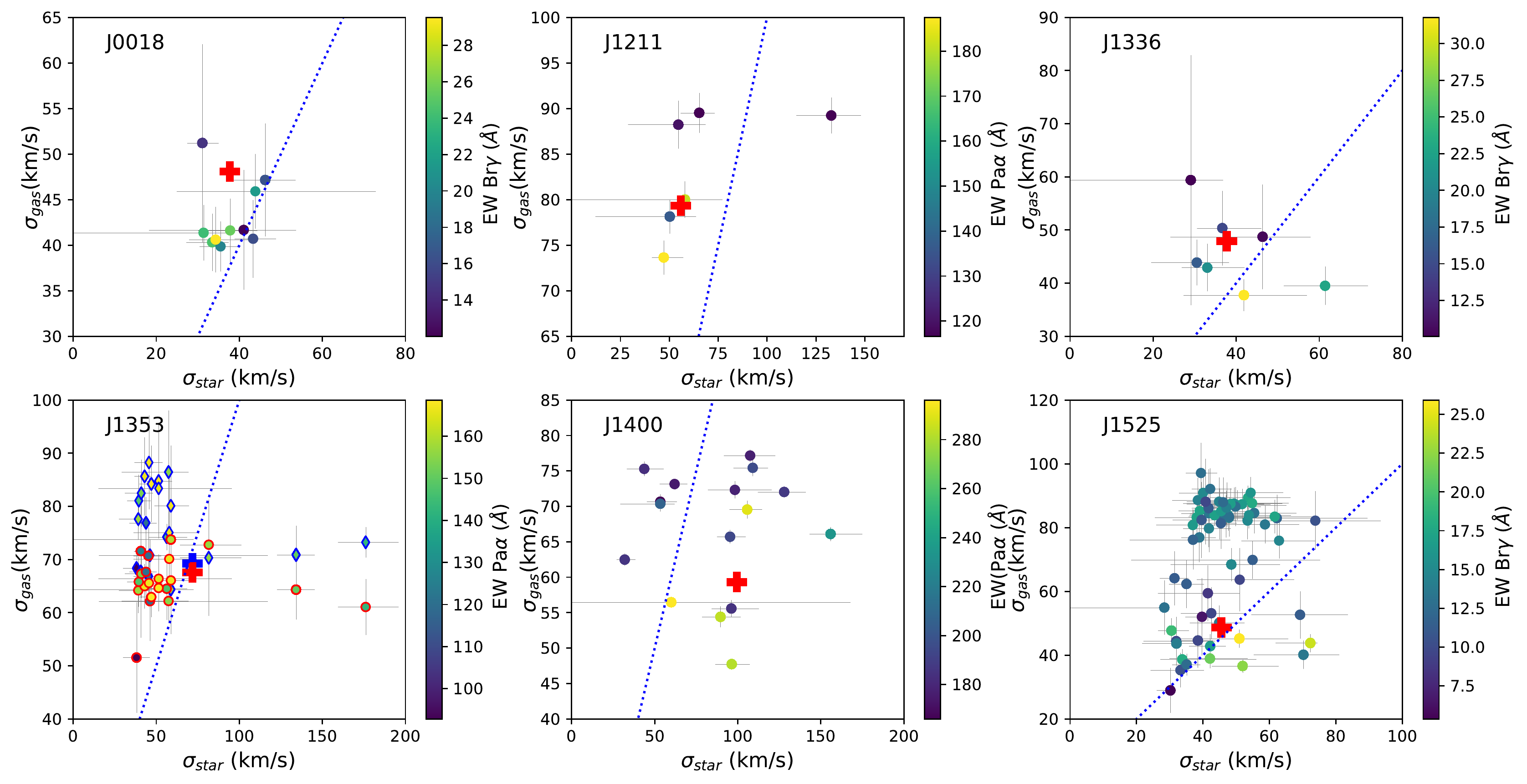}
      \caption{Velocity dispersion of the gas ($\sigma_{gas}$) versus that of the stars  ($\sigma_{stars}$)  of six galaxies with sufficient continuum S/N. The blue dashed line is the one-to-one line. Each data point represents a Voronoi cell. The colour coding represents the Equivalent Width of the \paa\ or \brg\ emission line in the Voronoi cell. For J1353 the blue-shifted nebular emission plotted as diamond shapes with a blue edge and the redshifted emission with circles with a red edge. The plus signs represent the relation between the averaged $\sigma_{gas}$ and $\sigma_{star}$.
          \label{fig:v_rms}}
\end{figure*}

\subsection{Spatially resolved kinematics}\label{sec:resolved_kinematics}
A spatially resolved comparison of the stellar and gas kinematics sheds more light on the differences found in the 1D kinematic analysis.  
Based on the continuum S/N map (Sect. \ref{sec:cont_snr}) of the galaxy we created a Voronoi pattern with a  minimum S/N = 25 and a maximum area of 2.5 $\square$\arcsec (10 $\square$ pixels), resulting in a minimum spatial resolution of 1.3 - 2.8 kpc. This Voronoi pattern is applied to the entire cube to have the same Voronoi pattern for the continuum and the emission line fits. Even though the S/N limit was set to 25, typically the cells scatter around that value, therefore we select the Voronoi cells with a S/N = 20 or higher for further analysis, the lower S/N cells are masked out before the fitting.  In order to get the stellar kinematics we fitted the spectrum of the high S/N cells using pPXF in the same way as done for the integrated spectrum. For each Voronoi cell, the errors are calculated using Monte-Carlo simulations by running pPXF 1000 times, varying the input spectrum randomly based on the measured S/N map of the stellar continuum. For further analysis we selected only those galaxies that have at least 5 Voronoi bins with S/N = 20 or higher, resulting in spatially resolved information of six out of the 15 galaxies.  The other nine galaxies have very faint continuum emission, resulting in only a few, or even no bins with the required S/N to fit the absorption lines. The gas kinematics are extracted using a Gaussian fit to the emission line spectrum of the same Voronoi cells.  For the more distant galaxies in our sample (z $>$ 0.036), we used the \paa\ emission line for analysis as \brg\ is becoming fainter and more difficult to observe. For each Voronoi cell we derived the velocity ($v$) and velocity dispersion ($\sigma$). 

 Figure \ref{fig:v_rms} shows the comparison between the stellar and gas velocity dispersion for  six galaxies where we have a sufficient number of Voronoi cells. Each point in the diagram represents the values in a Voronoi cell with continuum S/N above 20.  Galaxies where the stellar and gas kinematics are identical are located on the blue dotted lines in the graphs. Two of our galaxies show this behaviour; J0018 and J1336 do not show any measurable difference between the stellar and gas kinematics, the data points are at maximum 2 $\sigma$ away from the blue dotted line. 

The other four galaxies on the other hand, show large differences between  $\sigma_{gas}$  and  $\sigma_{star}$.  In J1400, which looks like a  spiral galaxies viewed almost face on, 
 we find that average the stars have a larger velocity dispersion than the gas. This could be caused by  asymmetric drift, where the velocity dispersion of the stars is increased due to random motions, resulting in a larger velocity dispersion than that of the gas \citep{Oh22}.

\begin{figure}[!t]
   \includegraphics[width=\hsize]{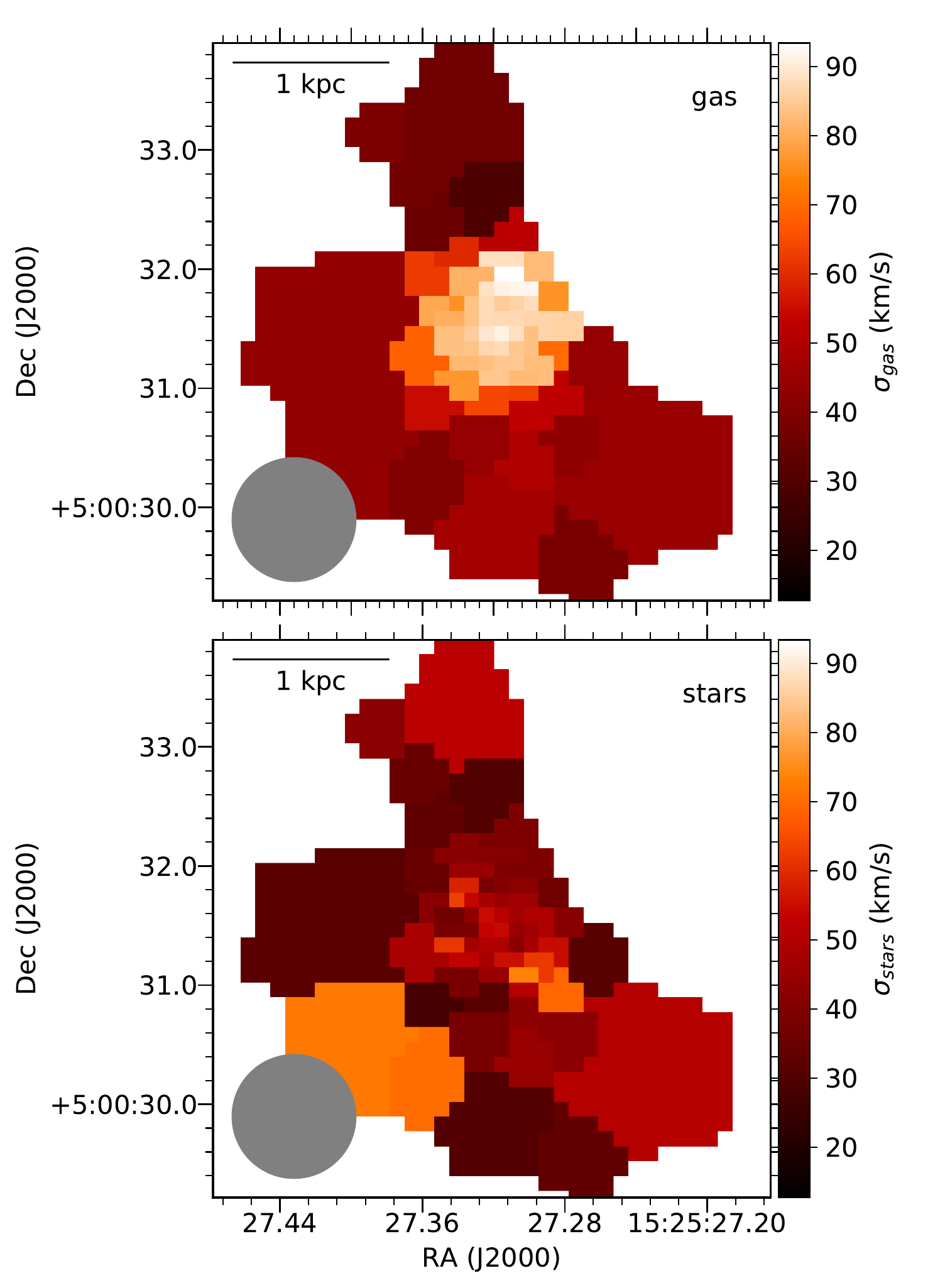}
      \caption{Velocity dispersion ($\sigma$) maps from the gas (top) and stars (bottom) for J1525+0500, showing a clear difference in the central area of the galaxy between the stellar and gas velocity dispersion. The grey circle represents the FWHM of the observed PSF as derived in Sect. \ref{subsec:resolved}. The 1 kpc scale bar at the top is derived using the observed redshift and the cosmological parameters of \citet{Planck15}.
          \label{fig:sigma_J1525}}
          \end{figure}

In J1525, we found that some of the Voronoi cells have a significant difference between the gas and stellar dispersion. Fig. \ref{fig:sigma_J1525} shows a spatially resolved map of the nebular (top) and stellar (bottom) velocity dispersion.  These maps show that the nebular velocity dispersion is enhanced in the central knot of the galaxy with values up to 90 \kms. This increase is not present in the stellar velocity map, where values of 50-60 \kms\ are measured in the same area.  The spatially-resolved emission line map of \paa\ (Fig. \ref{fig:mapJ1525}, see also Sect. \ref{sec:resolved}) shows that this area has the highest \paa\ flux. This suggest that this region is strongly star forming and that the increase in $\sigma_{gas}$ may be caused by turbulence induced by stellar feedback in this compact region. 

The galaxy J1353 has double peaked emission lines. We fitted the emission line data with a double Gaussian and derived the dispersion for both the blue- and red- component. For the stars we used the single component output from pPXF. In Fig. \ref{fig:v_rms} we show the relation between the dispersion of both the blue and the red component as a function of the total stellar velocity dispersion. The blue emission component is slightly broader then the red component. Due to the fact that we cannot disentangle the two components in the stellar kinematics a  comparison between the gas and stellar velocity dispersion becomes difficult. The stellar velocity dispersion shows a large increase from $\sim$50 \kms\ for most voronoi cells to above 100 for a few cells. These cells are located in the central area of the galaxy, where in the gas both components are equally strong.

In J1211, which does not show double-peaked emission, we also find that $\sigma_{star}$ is larger than the $\sigma_{gas}$ towards the central area of the galaxy. The rest of the data points in J1211 follow roughly the one-to-one line, The SDSS image stamp in  Fig. \ref{fig:SDSS_images} of J1211  suggests that this galaxy is a spiral viewed under moderate inclination. The central voronoi cell  is located towards the  bulge of the galaxy.  An AGN origin would be excluded as the gas does not show the high $\sigma$ typically seen in AGN emission lines.

The fact that  the central area of the galaxy has a much larger stellar velocity dispersion than the gas can be explained by the presence of a galaxy bulge, where the gas and the stars are decoupled.
Bulges are commonly observed in spiral galaxies \citep[e.g.][]{Kormendy04} and their kinematics are dominated by turbulent motions, in contrast to the disks of the spiral galaxies where rotational motion dominates \citep{Kormendy04,Falcon16}. This results in an increased stellar velocity dispersion. Bulges which consist of older stellar populations are likely gas free and therefore bulges become visible when observing the kinematics of the stellar component, while the gas  traces that the kinematics of the underlying disk component or halo around the galaxy \citep[e.g.][]{Oh20}.

\begin{figure*}[!ht]
   \includegraphics[width=\hsize]{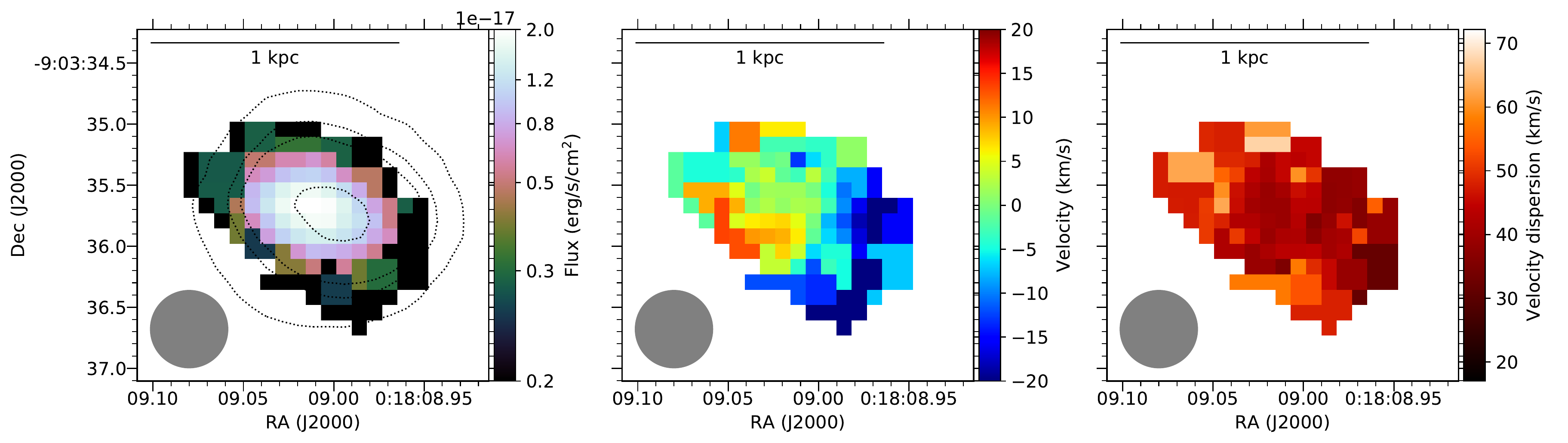}
      \caption{Spatially resolved maps of the \brg\ emission line of J0018-0903. \emph{Left:} \brg\ intensity map with as contours the \Ks\ band continuum flux created by convolving the datacube with the 2MASS \Ks\ response curve. The grey circle represents the FWHM of the observed PSF as derived in Sect. \ref{subsec:resolved}. The 1 kpc scale bar at the top is derived using the observed redshift and the cosmological parameters of  \citet{Planck15}. 
      \emph{Middle:} \brg\ velocity map constructed from a single Gaussian fit. \emph{Right:} Observed velocity dispersion map corrected for the instrumental resolution. This map is not corrected for beam smearing (Sect. \ref{subsec:resolved}). As the beam smearing correction only has a minor effect on the velocity dispersion maps we have chosen to show the observed maps before correction.
    \label{fig:2D_J0018}}
\end{figure*}

\begin{figure*}[!ht]
   \includegraphics[width=\hsize]{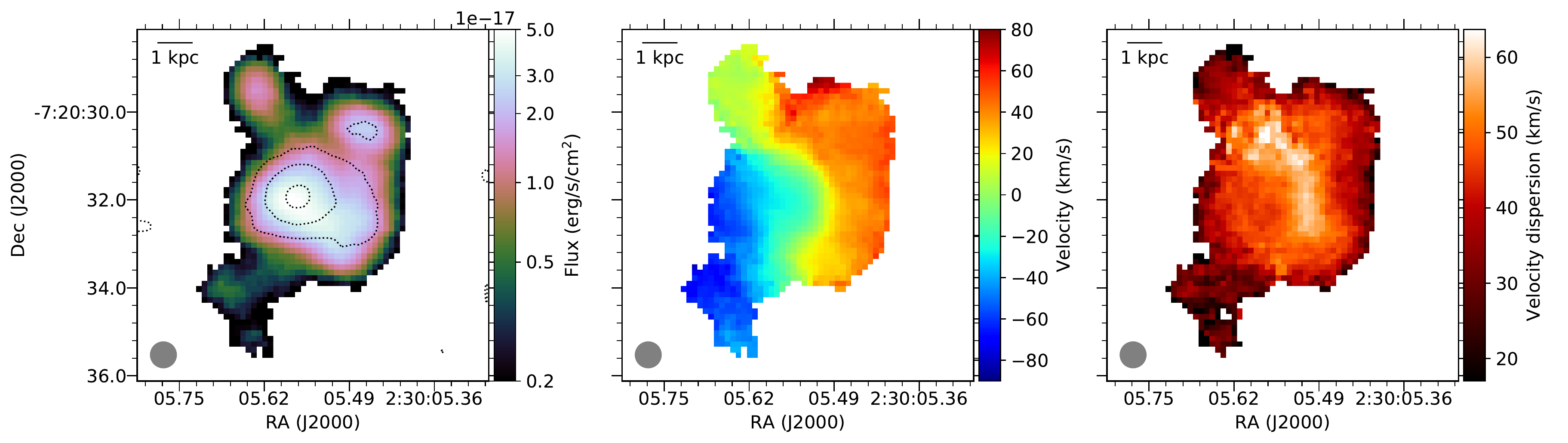}
      \caption{Spatially resolved maps of the \paa\ line of J0230-0720. See caption of Fig. \ref{fig:2D_J0018}.
    \label{fig:2D_J0230}}
\end{figure*}

\section{Spatially resolved emission line kinematics}\label{sec:resolved}

The emission lines are observed at high enough S/N such that we can construct spatially-resolved maps of the emission as well as velocity and velocity dispersion of all the galaxies, much further out and in much more detail than the stellar continuum. For the galaxies where \paa\ is visible we use the \paa\ emission line, for the others the \brg\ emission line. We use a minimum S/N of 30 and a maximum bin size of 25 pixels (1.6$\square\arcsec$) for the Voronoi tessellation of the emission-line map.  We used this rather high S/N to avoid the reduction artefacts  becoming dominant.
 The binning pattern is applied to the cube and a Gaussian is fitted to the emission lines, resulting in a velocity and velocity dispersion map. The maps are corrected for the offset in velocity found in Sect. \ref{sec:reduction} as well as the instrumental resolution ($\sigma_{instr}$). Integration of the flux under the emission line profile and subtracting the continuum results in the emission line intensity map. 

Two examples of the resulting maps are shown in Fig. \ref{fig:2D_J0018} and \ref{fig:2D_J0230} for J0018-0903 and J0230-0720 respectively. The left panel shows the intensity map of the \brg\ or \paa\ line with over-plotted in contours the pseudo-\Ks\ image discussed in Sect. \ref{sec:spec_extract}. The middle and right panel show the maps of the velocity and velocity dispersion respectively. The same figures for the other 13 galaxies are shown in Appendix \ref{sec:appendixA}. We find a large difference in morphology and velocity and dispersion patterns. Several galaxies show clear signs of rotation (e.g. J1211 and J1353), while others seem to be dominated by more turbulent motions or other complex kinematics (e.g. J1140  or J1155).

In this section we  used the effective radius and position angle  derived with the exponential profile fit (e.g. \verb|expRad_r|) in the SDSS r-band from  the \emph{PhotoObj} table. 
We chose the exponential fit over the de Vaucouleurs profile as the latter is applicable to elliptical galaxies, and non of our galaxies is classified as an elliptical galaxy.  To  estimate the inclination ($i$) of the galaxy, we used the thin-disk approximation and the ratio of the minor axis and major axis (\verb|expAB_r|) with the corresponding error from the SDSS table \emph{PhotoObj}. The parameters, including the applied correction factor $f_{corr} = \frac{1}{\sin(i)}$, are listed in Tab. \ref{tab:gal_prop}.

\subsection{Spatially resolved emission line kinematics}\label{subsec:resolved}

Based on the spatially resolved emission line and kinematics maps extracted from the SINFONI cubes, we calculated the flux-weighted velocity dispersion, a measure for the random velocities in the galaxy: 
\begin{equation}
\sigma_{\mathrm{m}} =  \frac{\Sigma_i \sigma_i  f_i}{\Sigma_i f_i}, \label{eq:sigma}
\end{equation}
where $\sigma_i$ is the measured velocity dispersion in each pixel in the dispersion map and $f_{i}$ represents the flux in that pixel. This value is lower than the $\sigma_{tot}$ calculated in Sect \ref{sec:integrated_kin} as $\sigma_{tot}$ is the $\sigma$ of the integrated-line profile, while $\sigma_{\mathrm{m}}$ is the average of the $\sigma$s at each pixel, hence  taking out the systemic motions in the galaxy. Therefore, $\sigma_{\mathrm{m}}$ is a much more reliable measure of the turbulence of the gas in a galaxy than  $\sigma_{tot}$. 

 The shearing velocity ($v_{shear}$) is a measure of the large scale gas bulk motion along the line of sight.  Following the approach of \citet{Herenz16}, we calculated this as follows:
 \begin{equation}
 v_{shear} = \frac{1}{2}(v_{max} - v_{min}).
 \end{equation}
 Following \citet{Herenz16} we calculated v$_{max}$ and v$_{min}$ by taking the median of the upper and lower fifth percentile of the distribution of values in the velocity maps with a S/N of 5 or more. This approach prevents single pixel outliers from dominating the derived  $v_{shear}$. The resulting values for our galaxies are listed in Tab. \ref{tab:2D_vel}. 

Due to the limited spatial resolution, velocity variations on scales less than a resolution elements  result in artificial broadening of the emission line (beam smearing). In order to investigate whether beam smearing significantly increases the observed $\sigma_{m}$ values of the emission-line maps, we calculated the correction using the procedure outlined by \citet{Green10}.  First, resample the flux and velocity maps to a five times higher spatial resolution using linear interpolation. Second, create a high resolution data cube with a Gaussian at each pixel, of which the central velocity is taken from the high resolution velocity map and the flux from the flux map. The velocity dispersion is the instrumental dispersion (34 \kms, Sect. \ref{sec:reduction}). Each wavelength plane in the high resolution cube is convolved with a Gaussian with the 
image quality (IQ)  of the observation, and binned back to the original resolution. Finally, a Gaussian fit is applied to that cube in order to derive the correction map, which is then quadratically subtracted from the observed $\sigma$ map.

\begin{table*}[!t]
\caption{Galaxy properties derived from SDSS imaging and the SINFONI emission line maps.}
\centering
\begin{tabular}{lrrrrrrr}
\hline\hline 
Name &  $i$  & P.A. & f$_{corr}$ & \verb|expRad_r| &$r_{\ion{H}{II}}$  & $r_{\Sigma_{SFR}}$ \\
	& $\degr$ & $\degr$ &     &  (kpc) &(kpc)  &(kpc) \\
          \hline
 J0018$-$0903 & 10.6 $\pm$ 4.3  & 133.4 & 5.5 $\pm$ 2.2 &  0.53  $\pm$ 0.01 &  0.53 $\pm$ 0.05 & 0.69 \\
 J0230$-$0720 & 44.2 $\pm$ 0.5 & 154.4 & 1.43 $\pm$ 0.01 & 2.65 $\pm$ 0.02& 4.48 $\pm$  0.4  & 2.44\\
 J1140$-$0024 & 62.2 $\pm$ 0.3 & 88.8 & 1.13 $\pm$ 0.003 &  0.73 $\pm$ 0.01 & 0.56 $\pm$  0.05 & 0.85\\
 J1155$-$0100 & 26.3 $\pm$ 1.0 & 166.3 & 2.26 $\pm$ 0.08 &  1.31 $\pm$ 0.01 & 1.94 $\pm$  0.2  & 2.76 \\
 J1211$+$1424 & 71.8 $\pm$ 0.2 & 85.5 & 1.05 $\pm$ 0.001 &  3.96$\pm$ 0.04 & 2.79 $\pm$  0.3 & 2.88 \\
 J1336$+$0748 & 57.1 $\pm$ 0.3 & 150.6 & 1.19 $\pm$ 0.004 &  1.03$\pm$ 0.01 & 0.94 $\pm$ 0.09 & 1.19\\
 J1339$-$0045 & 62.7 $\pm$ 0.3 & 33.5 & 1.26 $\pm$ 0.003 &  0.90$\pm$ 0.01 & 0.52 $\pm$ 0.05  & 0.48\\ 
 J1340$+$0616 & 67.4 $\pm$ 0.3 & 19.1 & 1.08 $\pm$ 0.002 &  1.53 $\pm$ 0.01 & 0.87 $\pm$ 0.09 & 1.22\\
 J1353$-$0231 & 60.6 $\pm$ 0.3 & 54.1 & 1.14 $\pm$ 0.004 &  4.88  $\pm$ 0.02 & 2.40 $\pm$ 0.2 & 2.82\\
 J1400$+$0404 &  35.7 $\pm$ 0.5 & 108.9 & 1.71 $\pm$ 0.02 &  2.63 $\pm$ 0.01 & 2.84 $\pm$  0.3 & 3.64 \\
 J1412$-$0049 &  48.2 $\pm$ 0.4 & 54.3 & 1.34 $\pm$ 0.009 &  1.39 $\pm$ 0.01 & 0.83 $\pm$ 0.08 & 0.79\\
 J1513$+$0431 & 68.4 $\pm$ 0.2 & 139.1 & 1.08 $\pm$ 0.001 &  1.58 $\pm$ 0.01  & 0.85 $\pm$  0.08  & 0.80\\
 J1525$+$0500 &  57.1 $\pm$ 0.2 & 17.6  & 1.19 $\pm$ 0.003 & 3.12 $\pm$ 0.02 &  2.74 $\pm$ 0.3 & 2.80\\
 J1545$+$0814 &  64.5 $\pm$ 0.3 & 176.5 & 1.11 $\pm$ 0.003 &  2.09 $\pm$ 0.02 & 2.13 $\pm$  0.1 & 1.67 \\
 J2256$+$1305 &   67.7 $\pm$ 0.2 & 167.0 & 1.08 $\pm$ 0.001 &  2.93 $\pm$ 0.01 & 1.04 $\pm$ 0.1  & 1.54\\
\hline
\label{tab:gal_prop}
\end{tabular}
\end{table*}

\begin{table*}[!ht]
\caption{Spatially resolved measurements}
\centering
\begin{tabular}{llrrrrrr}
\hline\hline 
 
Name &  Class\tablefootmark{a}  & v$_{\mathrm{shear}}$ & v$_{\mathrm{rot}}$ & $\sigma_m$ & v$_{\mathrm{shear}}$/$\sigma_{m}$ & $\sigma_{m,corr}$ & v$_{\mathrm{rot}}$/$\sigma_{m,corr}$\\
	&  & (\kms) & (\kms)    &  (\kms) &  \\
          \hline
 J0018$-$0903 & CK &18.2 $\pm$ 1.6 & 99.6 $\pm$ 41.2 & 44.5 $\pm$ 6.6 & 0.41 $\pm$ 0.07  & 43.0 $\pm$ 6.5 & 2.31 $\pm$ 1.02\\
 J0230$-$0720 & PR &61.4 $\pm$ 0.3 & 88.1 $\pm$  1.0 & 46.5 $\pm$ 3.6 & 1.32 $\pm$ 0.10 & 44.6 $\pm$   3.5 & 1.98 $\pm$  0.16\\
 J1140$-$0024 & CK &18.4 $\pm$ 2.0 & 	20.8 $\pm$ 2.3		& 43.3 $\pm$ 9.4 & 0.43 $\pm$ 0.10& 43.1  $\pm$  9.4 &  0.48  $\pm$  0.12\\
 J1155$-$0100 & PR &55.7 $\pm$ 0.5 & 	125.8 $\pm$  4.7		&44.7 $\pm$ 4.5 & 1.25 $\pm$ 0.12 & 43.8  $\pm$  4.4  & 2.87 $\pm$ 0.31\\
 J1211$+$1424 &RD& 132.6 $\pm$ 0.6 & 	139.6 $\pm$  0.6		&75.2 $\pm$ 7.8 & 1.76 $\pm$ 0.18&  69.2  $\pm$ 7.2  & 2.01  $\pm$ 0.21 \\
 J1336$+$0748 & CK* &30.3 $\pm$ 1.7 & 36.0 $\pm$  2.0		&42.1 $\pm$ 8.3 & 0.72 $\pm$ 0.15 &  41.2 $\pm$ 8.2 & 0.88 $\pm$ 0.18 \\
 J1339$-$0045 & CK* &19.1 $\pm$ 2.2 &	21.5 $\pm$  2.4		& 34.0 $\pm$ 6.6 &  0.56 $\pm$ 0.12 & 31.7 $\pm$ 6.4 & 0.68 $\pm$ 0.16\\ 
 J1340$+$0616 & M*& 57.9 $\pm$ 2.2 & 62.7 $\pm$  2.4		& 54.4 $\pm$ 9.6 & 1.06 $\pm$ 0.19 & 50.5 $\pm$  9.1 & 1.24 $\pm$ 0.22 \\
 J1353$-$0231 & M &119.3 $\pm$ 0.9 & 	137.0 $\pm$  1.1		&125.3 $\pm$ 8.2 & 0.95 $\pm$ 0.06 & 118.5 $\pm$  7.8 &1.15  $\pm$ 0.08\\
 J1400$+$0404 & RD &128.1 $\pm$ 0.2 &219.4 $\pm$  2.6		& 52.3 $\pm$ 4.5 & 2.45 $\pm$ 0.21 & 44.9 $\pm$ 3.9 & 4.89  $\pm$ 0.43 \\
 J1412$-$0049 &  M*&  28.4 $\pm$ 1.3 & 38.1 $\pm$  1.7		&37.8 $\pm$ 6.5 & 0.76 $\pm$ 0.13 & 36.2 $\pm$ 6.3 & 1.05 $\pm$ 0.19\\
 J1513$+$0431 & PR*&35.0 $\pm$ 2.0 & 	37.6 $\pm$  2.2		&45.3 $\pm$ 6.1 & 0.77 $\pm$ 0.11 & 43.1  $\pm$  5.9 & 0.87 $\pm$ 0.13 \\
 J1525$+$0500 & PR&143.9 $\pm$ 0.6 &	171.5$\pm$  0.9		& 55.8 $\pm$ 8.7 & 2.58 $\pm$ 0.40 &  48.5 $\pm$ 7.8 & 3.53 $\pm$ 0.56 \\
 J1545$+$0814 & M& 97.1 $\pm$ 1.3& 	107.5 $\pm$  1.5		&82.6 $\pm$ 9.1 & 1.18 $\pm$ 0.13 &79.8  $\pm$ 8.8 &  1.35 $\pm$ 0.15 \\
 J2256$+$1305 & PR*& 29.2 $\pm$ 1.7 & 31.5 $\pm$  1.8	&48.1 $\pm$ 8.5 & 0.61 $\pm$ 0.11  &47.7 $\pm$ 8.5 & 0.66 $\pm$ 0.12 \\
\hline
\label{tab:2D_vel}
\end{tabular}
\tablefoot{
\tablefoottext{a}{ Galaxy classification: RD: rotating disks , PR: perturbed rotators, CK: complex kinematics, M: mergers, evidence for interacting (see text). The sources denoted with a star have only part of the galaxy covered with ionised gas, so it is hard to see the entire galaxy kinematics.}
}

\end{table*}

As our observation do not have a point source in the observed field-of-view, we estimated the image quality (IQ) based on the Differential Image Motion Monitor (DIMM) seeing reported in the fits header by the observatory. However, the DIMM seeing is measured in the optical and is typically larger than the observed image quality in the infrared.  To estimate the full-width at half maximum (FWHM) of the observations, we use the formulae presented in the documentation of the ESO SINFONI ETC\footnote{https://www.eso.org/observing/etc/doc/helpsinfoni.html\#seeing} and described in more detail in \citet{Martinez10}. Here the FWHM of the atmospheric point spread function (PSF) is calculated from the DIMM seeing, taking into account the observed wavelength and airmass. The prediction of the observed FWHM is then calculating by adding in quadrature the telescope (0.003\arcsec) and instrument PSF FWHM (2 pixels: 0.5\arcsec).

\begin{figure}[!t]
   \includegraphics[width=\hsize]{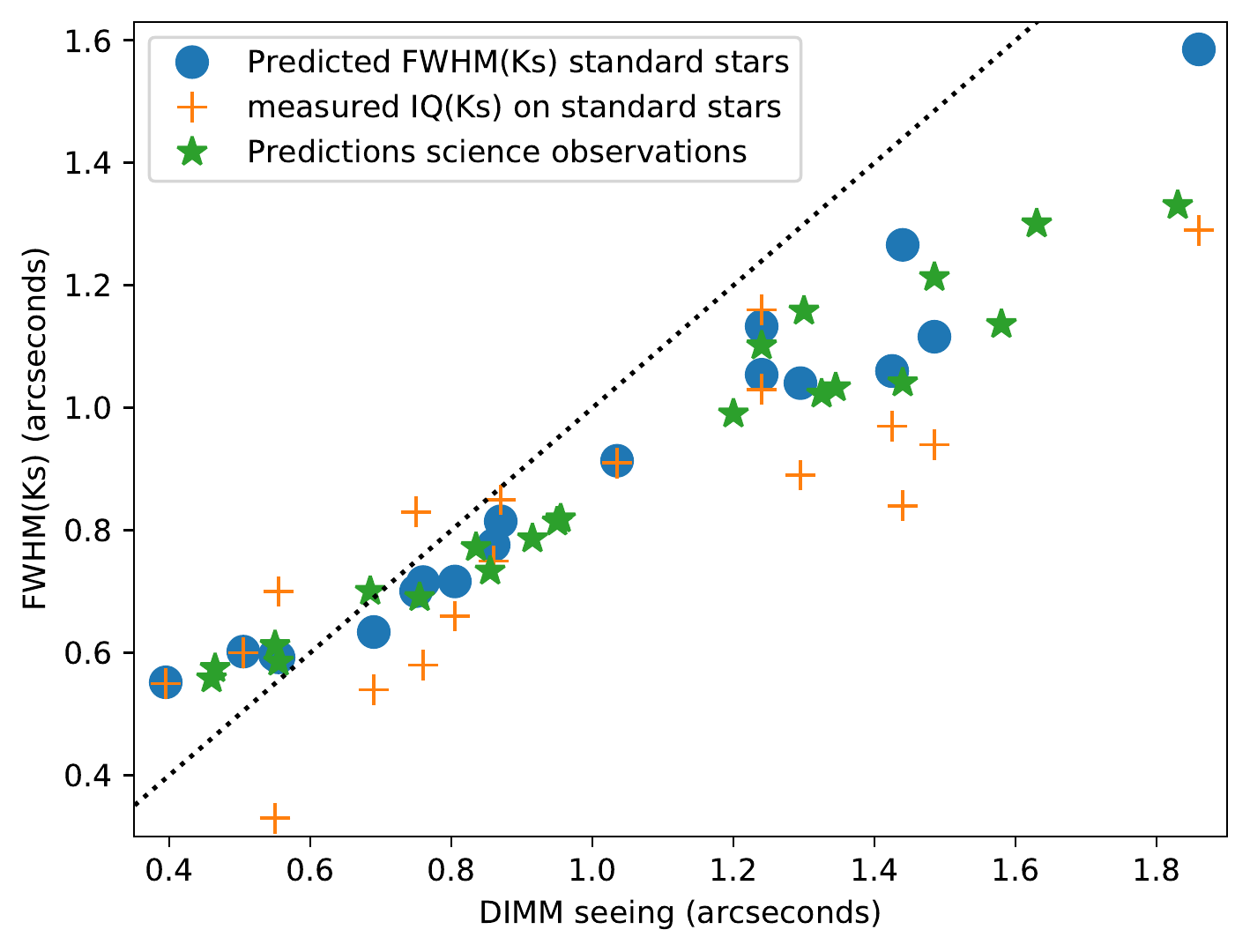}
      \caption{Prediction of the image quality of the data cubes of the galaxies based on the airmass and  DIMM seeing vs the DIMM seeing. The plus signs show the measured image quality in the reduced cubes of the standard star observations plotted vs the DIMM seeing in the fits headers. The filled circles show the predicted IQ using the formulas from the ETC with the airmass and DIMM seeing from the fits headers. The star signs show the predicted IQ values for the science observations. The dotted line shows the one-to-one relation.}
    \label{fig:seeing}
\end{figure}

As a test we used the standard star observations observed immediately before or after the science observations. We took the information about the airmass and DIMM seeing from the fits headers and measured the IQ in the reduced data cubes. The plus signs in Fig. \ref{fig:seeing} show the relation between the measured IQ and the DIMM seeing in the headers. The filled circles in Fig. \ref{fig:seeing}  are the prediction using the formulae describe above for the standard star observations. Comparing  the predicted values to the measured values, we found a similar trends with a spread of a few tenths of an arc second. Such a difference in beam size does not have strong consequences for the final corrected $\sigma$ values. Included in Fig. \ref{fig:seeing} as asterisk are the predicted values of the science observations as function of the DIMM seeing values in the fits headers, these values are used in calculating the effect of beam smearing in the observations.

From the beam smearing corrected $\sigma$ maps we calculated  the flux weighted $\sigma_{m,corr}$ (Eq. \ref{eq:sigma}, Tab. \ref{tab:2D_vel}). Comparing both values shows that beam smearing does not strongly effect our $\sigma$ maps. Most $\sigma_{m}$ values change less than 10\%, with a maximum difference of 15 \%. The galaxies which are the strongest affected are the galaxies with a large velocity gradient due to rotation (e.g. J1211, J1400).  During the further analysis we used the values derived from the corrected maps. 

\begin{figure}[!t]
   \includegraphics[width=\hsize]{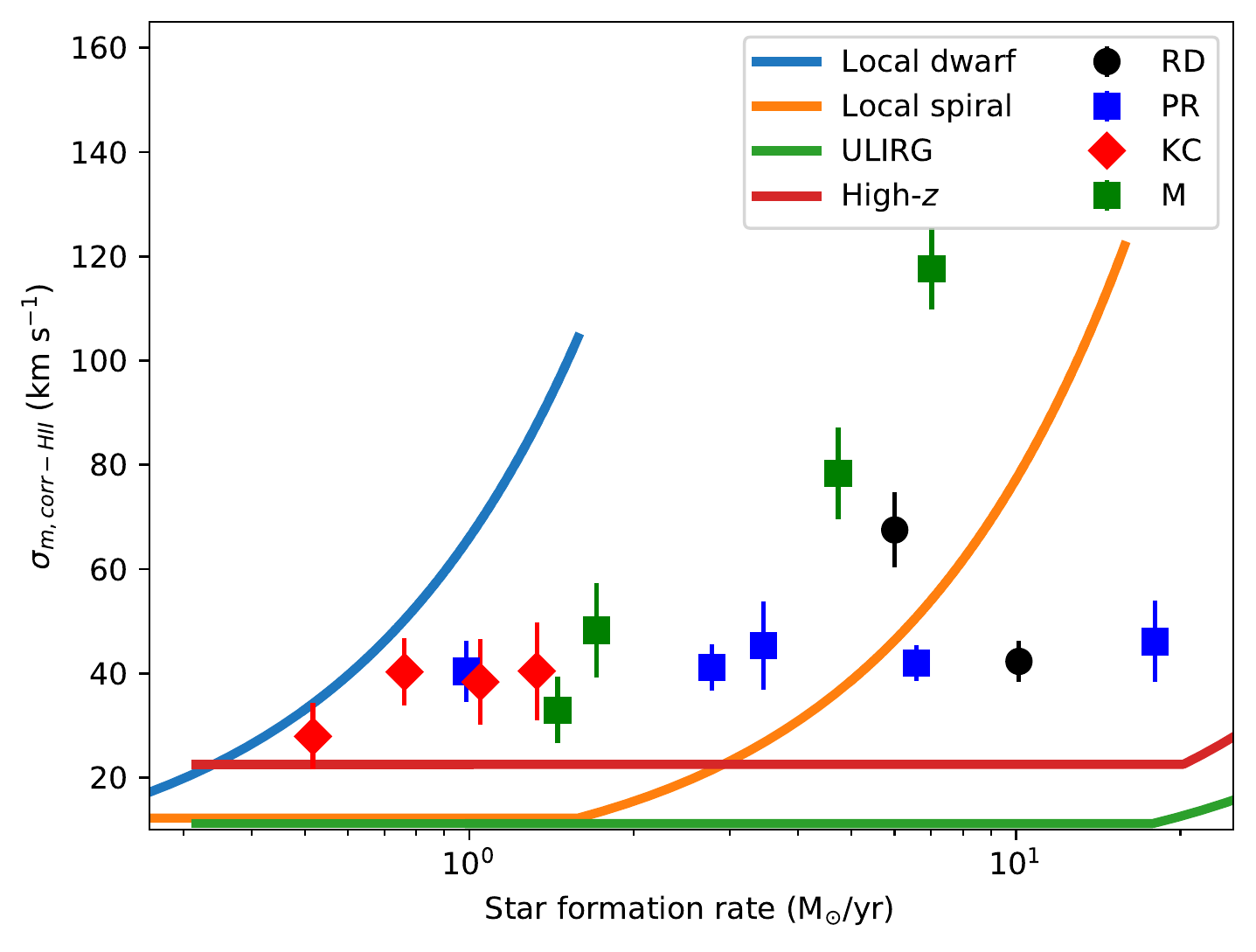}
      \caption{Flux weighted velocity dispersion corrected for beam smearing versus the extinction corrected SFR of the galaxies. Following \citet{Krumholz18}, we subtract 15 \kms\ in quadrature to remove the component caused by thermal broadening in \HII\ regions. Over-plotted are the fiducial model predictions of  \citet{Krumholz18}, including the effect of radial transport and stellar feedback for different galactic environments.
      The symbols represent  their morphological classification, identical to Fig. \ref{fig:sfr_mass}.}
    \label{fig:sigmavssfr}
\end{figure}

Using the inclination derived from the SDSS imaging (Tab. \ref{tab:gal_prop}), we corrected the measured v$_{shear}$  for the inclination in order to derive an estimate of the rotational velocity v$_{rot}$ (Tab. \ref{tab:2D_vel}).  From this we calculated the ratio v$_{rot}$/$\sigma_{m,corr} $ (Tab \ref{tab:2D_vel}). This ratio is a metric to quantify whether the kinematics in the galaxy is dominated by ordered (e.g. disks) or random motions (e.g. turbulence). Galaxies with  low  v$_{rot}$/$\sigma_{m,corr}$ are considered dispersion dominated galaxies \citep{Glazebrook13}, while high   v$_{rot}$/$\sigma_{m,corr}$ indicates a rotationally dominated kinematics. Following \citet{Schreiber20} we use v$_{rot}$/$\sigma_{m,corr}$ = $\sqrt{3.36}$ as the separation between dispersion and rotationally dominated galaxies. In our sample we have nine galaxies with   v$_{rot}$/$\sigma_{m,corr} < \sqrt{3.36}$ , the other six galaxies have  v$_{rot}$/$\sigma_{m,corr} > \sqrt{3.36}$, with the most extreme one having a value of 4.9. 

Based on their morphology and spatially-resolved kinematics,  we classified the galaxies in different categories following   \citet{Flores06} and used extensively after that \citep[see also][]{Glazebrook13, Herenz16}. Firstly, \emph{rotating disks (RD)}, galaxies with a symmetric dipolar velocity field, aligned with the morphological axis, with a symmetrically centrally peaked dispersion. In our sample, J1400 and J1211 show a regular rotating disk profile. 

Secondly, \emph{perturbed rotators (PR)} are  galaxies showing a dipolar velocity field, but with asymmetries, or not aligned with the morphological axis and/or with an offset peak in the dispersion map. J0230 and J1155 show asymmetries in the velocity and dispersion map, but show a velocity gradient.  J1513  and J2256 have only part of the galaxy disk covered in \brg\ emission. That looks regular, but too little area is covered to determine whether this is a rotating disk. J1525 has a semi regular velocity field with a big star forming clump in the centre increasing the velocity dispersion. 

Thirdly \emph{complex kinematic objects (CK)}; we found four galaxies with irregular and complex kinematics. J0018, J1140 J1336 and J1339 are compact galaxies with no clear velocity gradient and a uniform velocity-dispersion map. Note, however, these galaxies are among the more compact of the sample, making it more difficult to observe the same detail as in the larger galaxies. 

Finally, we added a forth class, \emph{mergers (M)}, of four galaxies showing signs of interaction in their morphology or emission lines. J1353 and J1545 show double peaked \paa\ emission lines, while J1340 shows signs of a merging event in the SDSS imaging (Fig \ref{fig:SDSS_images}).  J1412 shows two discrete emission blobs in \brg\, also in the \Ks\ image, signs of a double nucleus are visible.

From the total of 15 galaxies we found two galaxies ($\sim$13\%) with rotating disks (J1211 and J1400), four ($\sim$27\%) perturbed rotators, four galaxies ($\sim$27 \%) with complex kinematics and  four ($\sim$27\%) interacting galaxies. For some galaxies we only detected hydrogen emission in some regions of the galaxy. Those galaxies are marked with an asterisk in table \ref{tab:2D_vel} and their kinematics might be a little harder to interpret as we do not probe the entire velocity field of the galaxy. Considering our small sample of only 15 galaxies, these numbers are  consistent what is found for local low-mass starburst galaxies \citep{Green14,Herenz16} as well as higher redshift galaxies \citep[e.g.][]{Flores06}.

Now we discuss the derived quantities in more detail by comparing them to global galaxy properties and other studies in the literature. In Fig. \ref{fig:sigmavssfr} we plot the derived values for $\sigma_{m}$ versus SFR. The derived values for $\sigma_{m}$ range from 22 \kms\ (J1336) to as much as 125 \kms\ (J1353), with a median value of $\sigma_{m,corr}$ = 43 \kms.  The two galaxies with the largest velocity dispersion are classified as mergers. Especially  J1353 ( $\sigma_{m}$ = 125 \kms) and J1545 ( $\sigma_{m}$ = 80 \kms) are broadened due to their double peaked nature. The derived velocity dispersions of the galaxies are very similar to those derived in low-redshift studies with similar SFR \citep[e.g.][]{Green14,Moiseev15,Varidel20}.  Comparing it to the local galaxies with  higher SFR from the same authors and also \citet{Ostlin01} and \citet{Herenz16} shows, as expected, that our galaxies have typically lower velocity dispersion than the galaxies with higher SFR.

In order to compare our values with  \citet{Krumholz18}, we subtracted in quadrature 15 \kms, as an estimate for the thermal motions in \HII\ regions. We found that our $\sigma_{m}$ values are consistent with the trends between $\sigma_{m,corr}$ and SFR summarised in \citet{Krumholz18}, covering a much larger range of SFR than we observe. A comparison with the theoretical predictions of the fiducial model of \citet{Krumholz18} showed that the location of our galaxies compare well with the local dwarf galaxy track and has higher velocity dispersion than expected for local spiral galaxies. This is consistent with the nature of our galaxies as they are at the high-mass end of the dwarf galaxies (Fig. \ref{fig:sfr_mass}) and most of them do not show a clear spiral structure (Fig. \ref{fig:SDSS_images}).

\begin{figure*}[!t]
   \includegraphics[width=\hsize]{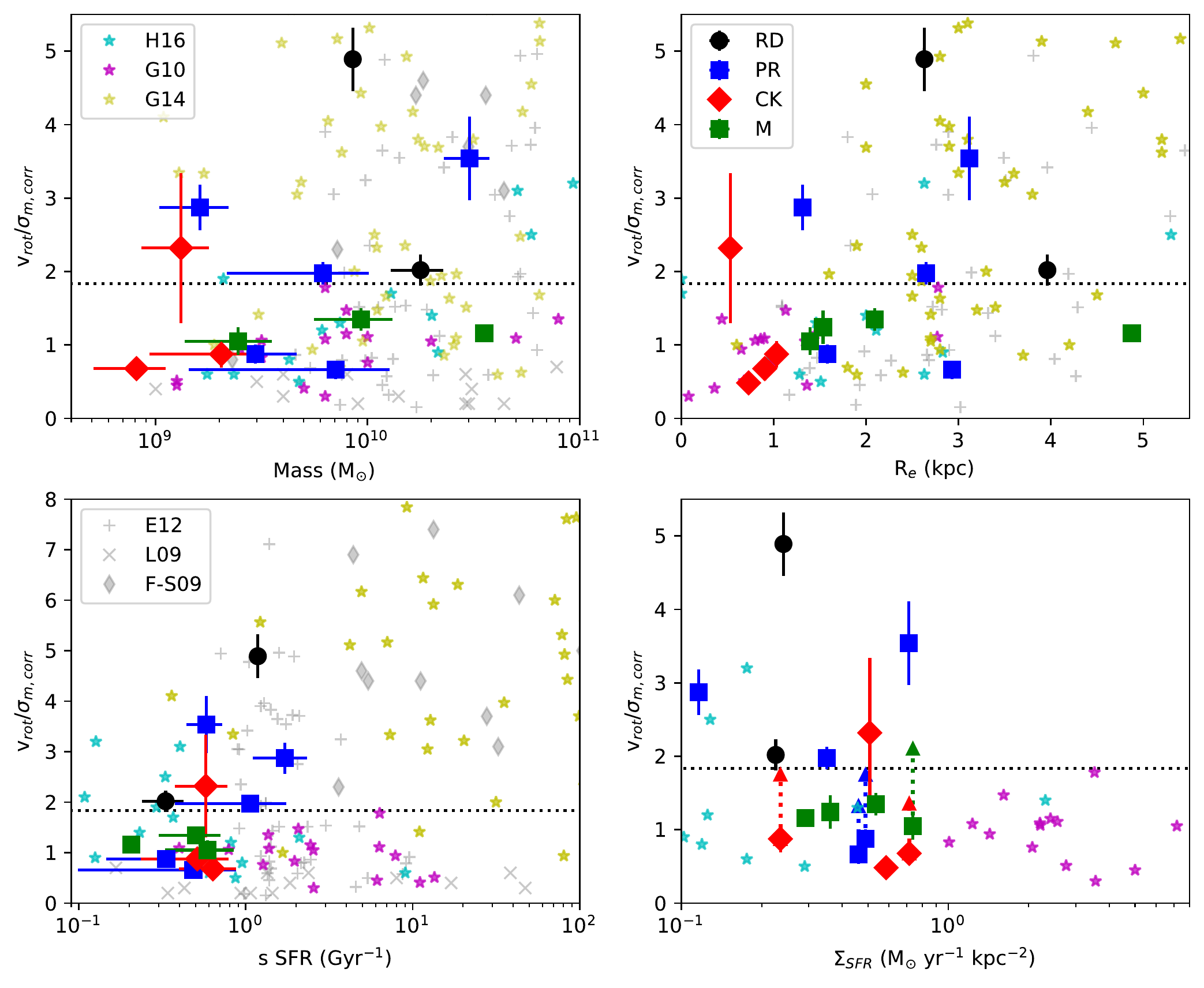}
      \caption{v/$\sigma_{m,corr}$ versus galaxy mass (\emph{Top left}) and effective radius (R$_e$, Top right),  specific star formation rate (sSFR, bottom left) and the surface density of star formation rate ($\Sigma_{SFR}$, bottom right).  Both the sSFR and  $\Sigma_{SFR}$ are corrected for extinction. The colours and symbol signs are the same as in Fig. \ref{fig:sigmavssfr}. Two galaxies, J1140 and J1340 are missing in the top- and bottom-left plots, as no masses are derived in the SDSS database. 
    The vertical dotted lines in the bottom right panel are those galaxies where the \paa\ or \brg\ emission does not cover the entire galaxy, therefor potentially resulting in underestimating  v$_{rot}$. The location of the triangle shows the position of the galaxies with 2 times higher v$_{rot}$ (Sect. \ref{sec:disc_kin}).   The horizontal dotted line shows the deviation between dispersion  and rotationally dominated galaxies \citep[v/$\sigma_{m,corr}$  = $\sqrt{3.36}$ = 1.83,][]{Schreiber20}. Over-plotted are sample from the literature, with symbols identical to Fig. \ref{fig:sfr_mass}. }
    \label{fig:voversigmavsmass}
\end{figure*}

 Figure \ref{fig:voversigmavsmass}  shows the relation between v$_{rot}$/$\sigma_{m,corr}$ and  galaxy mass (upper left,  Tab. \ref{tab:sfr_mass}), effective radius (R$_{e}$, upper right, Tab. \ref{tab:gal_prop}), specific star formation rate (sSFR, bottom left) and star formation rate surface density ($\Sigma_{SFR}$).  For calculating  $\Sigma_{SFR}$ we derived the radius of the line emitting area as follows; for each galaxy we calculated the physical surface brightness (Luminosity/kpc$^2$) map from the Voronoi binned emission-line map. We multiplied the emission-line maps by the \paa/\brg\ ratio for case B: 12.15  \citep[N$_e =  10^2$ cm$^{-3}$, T$_{e}$ = 10000K,][]{Osterbrockbook} in order to have a direct comparison between the surface-brightness maps. We selected a surface-brightness threshold of 2.1 $\times 10^{39}$ erg/s/kpc$^2$, above which we calculate the area where the SB is above the threshold. This value is chosen such that this threshold is reached in the observations of all the galaxies. This threshold corresponds to $\sim$ 0.1 \msun/ kpc$^2$ assuming an H$\alpha$/\paa\ ratio of 8.46 \citep[N$_e =  10^2$ cm$^{-3}$, T$_{e}$ = 10000K,][]{Osterbrockbook}, no attenuation and the SFR calibration of \citet{Kennicutt12}. The radius is calculated by $r_{\Sigma_{SFR}}$ = $\sqrt{(area*\pi)}$ (Tab. \ref{tab:gal_prop}). This radius best reflects the  area of which star formation is going on in the galaxies. The effective radius on the other hand is derived from the $r$-band continuum image and does not necessarily trace the ongoing star formation. Over plotted with various symbols are the reference samples at low and high redshift (Tab. \ref{tab:refsamples}). The properties of the galaxies in the  low redshift samples LARS \citep{Herenz16} and DYNAMO \citep{Green14} are the closest in mass and SFR to our galaxies. The galaxies at higher redshift have typically  higher SFR (factor of ten or more) and are much more extremely star forming, reflected in a much larger sSFR and $\Sigma_{SFR}$.  For the high redshift samples we also compare to the results of \citet{Newman13}, who combine the results of the high-redshift samples listed in Tab. \ref{tab:refsamples}.

The top left panel in Fig. \ref{fig:voversigmavsmass} shows that there is no or very little correlation between v$_{rot}$/$\sigma_{m,corr}$ ands stellar mass.  The two galaxies with the highest v$_{rot}$/$\sigma_{m,corr}$  are among the most massive galaxies, but the scatter in v$_{rot}$/$\sigma_{m,corr}$ at fixed mass is very large. Looking at all the reference samples in the plot shows that low-mass galaxies typically have a lower v$_{rot}$/$\sigma_{m,corr}$ than high mass galaxies, albeit with a very large scatter.  \citet{Herenz16} did find a correlation in the LARS sample, but over larger mass range and it is the most massive galaxies which consistently show a larger v$_{rot}$/$\sigma_{m,corr}$.

The top right panel shows that in our sample there is no correlation between  v$_{rot}$/$\sigma_{m,corr}$ and the effective radius. In size, our galaxies compare well with most of the reference samples.  \citet{Newman13} reported a positive correlation between v$_{rot}$/$\sigma_{m,corr}$ and R$_{e}$ for the collection of high redshift samples they study. 

The bottom left panel shows the relation between sSFR and v$_{rot}$/$\sigma_{m,corr}$. Comparison with the reference samples shows that especially the high-redshift samples have higher sSFR than our sample. The galaxies in  this paper overlap with the lower-end of the LARS and DYNAMO samples.  In our sample there is a suggestion that the larger v$_{rot}$/$\sigma_{m,corr}$ galaxies are among the more massive ones. The reference samples cover a much larger parameter space in sSFR, they show a uniform distribution of galaxies as function of sSFR with values of v$_{rot}$/$\sigma_{m,corr}$ between 0 and 10. However at the lower sSFR (0.1 to 1 Gyr$^{-1}$), the v$_{rot}$/$\sigma_{m,corr}$ is typically lower (the top-left corner of the plot is unpopulated). Our galaxies are in that low sSFR range and showing the same increasing trend with sSFR as the reference samples do in that specific sSFR bin.

The final plot in Fig. \ref{fig:voversigmavsmass} shows the relation between  $\Sigma_{SFR}$  and v$_{rot}$/$\sigma_{m,corr}$. Here we  compare with the sample of Lyman break analogues (LBA) of \citet{Goncalves10} and the LARS sample \citep{Herenz16,Hayes14}, as they also measure the radius of the emission line area from the emission line maps. When we look at our galaxies alone we see an inverse relationship between  v$_{rot}$/$\sigma_{m,corr}$ and $\Sigma_{SFR}$. The higher the $\Sigma_{SFR}$, the lower the v$_{rot}$/$\sigma_{m,corr}$.  This trend is also visible when adding the comparison samples. A similar trend is found by \citet{Newman13} in the high redshift sample, although at  higher $\Sigma_{SFR}$.

When we separated the galaxies in the three different morphological classes, we find that the galaxies showing complex kinematics (CK) have a somewhat lower $\sigma_{m,corr}$ than the other two classes. Additionally they have lower star formation rates, but also more compact and less massive. This makes the CK galaxies have among the higher $\Sigma_{SFR}$ measured.

\subsection{Double peaked emission lines}\label{sec:doublepeaked}

As shown in Sect. \ref{sec:resolved_kinematics} the integrated spectra of J1353  and J1545 show double peaked emission lines. \citet{Maschmann20} study a large sample of double  emission-line galaxies selected from SDSS. They show that these galaxies have a higher velocity dispersion and an enhancement of star formation in the central regions compared to a single-line control sample. They concluded that these properties suggest that the double emission lines come from different components in a galaxy merger.  This behaviour can also be seen in J1353 and J1545 which have among the highest $\sigma_{tot}$ in the sample. The SDSS image of J1353 in Fig. \ref{fig:SDSS_images} shows that this galaxy has a bright central region (targeted by the spectroscopy) with a bit of a warped morphology and a more diffuse component south of the bright region, suggestive of a tidal arm. This irregular appearance could be caused by the ongoing merger event. 

\begin{figure*}[!ht]
   \includegraphics[width=\hsize]{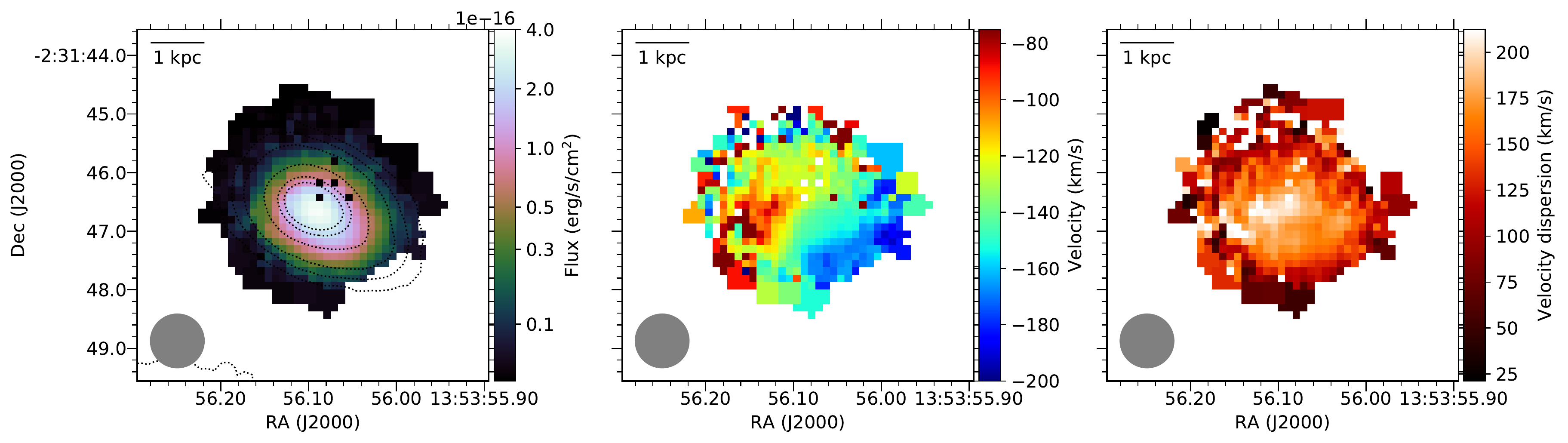}
   \includegraphics[width=\hsize]{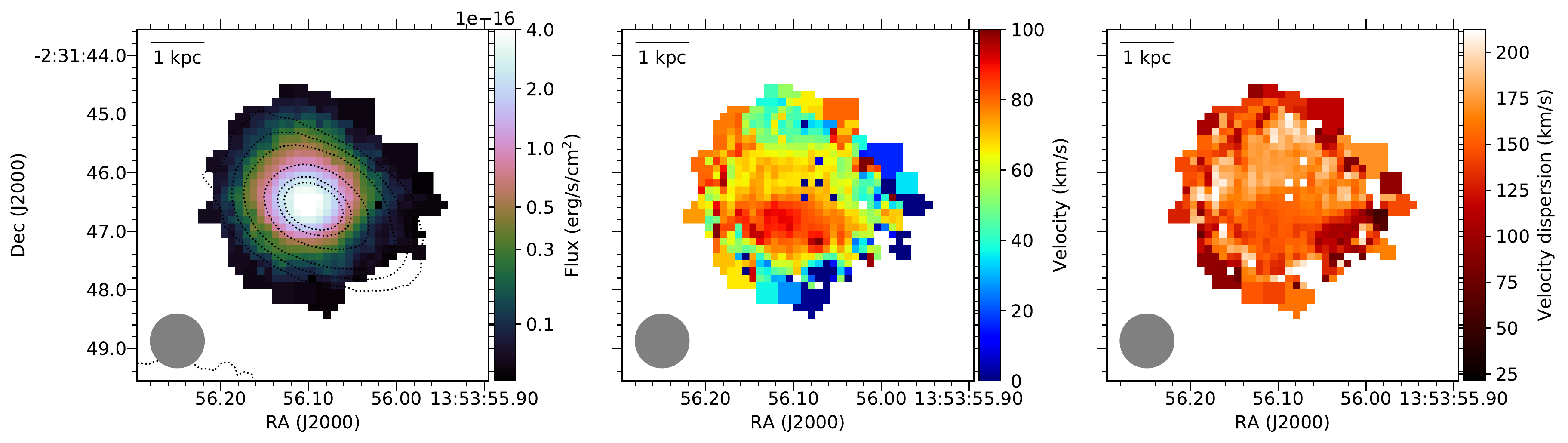}
   
      \caption{Spatially resolved maps of the blue-  (top panel) and  red component of the  \paa\ line of J1353. The first column shows the integrated flux of the Gaussian component. The contours in the flux maps are the contours of the \Ks\ image.  The middle column the velocity with respect to the redshift used in Tab. \ref{tab:sample} and in the last column the velocity dispersion is shown. The instrumental broadening is removed from the dispersion.
    \label{fig:J1353_twocomp}}
\end{figure*}

The two emission peak of J1353 are separated enough ($\sim $210 \kms) to be able to fit for each spatial pixel a double Gaussian profile to separate the two different components. We created a new  Voronoi pattern with a minimum S/N of 50, enabling us to reliable fit two Gaussians.  For each cell we simultaneously fitted two Gaussian components by forcing the blue component to remain at negative velocities and the red component at positive velocities.

Figure \ref{fig:J1353_twocomp} shows the results of the two component fitting. The flux maps of the two components reveal that the morphology of the emission line maps is slightly different. The blue component shows a more elongated  flux distribution, while the red component is more circular. Additionally, the red component is shifted with respect to the blue component towards the north-east by $\sim$0.7 kpc. The blue component shows the kinematics of a disk galaxy, with the velocity gradient aligned with the position angle of the emission and a peak in the velocity dispersion near the centre where the velocity gradient is the steepest. The nature of the red component is less clear and it's velocity maps shows a north-south velocity gradient, but no strong peak is detected in the dispersion map. This component could be a more dispersion dominated galaxy. 

The peak separation of J1545 is much smaller (120 \kms) and the emission line is of  lower S/N than in the case of J1353. Therefore a spatially resolved two component fitting was not possible. The SDSS image of J1545 shows a hint of two nuclear components, suggesting a merger origin.

\subsection{Rotation curves}

\begin{figure*}[!t]
   \includegraphics[width=\hsize]{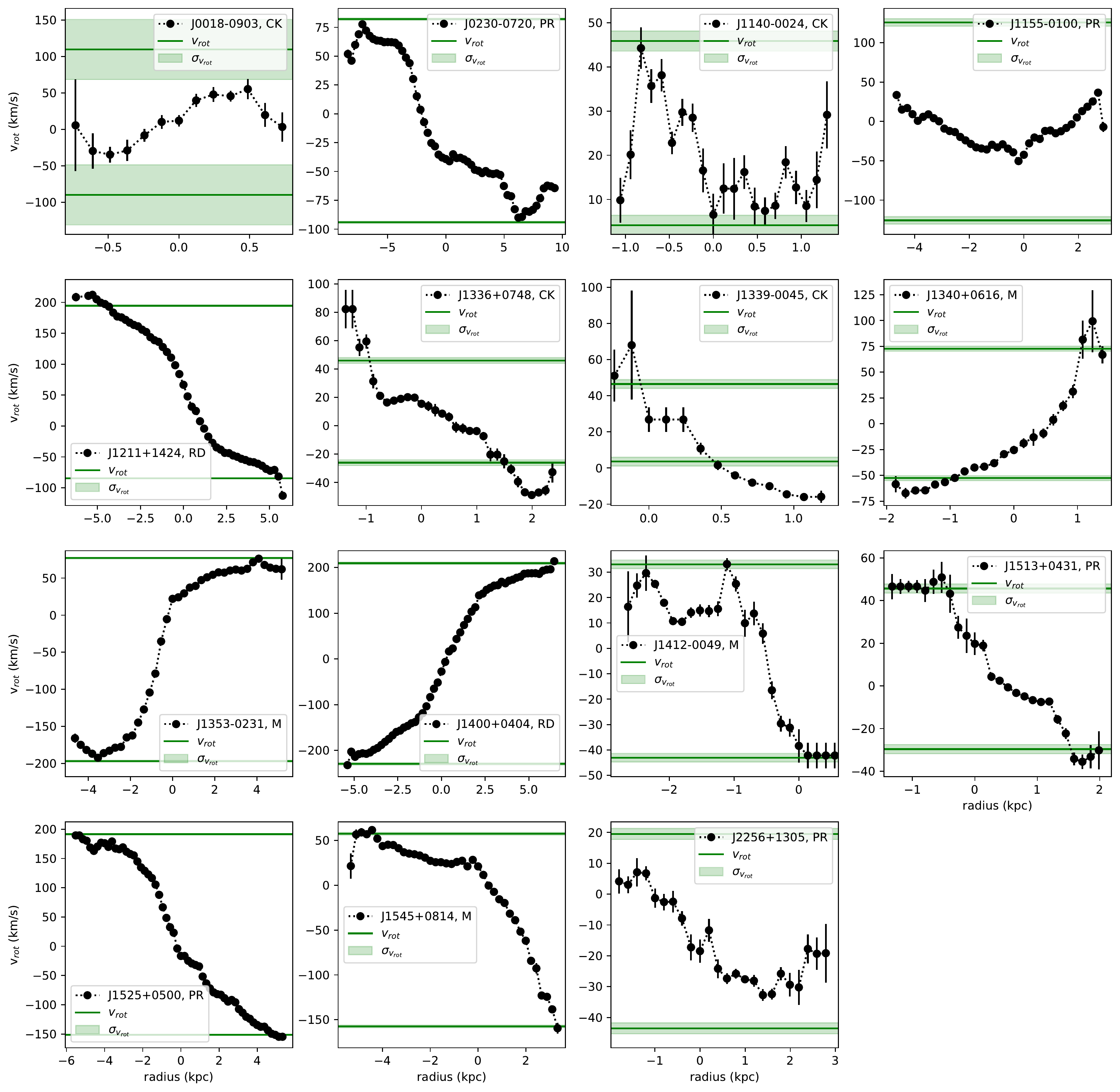}
      \caption{Velocity profiles galaxies derived by radially averaging a 30$\degr$ cone around the position angle derived from the SDSS images. The velocities are corrected for the inclination of the galaxies. The green lines show + and - $v_{rot}$ (Tab. \ref{tab:m_dyn}). The values for  $v_{rot}$  are shifted so they are symmetric around the observed velocity profiles. The shaded green areas show the errors on $v_{rot}$. Their classification as derived in Sect. \ref{subsec:resolved} is given.}
    \label{fig:velcurves}
\end{figure*}

We constructed rotation curves of the galaxies from the emission line maps. We first aligned the maps to the position angle (P.A.) obtained from the exponential disk fit from the SDSS table \emph{PhotoObj} (\verb|expPhi_r|, Table \ref{tab:gal_prop}). Then the velocity as function of radius is determined by summing up the flux for each radial bin in an opening angle of 30$\degr$ ($\pm$15$\degr$ from the semi major axis). The centre of the galaxy is derived by fitting a Gaussian profile to the \Ks\ continuum map constructed from the SINFONI cube.  The resulting velocity profile, corrected for inclination are plotted in Fig. \ref{fig:velcurves}. Also shown are the values of $v_{rot}$ shifted such that they are symmetric around the observed rotation curves.

Some galaxies show rotation curves consistent with that of rotation dominated galaxies (e.g. J1211 and J1400), while other show very irregular rotation curves with large difference between the blue and the red side of the rotation curve, some even lack signs of rotation (e.g. J1140 and J1155).  This is confirming the results in the previous section on the spatially resolved velocity profiles.

The comparison with the over-plotted v$_{rot}$ shows in general  that the rotation curves probe the highest velocities present in the observations. This means that the  emission along the position angle determined from optical continuum images (used for the determination of the P.A.) traces the largest velocity gradients in the galaxy. This is expected in the case of rotating disks, where the position angle derived from continuum imaging and emission lines typically agree within 20$\degr$ \citep{Epinat08,Rodrigues16}.  Some galaxies show larger velocities than the values derived for v$_{rot}$; this is caused by the way v$_{shear}$ is measured. We rejected the 5\% highest and lowest velocity values in order to reject outliers, this causes also to remove the highest velocities in case of a rotating profile. 

The galaxies showing some of the less regular rotation curves (e.g. J1155, J2256) have maximum velocities in the rotation curves which are smaller than the derived values for v$_{rot}$. This could be caused by larger differences between the position angles of the continuum and gas. As these galaxies also have an irregular morphology, they are not well behaved rotating disk galaxies but show more complex kinematics. 

Typically rotation curves are characterised analytically as an increasing relation as function of radius, until the maximum velocity (v$_{max}$) is reached, after that the rotation curve flattens \citep[e.g.][]{Courteau97, Epinat12,Glazebrook13}. For our galaxies only a few rotation curves (J0230, J1353) show that v$_{max}$ is reached. Most rotation curves are still rising and the maximum velocity cannot be observed with the data we have. Deeper data would be needed to observe the fainter surface brightness areas where v$_{max}$ is reached.

\subsection{Dynamical masses}

\begin{table}[!t]
\caption{Dynamical mass determinations}
\centering
\begin{tabular}{lrr}
\hline\hline 
Name & log M$_{vrot}$ & log M$_{\sigma}$\\
	&   (\msun) & (\msun) \\
          \hline
 J0018$-$0903 & 9.0 $\pm$ 0.8  & 9.0 $\pm$ 0.3 \\
 J0230$-$0720  & 9.8 $\pm$ 0.1 & 9.8 $\pm$ 0.2\\
 J1140$-$0024 & 7.7 $\pm$ 0.2 & 9.2 $\pm$ 0.4\\
 J1155$-$0100 &  9.8 $\pm$ 0.1 & 9.4 $\pm$ 0.2 \\
 J1211$+$1424 &10.0 $\pm$ 0.1 & 10.3 $\pm$ 0.2\\
 J1336$+$0748 & 8.4 $\pm$ 0.2 & 9.3 $\pm$  0.4\\
 J1339$-$0045 & 7.7 $\pm$ 0.3 & 9.0 $\pm$ 0.4 \\ 
 J1340$+$0616 & 8.8 $\pm$ 0.1 & 9.6 $\pm$  0.4\\
 J1353$-$0231 & 9.9 $\pm$ 0.1 & 10.9 $\pm$ 0.1\\
 J1400$+$0404 & 10.4 $\pm$ 0.1 &  9.8 $\pm$ 0.2\\
 J1412$-$0049  & 8.4 $\pm$ 0.1 & 9.3 $\pm$ 0.4\\
 J1513$+$0431 & 8.4 $\pm$  0.2 & 9.5 $\pm$ 0.3\\
 J1525$+$0500& 10.2 $\pm$ 0.1 & 9.9 $\pm$ 0.3\\
 J1545$+$0814 & 9.7 $\pm$ 0.1 & 10.2 $\pm$  0.2\\
 J2256$+$1305 & 8.3 $\pm$ 0.2 & 9.9 $\pm$ 0.4\\
\hline
\label{tab:m_dyn}
\end{tabular}
\end{table}

\begin{figure}[!t]
   \includegraphics[width=\hsize]{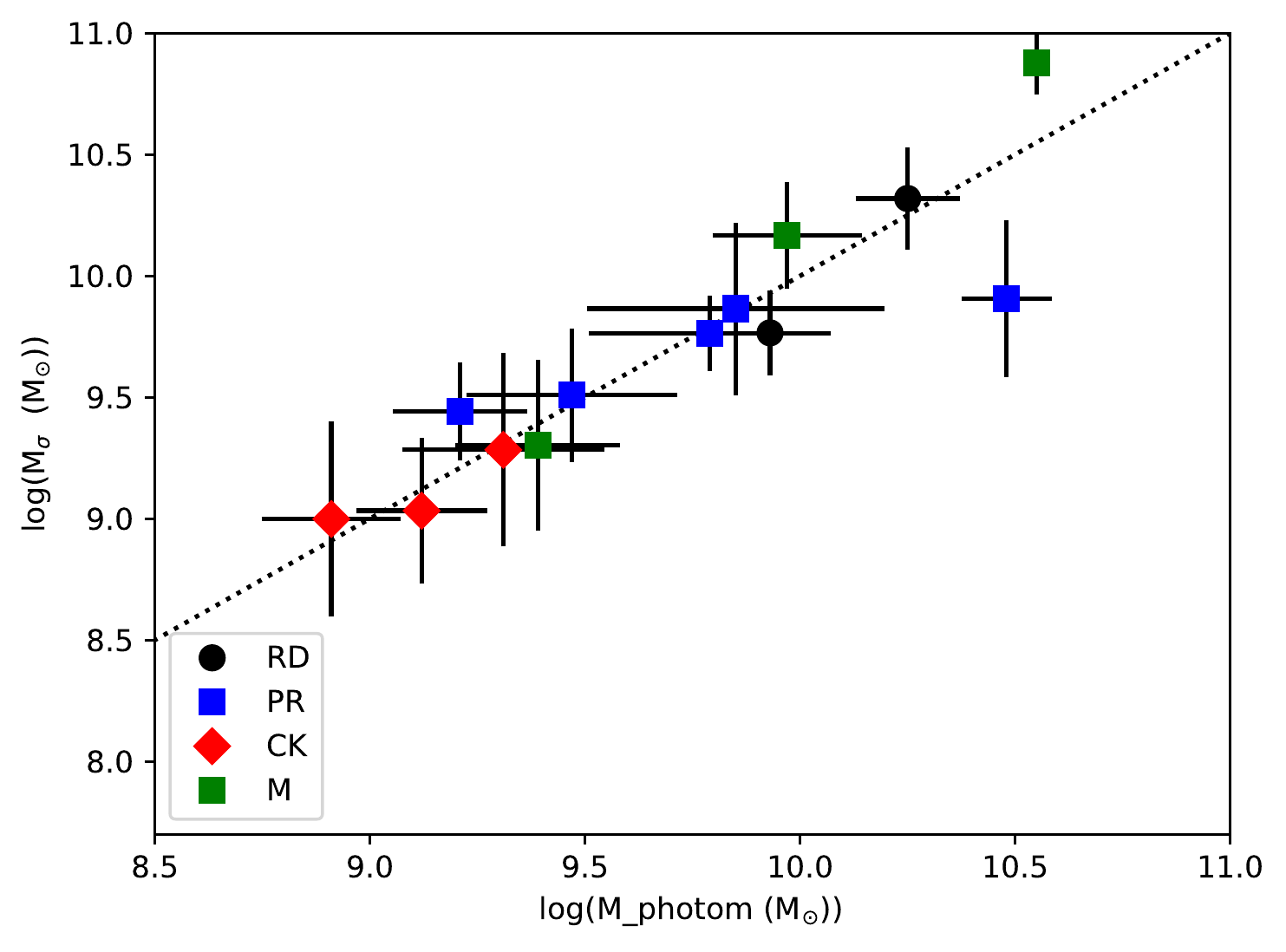}
      \caption{Dynamical mass derived from the velocity dispersion ($\sigma_{m,corr}$) vs the mass derived from the SED fitting, taken from SDSS (Tab. \ref{tab:sfr_mass}. The colours and symbol signs are the same as in Fig. \ref{fig:sigmavssfr}. The dotted line is one-to-one line.}
    \label{fig:mass_sigma}
\end{figure}

\begin{figure}[!t]
   \includegraphics[width=\hsize]{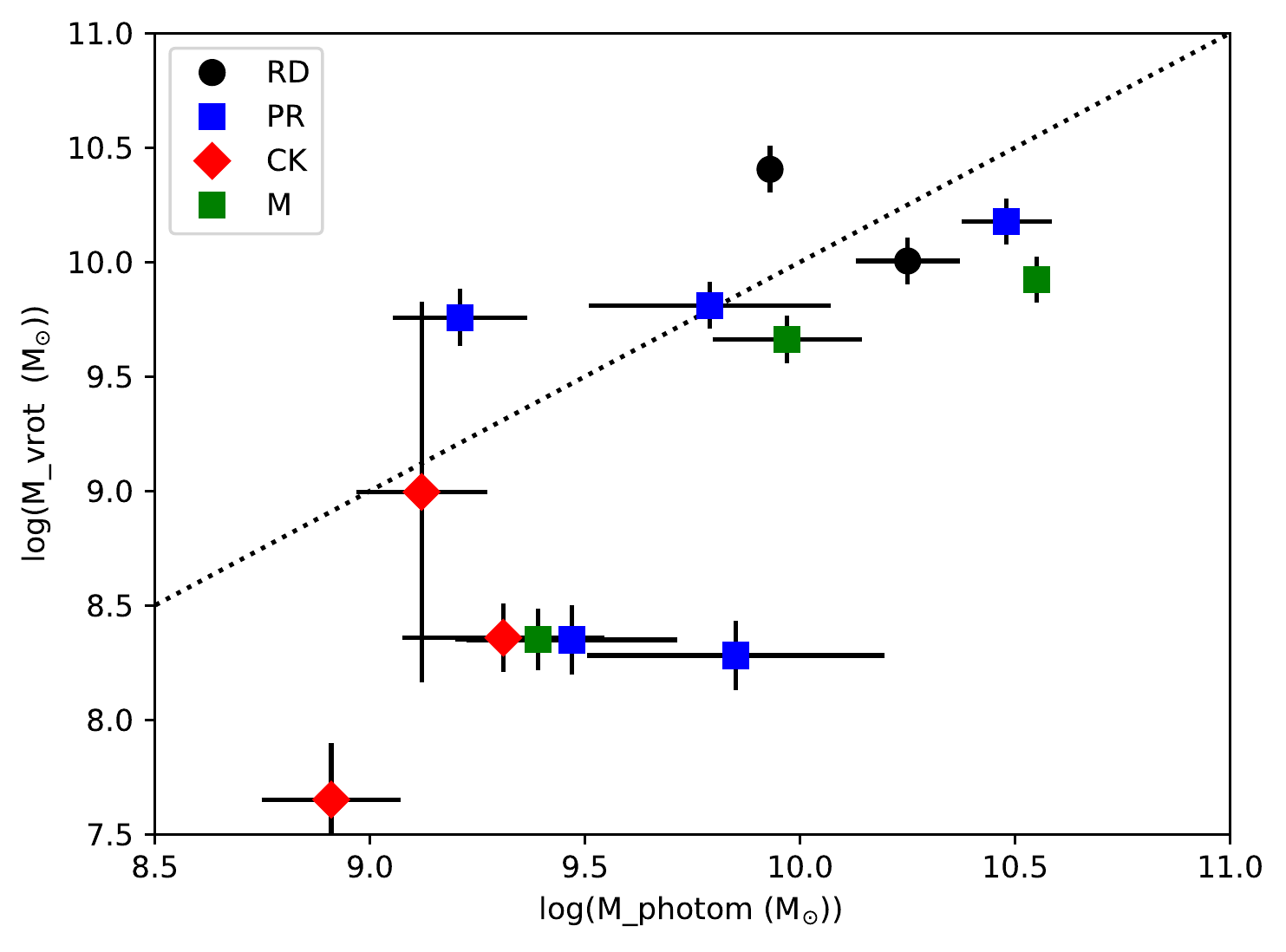}
      \caption{Dynamical mass derived from inclination corrected $v_{shear}$ as proxy for the rotational velocity. The colours and symbol signs are the same as in Fig. \ref{fig:sigmavssfr}. The dotted line is one-to-one line.}
    \label{fig:mass_vrot}
\end{figure}

The mass of a galaxy can be derived via several independent ways, each with its own assumptions. In Table \ref{tab:sfr_mass}, the stellar masses derived from the continuum spectra using the method of  \citet{Chen12} and the stellar population models of \citet{Maraston11} are listed.  With the kinematic information  obtained from  the spatially resolved emission lines in our SINFONI observations we can derive mass estimates from the velocity dispersion and the rotational velocity. Comparison between the dynamical mass and the stellar mass gives us insight how well the galaxy dynamics trace the stellar mass as well as about the contribution of the dark matter.

\subsubsection{M$_{\sigma}$}

The determination of the galaxy mass from the velocity dispersion  relies on the velocity dispersion being dominated by virial motions and trace the gravitational potential of the galaxies \citep{Terlevich81}. Using the virial theorem, there is a direct relation between the velocity dispersion, effective radius and the mass of the galaxy \citep{Guzman96}. We used the formula given by \citet{Ostlin01} to calculate the dynamical mass from the spatially resolved velocity dispersion measurement, corrected for beam smearing ($\sigma_{m,corr}$):
\begin{equation}
M_{\sigma} = 4.8 \cdot \frac{r_{e} \cdot \sigma_{m,corr}^2}{G},
\end{equation}
 where $M_{\sigma}$ is expressed in \msun, and $r_{e}$ represents the effective radius of the galaxy and $G$ the gravitational constant. The value of the constant (4.8) depends on the mass distribution and geometry of the galaxy. \citet{Epinat09} on the other hand uses 2.25 as an average value of known galactic mass distribution models.  This constant is a strong function of the underlying spatial distribution of the stars and varies with Sers\'ic index $n$, where for low values of $n$, the constant could be a factor of two higher than the value we adopted here. This would shift the relation up with a factor of two. As the dynamical mass traces also the dark matter contribution, it would not be unexpected that the dynamical-mass estimate are higher than the stellar mass estimate. 
 
Fig. \ref{fig:mass_sigma} shows the comparison between the dynamical mass from the velocity dispersion and the mass derived from the photometric data taken from SDSS.  The derived masses agree in most galaxies within one or two $\sigma$. If on the other hand, the value of 2.25 is used for the constant, then the derived M$_{\sigma}$ values are systematically less than the photometric mass,  while a larger constant would have a higher M$_{\sigma}$, suggestive of a significant contribution of dark matter. As galaxies are expected to contain a substantial amount of dark matter, a value of 2.25 for the constant would be unlikely for the galaxies in our sample.
The overall agreement of the dynamical and photometric mass estimates  suggests that  the velocity dispersion is mostly driven by gravity and that stellar feedback has only localised effect, but does not affect the overall velocity dispersion. 

\subsubsection{M$_{vrot}$}\label{sec:Mvrot}

Alternatively, we can derive the dynamical mass from our estimates of the rotational velocity using for example \citet{Lequeux83}:
\begin{equation}
M(R) = f\cdot \frac{R \cdot V^2(R)}{G},
\label{eq:mvrot}
\end{equation}
where M(R) represents the mass inside radius R, V(R), the velocity at that radius R and G is the gravitational constant.  The constant f has a value between 0.5 and 1.0, therefore only varying the resulting galaxy mass by a factor 2. For a disk with a flat rotation curve, $f$ would be 0.6, while a galaxy with a declining, Keplerian rotation curve, $f$ equals 0.5. For a spherical distribution (e.g. dominated by dark matter), $f$ gets a value of 1.0. Following \citet{Ostlin99}, we have chosen a value for $f$ of 0.8 as an average value. 

We used the derived rotational velocity from Table \ref{tab:2D_vel}. For the radius in equation \ref{eq:mvrot} we estimated the radius as the  spatial extent of the \brg\ or \paa\ emission from which we have measured the $v_{shear}$ (and derived v$_{rot}$). The final masses derived from the rotational velocities are listed in Tab. \ref{tab:m_dyn} and plotted as comparison to the photometry mass in Fig. \ref{fig:mass_vrot}.

 The dynamical mass derived from  v$_{rot}$ shows a much larger discrepancy. We observe a large scatter, with  eight out of the 13 galaxies are roughly following the one-to-one relation. The other five galaxies, however, are 0.5 dex or more below. Most of them show complicated kinematics, but also some perturbed rotators are showing a large offset.  Inspecting the emission line maps of these five galaxies shows that the nebular emission does not cover the full extend of the galaxy as measured in the \Ks\ continuum.  These galaxies are marked in Tab. \ref{tab:2D_vel} with an asterisk. The sixth galaxy marked with asterisk, J1340, does not have a photometric mass determined, but M$_{vrot}$ is much lower than M$_{\sigma}$ (Tab. \ref{tab:m_dyn}). As the gas does not cover the full extent of the galaxy, the measured value for v$_{shear}$ is under estimated, resulting in a lower estimate for M$_{vrot}$. The velocity dispersion is not affected that strongly as this varies less strongly over the galaxy.  As shown in the previous section, for several galaxies, especially the galaxies located far below the one-to-one line, the rotation curves did not reach their maximum yet in our observations, this would result in a lower estimate of the mass and could the reason that even the more massive galaxies are located below the one-to-one line.

\section{Discussion}
In this section we compare the results of the integrated and spatially resolved kinematics and discuss the origin of the observed kinematics in our galaxy sample. We also discuss the relevance of this study for future and high-redshift observations.

\subsection{What drives the kinematics in galaxies, comparing gas and  stellar kinematics}\label{sec:disc_kin}

 Based on the observed gas $\sigma_{m,corr}$ versus SFR relation, \citet{Krumholz18} showed that a combination of gravity induced turbulence and stellar feedback explain the increase of  $\sigma_{m,corr}$ with SFR.  The model predictions show that at low SFR values (below 0.1 - 1 \msun\ yr$^{-1}$) mostly stellar feedback is responsible for the observed 'floor' of a few tens of km/s. After that the predicted $\sigma_{m,corr}$  rises steeply with SFR, attributed to the gravity induced turbulence.  A similar picture arises from our analysis of the  spatially resolved  emission lines. 

Figure \ref{fig:sigmavssfr} shows that the galaxies in this paper are consistent with the local dwarf and spiral scenario of \citet{Krumholz18} and observed values of $\sim$ 40 - 60 \kms\ suggest a combination of feedback and turbulence as origin for the observed  $\sigma_{m,corr}$.  These findings are corroborated by our comparison between the measured v/$\sigma_{m,corr}$ and different galaxy properties (Fig. \ref{fig:voversigmavsmass}). We find a (negative) correlation between v/$\sigma_{m,corr}$  and the star formation rate surface density. The galaxies with higher values of $\Sigma_{SFR}$ are more dispersion dominated. 

As Fig. \ref{fig:sigmavssfr} shows that the measured $\sigma_{m,corr}$ is pretty constant with SFR and does not vary a lot (ignoring the galaxies identified as mergers). This means that the variation of $v_{shear}$ is mainly responsible for the observed trend with  $\Sigma_{SFR}$. As discussed in Sect. \ref{sec:Mvrot}, several galaxies with low v/$\sigma_{m,corr}$  show \brg\ or \paa\ emission originating from only part of the galaxy as observed in the \Ks\ continuum. For J1336, J1339, J1412, J1513 and J2256 the emission line maps only cover one hemisphere of the galaxy. This results in under-estimating the proper values of  $v_{shear}$. As a test we  increased the   $v_{shear}$ by a factor of two, by assuming that the other hemisphere is a mirror of the hemisphere detected in the emission lines. Assuming the measured $\sigma_{m,corr}$ is representative for the entire galaxy, this would result in an increase of  v/$\sigma_{m,corr}$ with a factor two. This would increase the values of the low  v/$\sigma_{m,corr}$  galaxies to values between 1.2 and 1.5. These are still lower than the observed values for the low $\Sigma_{SFR}$ galaxies and the negative correlation between v/$\sigma_{m,corr}$   and  $\Sigma_{SFR}$ remains present.

By including the stellar kinematics, we can try to separate the effects as the stellar kinematics are not influenced by  stellar feedback. However they are sensitive to the gravitational motions and gravitationally driven turbulence. When comparing the integrated velocity dispersion of the gas and the stars (Fig. \ref{fig:sigmaplot}), we found that for most galaxies the two values do not differ significantly. Several galaxies have rather low values and the accuracy is hampered by the limited spectral resolution of our observations ($\sigma_{instr}$ = 34 \kms). 
 
 Of those galaxies where we find deviations, the spatially resolved comparison of the velocity dispersions (Fig. \ref{fig:v_rms}) provides more clues on why the integrated velocity dispersions do not agree. In two galaxies, J1211 and J1400, the large stellar velocity dispersion is driven by a small area in the centre  of the galaxy. They are the brightest pixels and therefore dominate the integrated measurements. The decoupling between the gas and the stars in galaxy centres \citep[e.g.][]{Marquart07}  could be related to the presence of a galaxy bulge, which is known to be dispersion dominated \citep{Kormendy04}. 
 
Only in one galaxy, J1525,  we found direct evidence that stellar feedback affects the velocity dispersion of the gas (Fig. \ref{fig:sigma_J1525}), where the central starburst in the galaxy as a higher dispersion in the gas than in the stars. However, these effects are small scale effects compared to the size of the galaxy, and the integrated kinematics of J1525 do now show a strong discrepancy between the stellar and gas velocity dispersion.

Summarising, our analysis suggests that the measured integrated kinematics of the gas and the stars are typically very similar, suggesting that on galaxy-size scales, the gas and stellar velocity fields are not decoupled and the effect of stellar feedback on the total gas dynamics is not dominant in the galaxies studied in this paper. However, on the other hand, the negative correlation between  v/$\sigma_{m,corr}$  and $\Sigma_{SFR}$ suggests that stellar feedback does have an effect and makes the galaxies more dispersion dominated. We can reconcile both effects by the scenario that these galaxies have a more turbulent ISM. This turbulence could be caused by gravitational instabilities  due to for example a merger or an accretion event \citep{Cumming08,Bik18}.  Such an event would increase the star formation, but also increase the stellar-velocity dispersion as well as that of the giant molecular clouds, resulting in a young stellar population with a larger velocity dispersion. Similar results were found by \citet{Bergvall16} who studied a large sample of SDSS selected starburst galaxies and found that the M$_{\sigma}$ derived from H$\alpha$ is very similar to the photometric mass, suggesting that the gas dynamics trace that of the stars. 

 With our spatially resolved data, the localised effect of stellar feedback on the ionised gas can be easier identified than on spatially unresolved data. The comparison between stellar and gas kinematics is also an effective way to identify kinematically decoupled cores in the galaxy, due to for instance the presence of a stellar bulge in the centre. This is caused by the fact that the centre of galaxies are the brightest in the continuum emission and contribute therefore strongly to the integrated light. 
 
Observations at higher spectral resolution will shed light on difference between the stellar and gas velocity dispersion for galaxies with low velocity dispersion, where currently the error is dominated by the limited spectral resolution. Observing stronger starbursts with significantly higher SFR, where feedback becomes more dominant might provide a different picture by revealing galaxies whose emission lines are affected by stellar feedback.

\subsection{Relevance for  future observations and high redshift observations}

\subsubsection{Near-infrared diagnostics}

In this paper we measured the kinematics of starburst galaxies using the CO absorption, \brg\ and \paa\ emission in the near-infrared $K$-band. Typically this type of galaxy kinematic studies are performed in the optical where \halpha\ is used for the gas kinematics and strong absorption lines such as the Ca triplet for stellar absorption. Using ground-based telescopes, $K$-band observations are clearly less sensitive due to the high thermal background longwards of 2 micron. Additionally, the strong telluric absorption lines around 2.3 \micron\ make it challenging. We selected very specific redshift bins (Sect. \ref{sec:obs}), to avoid strong contamination of the absorption lines. 

However, for starburst galaxies, the CO absorption bands have clear advantages above the stronger optical absorption lines to derive stellar kinematics. The CO absorption lines originate in cool stars, red super giants or main sequence stars, which  have the peak of their spectral energy distribution in the near-infrared. This makes the CO lines easily observable when red super giants are present in a galaxy. Additionally, the CO absorption is not contaminated by nebular emission lines. This in contrast to the Ca triplet, which is contaminated by high order Paschen emission lines \citep{Cumming08}, making them less reliable to derive stellar kinematics in starburst galaxies where these Paschen lines are bright. 

With future, mostly infrared telescopes, such as JWST and ELT we demonstrate in this paper the power of the infrared CO and hydrogen lines as tracers of the galaxy kinematics. Specially for NIRSPEC  on board of JWST, where  sky background and telluric absorption are not present, this diagnostic becomes very powerful to derive the stellar and gas kinematics of galaxies in the local universe with lower mass than the current sample, allowing spatially resolved observations of nearby galaxies similar to galaxies in the early universe which are impossible to  observe in detail. 
The intermediate redshift universe becomes accessible both with NIRSPEC (up to z $\sim$ 1.2) and MIRI  (z >1), where galaxies with masses and SFRs similar to the current surveys in the local universe can be observed and extend the parameter space of the high-redshift surveys from the ground.

\subsubsection{High-z galaxy kinematics}

\begin{figure}[!t]
   \includegraphics[width=\hsize]{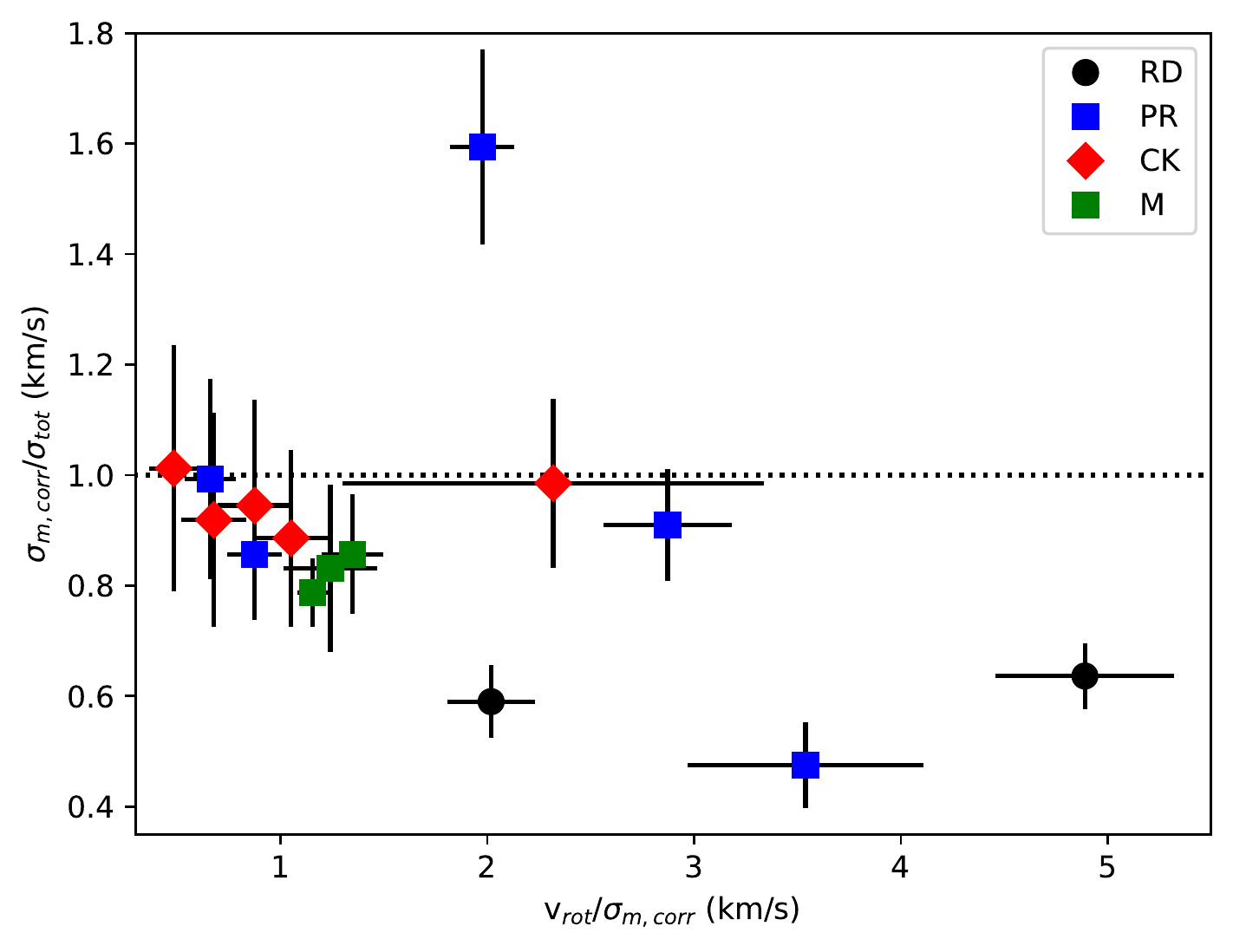}
      \caption{$\sigma_{m,corr}$ versus $\sigma_{tot}$.  The colours and symbol signs are the same as in Fig. \ref{fig:sigmavssfr}. The dotted line is one-to-one line.}
    \label{fig:sigma_comparison}
\end{figure}

In Sect. \ref{sec:integrated_kin} we derived the integrated gas and stellar kinematics, while in Sect. \ref{sec:resolved_kinematics}  and Sect. \ref{sec:resolved} we analysed the spatially resolved kinematics. In the context of high redshift studies where beam smearing becomes important and obtaining spatially resolved kinematics is not always possible, in this study we can compare the spatially resolved analysis with that of the integrated spectra.

The velocity derived from fitting the integrated spectra is useful to determine the redshift, but has no information anymore on the nature of the kinematics in the galaxies. In contrast the velocity map determined from the spatially resolved kinematics contains a lot of information on the galaxy properties and can be used to determine whether  the galaxy kinematics is  dominated by random motions or ordered motions such as disk galaxies (e.g. Fig.\ref{fig:velcurves}). 

The integrated spectra provide $\sigma_{tot}$ and the line shape to analyse. In Fig. \ref{fig:sigma_comparison} we compare  $\sigma_{tot}$ from the integrated spectra with $\sigma_{m,corr}$, derived from the spatially resolved spectra. We plot the ratio of $\sigma_{m,corr}$ and $\sigma_{tot}$ vs  v/$\sigma_{m,corr}$ in Fig. \ref{fig:sigma_comparison}. This comparison shows that for high  v/$\sigma_{m,corr}$ values, the values for $\sigma_{tot}$ can deviate up to a factor of two. The measurement for $\sigma_{tot}$ is twice as large as the measurement of $\sigma_{m,corr}$. This is caused by unresolved velocity gradients, which dominate the kinematics in galaxies with high v/$\sigma_{m,corr}$. By using $\sigma_{tot}$ to derive the dynamical mass of rotationally dominated galaxies we would over estimate the mass of the galaxy by a factor of two. The only outlier, where $\sigma_{m,corr}$  is larger than  $\sigma_{tot}$ is J0230 (Fig. \ref{fig:2D_J0230}), where both $\sigma$ values are low.

For dispersion-dominated galaxies, this discrepancy becomes less and less and the $\sigma_{tot}$  is a good tracer of $\sigma_{m,corr}$. At high redshift, where galaxies become more and more dispersion dominated, $\sigma_{tot}$ works better and better  as a  tracer of the dynamical mass, even if due to beam smearing the  $\sigma_{m,corr}$ is hard, or impossible to derive.

 \section{Conclusions}

We obtained near-infrared integral-field spectroscopy of 15  SDSS-selected star-forming galaxies with SINFONI. We measured the  integrated and spatially-resolved kinematics of both the gas and the stars in those galaxies using the near-infrared  \brg\ and \paa\ emission lines and  CO absorption bands.  The absorption spectra are fitted with pPXF in order to derive the velocity and velocity dispersion. The emission lines are fitted with single Gaussian profiles. Based on the analysis of the integrated spectra and spatially resolved data we derive the following conclusions.

\begin{itemize}
\item{The kinematics analysis of the integrated spectra shows that most galaxies have, within the error bars, the same velocity dispersion in the stars and gas. From the 15 galaxies we found two outliers with $\sigma_{tot,star} > \sigma_{tot,gas}$ and three with $\sigma_{tot,gas} > \sigma_{tot,stars}$}.
\item{Two galaxies with  $\sigma_{tot,gas} > \sigma_{tot,stars}$ show a double-peaked velocity profile. These galaxies are in the process of a merger event. A two component fit on one of them reveals a blue shifted event consistent with a rotation dominated galaxy and a redshift component which is more dispersion dominated. Two additional merger is identified based on the continuum emission and SDSS imaging.}
\item{The two dimensional comparison between gas and stellar kinematics shows that the stellar kinematics of the two galaxies with $\sigma_{tot,star} > \sigma_{tot,gas}$ are dominated by a decoupled core or stellar bulge.}
\item{In one galaxy we find the evidence that stellar feedback of the central starburst has increased the velocity dispersion of the gas.}
\item{All galaxies show spatially-resolved emission of ionised gas (\brg, \paa). Their flux-weighted, for beam-smearing corrected, velocity dispersion is consistent with reports in the literature of galaxies with similar star formation rate. We find $v_{rot}/\sigma_{m,corr}$ values showing that our galaxy sample as a range of galaxies varying from dispersion dominated to rotationally dominated.}
\item{We find a negative trend between v/$\sigma_{m,corr}$ and $\Sigma_{SFR}$, corroborated by other studies in the literature.}
\item{The global gas and stellar kinematics are consistent with being dominated by gravitational instabilities, possibly caused by merger or accretion events.}
\item{We find a reasonable agreement between the  M$_{\sigma}$ and the stellar mass, while the M$_{vrot}$ is under estimated in those galaxies where only one hemisphere is emitting \paa\ or \brg, as the gas lines do not trace the full velocity field of the galaxy.}
\end{itemize}

The near-infrared diagnostics used in this paper show strong potential for studying starburst galaxies as the CO bandheads are not affected by other emission lines. The limitations due to the sky emission and telluric absorption are removed with future space-based instrumentation such as  NIRSPEC and MIRI on board of JWST.

\begin{acknowledgements}
We thank the anonymous referee for their constructive feedback, improving the quality of the paper. The authors thank Nils Bergvall for feedback on an earlier version of the manuscript. This work was supported by the Swedish Research Council (Vetenskapsr\aa det).
This research has made use of the NASA/IPAC Extragalactic Database (NED), which is operated by the Jet Propulsion Laboratory, California Institute of Technology, under contract with the National Aeronautics and Space Administration. This research has made use of NASA’s Astrophysics Data System Bibliographic Services (ADS). This research made use of Astropy,\footnote{http://www.astropy.org} a community-developed core Python package for Astronomy \citep{astropy13, astropy18}. This research made use of APLpy, an open-source plotting package for Python  \citep{Robitaille12}.

Funding for SDSS-III has been provided by the Alfred P. Sloan Foundation, the Participating Institutions, the National Science Foundation, and the U.S. Department of Energy Office of Science. The SDSS-III web site is http://www.sdss3.org/. SDSS-III is managed by the Astrophysical Research Consortium for the Participating Institutions of the SDSS-III Collaboration including the University of Arizona, the Brazilian Participation Group, Brookhaven National Laboratory, Carnegie Mellon University, University of Florida, the French Participation Group, the German Participation Group, Harvard University, the Instituto de Astrofisica de Canarias, the Michigan State/Notre Dame/JINA Participation Group, Johns Hopkins University, Lawrence Berkeley National Laboratory, Max Planck Institute for Astrophysics, Max Planck Institute for Extraterrestrial Physics, New Mexico State University, New York University, Ohio State University, Pennsylvania State University, University of Portsmouth, Princeton University, the Spanish Participation Group, University of Tokyo, University of Utah, Vanderbilt University, University of Virginia, University of Washington, and Yale University.

  \end{acknowledgements}

\bibliographystyle{aa}
\bibliography{sinfo_sbtest.bib}

\begin{thebibliography}{104}
\expandafter\ifx\csname natexlab\endcsname\relax\def\natexlab#1{#1}\fi

\bibitem[{Alam {et~al.}(2015)Alam, Albareti, Prieto, Anders, Anderson,
  Anderton, Andrews, Armengaud, Aubourg, Bailey, Basu, Bautista, Beaton, Beers,
  Bender, Berlind, Beutler, Bhardwaj, Bird, Bizyaev, Blake, Blanton, Blomqvist,
  Bochanski, Bolton, Bovy, Bradley, Brandt, Brauer, Brinkmann, Brown,
  Brownstein, Burden, Burtin, Busca, Cai, Capozzi, Rosell, Carr, Carrera,
  Chambers, Chaplin, Chen, Chiappini, Chojnowski, Chuang, Clerc, Comparat,
  Covey, Croft, Cuesta, Cunha, Costa, Rio, Davenport, Dawson, Lee, Delubac,
  Deshpande, Dhital, Ferreira, Dwelly, Ealet, Ebelke, Edmondson, Eisenstein,
  Ellsworth, Elsworth, Epstein, Eracleous, Escoffier, Esposito, Evans, Fan,
  Fernández-Alvar, Feuillet, Ak, Finley, Finoguenov, Flaherty, Fleming,
  Font-Ribera, Foster, Frinchaboy, Galbraith-Frew, García, García-Hernández,
  Pérez, Gaulme, Ge, Génova-Santos, Georgakakis, Ghezzi, Gillespie, Girardi,
  Goddard, Gontcho, Hernández, Grebel, Green, Grieb, Grieves, Gunn, Guo,
  Harding, Hasselquist, Hawley, Hayden, Hearty, Hekker, Ho, Hogg,
  Holley-Bockelmann, Holtzman, Honscheid, Huber, Huehnerhoff, Ivans, Jiang,
  Johnson, Kinemuchi, Kirkby, Kitaura, Klaene, Knapp, Kneib, Koenig, Lam, Lan,
  Lang, Laurent, Goff, Leauthaud, Lee, Lee, Licquia, Liu, Long,
  López-Corredoira, Lorenzo-Oliveira, Lucatello, Lundgren, Lupton, Mack,
  Mahadevan, Maia, Majewski, Malanushenko, Malanushenko, Manchado, Manera, Mao,
  Maraston, Marchwinski, Margala, Martell, Martig, Masters, Mathur, McBride,
  McGehee, McGreer, McMahon, Ménard, Menzel, Merloni, Mészáros, Miller,
  Miralda-Escudé, Miyatake, Montero-Dorta, More, Morganson, Morice-Atkinson,
  Morrison, Mosser, Muna, Myers, Nandra, Newman, Neyrinck, Nguyen, Nichol,
  Nidever, Noterdaeme, Nuza, O'Connell, O'Connell, O'Connell, Ogando, Olmstead,
  Oravetz, Oravetz, Osumi, Owen, Padgett, Padmanabhan, Paegert,
  Palanque-Delabrouille, Pan, Parejko, Pâris, Park, Pattarakijwanich,
  Pellejero-Ibanez, Pepper, Percival, Pérez-Fournon, fols, Petitjean, Pieri,
  Pinsonneault, Mello, Prada, Prakash, Price-Whelan, Protopapas, Raddick,
  Rahman, Reid, Rich, Rix, Robin, Rockosi, Rodrigues, Rodríguez-Torres, Roe,
  Ross, Ross, Rossi, Ruan, Rubiño-Martín, Rykoff, Salazar-Albornoz, Salvato,
  Samushia, Sánchez, Santiago, Sayres, Schiavon, Schlegel, Schmidt, Schneider,
  Schultheis, Schwope, Scóccola, Scott, Sellgren, Seo, Serenelli, Shane, Shen,
  Shetrone, Shu, Aguirre, Sivarani, Skrutskie, Slosar, Smith, Sobreira, Souto,
  Stassun, Steinmetz, Stello, Strauss, Streblyanska, Suzuki, Swanson, Tan,
  Tayar, Terrien, Thakar, Thomas, Thomas, Thompson, Tinker, Tojeiro, Troup,
  Magaña, Vazquez, Verde, Viel, Vogt, Wake, Wang, Weaver, Weinberg, Weiner,
  White, Wilson, Wisniewski, Wood-Vasey, che, York, Zakamska, Zamora, Zasowski,
  Zehavi, Zhao, Zheng, Zhou, Zhou, Zou, \& Zhu}]{Alam15}
Alam, S., Albareti, F.~D., Prieto, C.~A., {et~al.} 2015, \apjs, 219, 12

\bibitem[{Bacon {et~al.}(2001)Bacon, Copin, Monnet, Miller, Allington-Smith,
  Bureau, Carollo, Davies, Emsellem, Kuntschner, Peletier, Verolme, \&
  Zeeuw}]{Bacon01}
Bacon, R., Copin, Y., Monnet, G., {et~al.} 2001, \mnras, 326, 23

\bibitem[{Barbary(2016)}]{Barbary16}
Barbary, K. 2016, Zenodo, 804967

\bibitem[{Bergvall {et~al.}(2016)Bergvall, Marquart, Way, Blomqvist, Holst,
  Östlin, \& Zackrisson}]{Bergvall16}
Bergvall, N., Marquart, T., Way, M.~J., {et~al.} 2016, \aap, 587, A72

\bibitem[{Bik {et~al.}(2010)Bik, Puga, Waters, Horrobin, Henning, Vasyunina,
  Beuther, Linz, Kaper, Ancker, Lenorzer, Churchwell, Kurtz, Kouwenhoven,
  Stolte, Koter, Thi, Comerón, \& Waelkens}]{Bik10}
Bik, A., Puga, E., Waters, L., {et~al.} 2010, \apj, 713, 883

\bibitem[{Bik {et~al.}(2015)Bik, Östlin, Hayes, Adamo, Melinder, \&
  Amram}]{Bik15}
Bik, A., Östlin, G., Hayes, M., {et~al.} 2015, \aap, 576, L13

\bibitem[{Bik {et~al.}(2018)Bik, Östlin, Menacho, Adamo, Hayes, Herenz, \&
  Melinder}]{Bik18}
Bik, A., Östlin, G., Menacho, V., {et~al.} 2018, \aap, 619, A131

\bibitem[{Bonnet {et~al.}(2004)Bonnet, Abuter, Baker, Bornemann, Brown,
  Castillo, Conzelmann, Damster, Davies, Delabre, Donaldson, Dumas, Eisenhauer,
  Elswijk, Fedrigo, Finger, Gemperlein, Genzel, Gilbert, Gillet, Goldbrunner,
  Horrobin, Horst, Huber, Hubin, Iserlohe, Kaufer, Kissler-Patig, Kragt, Kroes,
  Lehnert, Lieb, Liske, Lizon, Lutz, Modigliani, Monnet, Nesvadba, Patig,
  Pragt, Reunanen, Röhrle, Rossi, Schmutzer, Schoenmaker, Schreiber,
  Stroebele, Szeifert, Tacconi, Tecza, Thatte, Tordo, Werf, \&
  Weisz}]{Bonnet04}
Bonnet, H., Abuter, R., Baker, A., {et~al.} 2004, The Messenger, 117, 17

\bibitem[{Bosma(1989)}]{Bosma89}
Bosma, A. 1989, ASSL, 151, 65

\bibitem[{Bundy {et~al.}(2015)Bundy, Bershady, Law, Yan, Drory, MacDonald,
  Wake, Cherinka, Sánchez-Gallego, Weijmans, Thomas, Tremonti, Masters,
  Coccato, Diamond-Stanic, Aragón-Salamanca, Avila-Reese, Badenes,
  Falcón-Barroso, Belfiore, Bizyaev, Blanc, Bland-Hawthorn, Blanton,
  Brownstein, Byler, Cappellari, Conroy, Dutton, Emsellem, Etherington,
  Frinchaboy, Fu, Gunn, Harding, Johnston, Kauffmann, Kinemuchi, Klaene,
  Knapen, Leauthaud, Li, Lin, Maiolino, Malanushenko, Malanushenko, Mao,
  Maraston, McDermid, Merrifield, Nichol, Oravetz, Pan, Parejko, Sanchez,
  Schlegel, Simmons, Steele, Steinmetz, Thanjavur, Thompson, Tinker, Bosch,
  Westfall, Wilkinson, Wright, Xiao, \& Zhang}]{Bundy15}
Bundy, K., Bershady, M.~A., Law, D.~R., {et~al.} 2015, \apj, 798, 7

\bibitem[{Böker {et~al.}(2008)Böker, Falcón-Barroso, Schinnerer, Knapen, \&
  Ryder}]{Boker08}
Böker, T., Falcón-Barroso, J., Schinnerer, E., Knapen, J.~H., \& Ryder, S.
  2008, \aj, 135, 479

\bibitem[{Calzetti {et~al.}(1994)Calzetti, Kinney, \&
  Storchi-Bergmann}]{Calzetti94}
Calzetti, D., Kinney, A.~L., \& Storchi-Bergmann, T. 1994, \apj, 429, 582

\bibitem[{Cappellari(2017)}]{Cappellari17}
Cappellari, M. 2017, \mnras, 466, 798

\bibitem[{Cappellari \& Copin(2003)}]{Cappellari03}
Cappellari, M. \& Copin, Y. 2003, \mnras, 342, 345

\bibitem[{Cappellari \& Emsellem(2004)}]{Cappellari04}
Cappellari, M. \& Emsellem, E. 2004, \pasp, 116, 138

\bibitem[{Cardelli {et~al.}(1989)Cardelli, Clayton, \& Mathis}]{Cardelli89}
Cardelli, J.~A., Clayton, G.~C., \& Mathis, J.~S. 1989, \apj, 345, 245

\bibitem[{Catalán-Torrecilla {et~al.}(2020)Catalán-Torrecilla,
  Castillo-Morales, Paz, Gallego, Carrasco, Iglesias-Páramo, Cedazo,
  Chamorro-Cazorla, Pascual, García-Vargas, Cardiel, Gómez-Alvarez,
  Pérez-Calpena, Martínez-Delgado, Dullo, Coelho, Bruzual, \&
  Charlot}]{CatalanTorrecilla20}
Catalán-Torrecilla, C., Castillo-Morales, A., Paz, A. G.~d., {et~al.} 2020,
  \apj, 890, 5

\bibitem[{Chang {et~al.}(2015)Chang, Wel, Cunha, \& Rix}]{Chang15}
Chang, Y.-Y., Wel, A. v.~d., Cunha, E.~d., \& Rix, H.-W. 2015, \apjs, 219, 8

\bibitem[{Chen {et~al.}(2012)Chen, Kauffmann, Tremonti, White, Heckman, Kovač,
  Bundy, Chisholm, Maraston, Schneider, Bolton, Weaver, \& Brinkmann}]{Chen12}
Chen, Y.-M., Kauffmann, G., Tremonti, C.~A., {et~al.} 2012, \mnras, 421, 314

\bibitem[{Chevalier \& Clegg(1985)}]{Chevalier85}
Chevalier, R.~A. \& Clegg, A.~W. 1985, Nature, 317, 44

\bibitem[{Cleri {et~al.}(2022)Cleri, Trump, Backhaus, Momcheva, Papovich,
  Simons, Weiner, Estrada-Carpenter, Finkelstein, Giavalisco, Ji, Jung,
  Matharu, Martinez, \& Sturm}]{Cleri22}
Cleri, N.~J., Trump, J.~R., Backhaus, B.~E., {et~al.} 2022, \apj, 929, 3

\bibitem[{Cohen {et~al.}(2003)Cohen, Wheaton, \& Megeath}]{Cohen03}
Cohen, M., Wheaton, W.~A., \& Megeath, S.~T. 2003, \aj, 126, 1090, (c) 2003:
  The American Astronomical Society

\bibitem[{Colina {et~al.}(2005)Colina, Arribas, \& Monreal-Ibero}]{Colina05}
Colina, L., Arribas, S., \& Monreal-Ibero, A. 2005, \apj, 621, 725

\bibitem[{Contini {et~al.}(2012)Contini, Garilli, Fèvre, Kissler-Patig, Amram,
  Epinat, Moultaka, Paioro, Queyrel, Tasca, Tresse, Vergani, López-Sanjuan, \&
  Perez-Montero}]{Contini12}
Contini, T., Garilli, B., Fèvre, O.~L., {et~al.} 2012, \aap, 539, A91

\bibitem[{Courteau(1997)}]{Courteau97}
Courteau, S. 1997, \aj, 114, 2402

\bibitem[{Cresci {et~al.}(2009)Cresci, Hicks, Genzel, Schreiber, Davies,
  Bouché, Buschkamp, Genel, Shapiro, Tacconi, Sommer-Larsen, Burkert,
  Eisenhauer, Gerhard, Lutz, Naab, Sternberg, Cimatti, Daddi, Erb, Kurk, Lilly,
  Renzini, Shapley, Steidel, \& Caputi}]{Cresci09}
Cresci, G., Hicks, E. K.~S., Genzel, R., {et~al.} 2009, \apj, 697, 115

\bibitem[{Cresci {et~al.}(2017)Cresci, Vanzi, Telles, Lanzuisi, Brusa,
  Mingozzi, Sauvage, \& Johnson}]{Cresci17}
Cresci, G., Vanzi, L., Telles, E., {et~al.} 2017, \aap, 604, A101

\bibitem[{Cumming {et~al.}(2008)Cumming, Fathi, Östlin, Marquart, Márquez,
  Masegosa, Bergvall, \& Amram}]{Cumming08}
Cumming, R.~J., Fathi, K., Östlin, G., {et~al.} 2008, \aap, 479, 725

\bibitem[{Davies(2007)}]{Davies07}
Davies, R.~I. 2007, \mnras, 375, 1099

\bibitem[{Diehl \& Statler(2006)}]{Diehl06}
Diehl, S. \& Statler, T.~S. 2006, \mnras, 368, 497

\bibitem[{Eisenhauer {et~al.}(2003)Eisenhauer, Abuter, Bickert,
  Biancat-Marchet, Bonnet, Brynnel, Conzelmann, Delabre, Donaldson, Farinato,
  Fedrigo, Genzel, Hubin, Iserlohe, Kasper, Kissler-Patig, Monnet, Roehrle,
  Schreiber, Stroebele, Tecza, Thatte, \& Weisz}]{Eisenhauer03}
Eisenhauer, F., Abuter, R., Bickert, K., {et~al.} 2003, \procspie, 4841, 1548

\bibitem[{Eisenstein {et~al.}(2011)Eisenstein, Weinberg, Agol, Aihara, Prieto,
  Anderson, Arns, Aubourg, Bailey, Balbinot, Barkhouser, Beers, Berlind,
  Bickerton, Bizyaev, Blanton, Bochanski, Bolton, Bosman, Bovy, Brandt,
  Breslauer, Brewington, Brinkmann, Brown, Brownstein, Burger, Busca, Campbell,
  Cargile, Carithers, Carlberg, Carr, Chang, Chen, Chiappini, Comparat,
  Connolly, Cortes, Croft, Cunha, Costa, Davenport, Dawson, Lee, Mello, Simoni,
  Dean, Dhital, Ealet, Ebelke, Edmondson, Eiting, Escoffier, Esposito, Evans,
  Fan, Castellá, Ferreira, Fitzgerald, Fleming, Font-Ribera, Ford, Frinchaboy,
  Pérez, Gaudi, Ge, Ghezzi, Gillespie, Gilmore, Girardi, Gott, Gould, Grebel,
  Gunn, Hamilton, Harding, Harris, Hawley, Hearty, Hennawi, Hernández, Ho,
  Hogg, Holtzman, Honscheid, Inada, Ivans, Jiang, Jiang, Johnson, Jordan,
  Jordan, Kauffmann, Kazin, Kirkby, Klaene, Knapp, Kneib, Kochanek, Koesterke,
  Kollmeier, Kron, Lampeitl, Lang, Lawler, Goff, Lee, Lee, Leisenring, Lin,
  Liu, Long, Loomis, Lucatello, Lundgren, Lupton, Ma, Ma, MacDonald, Mack,
  Mahadevan, Maia, Majewski, Makler, Malanushenko, Malanushenko, Mandelbaum,
  Maraston, Margala, Maseman, Masters, McBride, McDonald, McGreer, McMahon,
  Requejo, Ménard, Miralda-Escudé, Morrison, Mullally, Muna, Murayama, Myers,
  Naugle, Neto, Nguyen, Nichol, Nidever, O'Connell, Ogando, Olmstead, Oravetz,
  Padmanabhan, Paegert, Palanque-Delabrouille, Pan, Pandey, Parejko, Pâris,
  Pellegrini, Pepper, Percival, Petitjean, Pfaffenberger, Pforr, Phleps,
  Pichon, Pieri, Prada, Price-Whelan, Raddick, Ramos, Reid, Reyle, Rich,
  Richards, Rieke, Rieke, Rix, Robin, Rocha-Pinto, Rockosi, Roe, Rollinde,
  Ross, Ross, Rossetto, Sánchez, Santiago, Sayres, Schiavon, Schlegel,
  Schlesinger, Schmidt, Schneider, Sellgren, Shelden, Sheldon, Shetrone, Shu,
  Silverman, Simmerer, Simmons, Sivarani, Skrutskie, Slosar, Smee, Smith,
  Snedden, Stassun, Steele, Steinmetz, Stockett, Stollberg, Strauss, Szalay,
  Tanaka, Thakar, Thomas, Tinker, Tofflemire, Tojeiro, Tremonti, Magaña,
  Verde, Vogt, Wake, Wan, Wang, Weaver, White, White, Wilson, Wisniewski,
  Wood-Vasey, Yanny, Yasuda, Yèche, York, Young, Zasowski, Zehavi, \&
  Zhao}]{Eisenstein11}
Eisenstein, D.~J., Weinberg, D.~H., Agol, E., {et~al.} 2011, \aj, 142, 72

\bibitem[{Elias {et~al.}(1998)Elias, Vukobratovich, Andrew, Cho, Cuberly, Don,
  Gerzoff, Harmer, Harris, Heynssens, Hicks, Kovacs, Li, Liang, Moon, Pearson,
  Plum, Roddier, Tvedt, Wolff, \& Wong}]{Elias98}
Elias, J.~H., Vukobratovich, D., Andrew, J.~R., {et~al.} 1998, Society of
  Photo-Optical Instrumentation Engineers (SPIE) Conference Series, 3354, 555

\bibitem[{Emsellem {et~al.}(2004)Emsellem, Cappellari, Peletier, McDermid,
  Bacon, Bureau, Copin, Davies, Krajnović, Kuntschner, Miller, \&
  Zeeuw}]{Emsellem04}
Emsellem, E., Cappellari, M., Peletier, R.~F., {et~al.} 2004, \mnras, 352, 721

\bibitem[{Epinat {et~al.}(2010)Epinat, Amram, Balkowski, \&
  Marcelin}]{Epinat10}
Epinat, B., Amram, P., Balkowski, C., \& Marcelin, M. 2010, \mnras, 401, 2113

\bibitem[{Epinat {et~al.}(2008)Epinat, Amram, \& Marcelin}]{Epinat08}
Epinat, B., Amram, P., \& Marcelin, M. 2008, \mnras, 390, 466

\bibitem[{Epinat {et~al.}(2009)Epinat, Contini, Fèvre, Vergani, Garilli,
  Amram, Queyrel, Tasca, \& Tresse}]{Epinat09}
Epinat, B., Contini, T., Fèvre, O.~L., {et~al.} 2009, \aap, 504, 789

\bibitem[{Epinat {et~al.}(2012)Epinat, Tasca, Amram, Contini, Fèvre, Queyrel,
  Vergani, Garilli, Kissler-Patig, Moultaka, Paioro, Tresse, Bournaud,
  López-Sanjuan, \& Perret}]{Epinat12}
Epinat, B., Tasca, L., Amram, P., {et~al.} 2012, \aap, 539, A92

\bibitem[{Falcón-Barroso(2016)}]{Falcon16}
Falcón-Barroso, J. 2016, Galactic Bulges, 418, 161

\bibitem[{Falcón-Barroso {et~al.}(2006)Falcón-Barroso, Bacon, Bureau,
  Cappellari, Davies, Zeeuw, Emsellem, Fathi, Krajnović, Kuntschner, McDermid,
  Peletier, \& Sarzi}]{FalconBarroso06}
Falcón-Barroso, J., Bacon, R., Bureau, M., {et~al.} 2006, \mnras, 369, 529

\bibitem[{Falcón-Barroso {et~al.}(2017)Falcón-Barroso, Lyubenova, Ven,
  Méndez-Abreu, Aguerri, García-Lorenzo, Bekeraité, Sánchez, Husemann,
  García-Benito, Mast, Walcher, Zibetti, Barrera-Ballesteros, Galbany,
  Sánchez-Blázquez, Singh, Bosch, Wild, Zhu, Bland-Hawthorn, Fernandes,
  Lorenzo-Cáceres, Gallazzi, Delgado, Marino, Márquez, Pérez, Pérez, Roth,
  Rosales-Ortega, Ruiz-Lara, Wisotzki, Ziegler, \&
  Collaboration}]{FalconBarroso17}
Falcón-Barroso, J., Lyubenova, M., Ven, G. v.~d., {et~al.} 2017, \aap, 597,
  A48

\bibitem[{Flores {et~al.}(2006)Flores, Hammer, Puech, Amram, \&
  Balkowski}]{Flores06}
Flores, H., Hammer, F., Puech, M., Amram, P., \& Balkowski, C. 2006, \aap, 455,
  107

\bibitem[{Förster-Schreiber {et~al.}(2009)Förster-Schreiber, Genzel, Bouché,
  Cresci, Davies, Buschkamp, Shapiro, Tacconi, Hicks, Genel, Shapley, Erb,
  Steidel, Lutz, Eisenhauer, Gillessen, Sternberg, Renzini, Cimatti, Daddi,
  Kurk, Lilly, Kong, Lehnert, Nesvadba, Verma, McCracken, Arimoto, Mignoli, \&
  Onodera}]{Forster09}
Förster-Schreiber, N.~M., Genzel, R., Bouché, N., {et~al.} 2009, \apj, 706,
  1364

\bibitem[{Ganda {et~al.}(2006)Ganda, Falcón-Barroso, Peletier, Cappellari,
  Emsellem, McDermid, Zeeuw, \& Carollo}]{Ganda06}
Ganda, K., Falcón-Barroso, J., Peletier, R.~F., {et~al.} 2006, \mnras, 367, 46

\bibitem[{Glazebrook(2013)}]{Glazebrook13}
Glazebrook, K. 2013, \pasa, 30, 7

\bibitem[{Gonçalves {et~al.}(2010)Gonçalves, Basu-Zych, Overzier, Martin,
  Law, Schiminovich, Wyder, Mallery, Rich, \& Heckman}]{Goncalves10}
Gonçalves, T.~S., Basu-Zych, A., Overzier, R., {et~al.} 2010, \apj, 724, 1373

\bibitem[{Green {et~al.}(2010)Green, Glazebrook, McGregor, Abraham, Poole,
  Damjanov, McCarthy, Colless, \& Sharp}]{Green10}
Green, A.~W., Glazebrook, K., McGregor, P.~J., {et~al.} 2010, Nature, 467, 684

\bibitem[{Green {et~al.}(2014)Green, Glazebrook, McGregor, Damjanov, Wisnioski,
  Abraham, Colless, Sharp, Crain, Poole, \& McCarthy}]{Green14}
Green, A.~W., Glazebrook, K., McGregor, P.~J., {et~al.} 2014, \mnras, 437, 1070

\bibitem[{Guaita {et~al.}(2015)Guaita, Melinder, Hayes, Östlin, Gonzalez,
  Micheva, Adamo, Mas-Hesse, Sandberg, Otí-Floranes, Schaerer, Verhamme,
  Freeland, Orlitová, Laursen, Cannon, Duval, Rivera-Thorsen, Herenz, Kunth,
  Atek, Puschnig, Gruyters, \& Pardy}]{Guaita15}
Guaita, L., Melinder, J., Hayes, M., {et~al.} 2015, \aap, 576, A51

\bibitem[{Guzman {et~al.}(1996)Guzman, Koo, Faber, Illingworth, Takamiya, Kron,
  \& Bershady}]{Guzman96}
Guzman, R., Koo, D.~C., Faber, S.~M., {et~al.} 1996, \apjl, 460, L5

\bibitem[{Hayes {et~al.}(2014)Hayes, Östlin, Duval, Sandberg, Guaita,
  Melinder, Adamo, Schaerer, Verhamme, Orlitová, Mas-Hesse, Cannon, Atek,
  Kunth, Laursen, Otí-Floranes, Pardy, Rivera-Thorsen, \& Herenz}]{Hayes14}
Hayes, M., Östlin, G., Duval, F., {et~al.} 2014, \apj, 782, 6

\bibitem[{Herenz {et~al.}(2016)Herenz, Gruyters, Orlitová, Hayes, Östlin,
  Cannon, Roth, Bik, Pardy, Otí-Floranes, Mas-Hesse, Adamo, Atek, Duval,
  Guaita, Kunth, Laursen, Melinder, Puschnig, Rivera-Thorsen, Schaerer, \&
  Verhamme}]{Herenz16}
Herenz, E.~C., Gruyters, P., Orlitová, I., {et~al.} 2016, \aap, 587, A78

\bibitem[{Horne(1986)}]{Horne86}
Horne, K. 1986, \pasp, 98, 609

\bibitem[{Johnson {et~al.}(2018)Johnson, Harrison, Swinbank, Tiley, Stott,
  Bower, Smail, Bunker, Sobral, Turner, Best, Bureau, Cirasuolo, Jarvis,
  Magdis, Sharples, Bland-Hawthorn, Catinella, Cortese, Croom, Federrath,
  Glazebrook, Sweet, Bryant, Goodwin, Konstantopoulos, Lawrence, Medling,
  Owers, \& Richards}]{Johnson18}
Johnson, H.~L., Harrison, C.~M., Swinbank, A.~M., {et~al.} 2018, \mnras, 474,
  5076

\bibitem[{Kennicutt \& Evans(2012)}]{Kennicutt12}
Kennicutt, R.~C. \& Evans, N.~J. 2012, \araa, 50, 531

\bibitem[{Kobulnicky \& Gebhardt(2000)}]{Kobulnicky00}
Kobulnicky, H.~A. \& Gebhardt, K. 2000, \aj, 119, 1608

\bibitem[{Kormendy \& Kennicutt(2004)}]{Kormendy04}
Kormendy, J. \& Kennicutt, R. C.~J. 2004, \araa, 42, 603

\bibitem[{Krumholz \& Burkhart(2016)}]{Krumholz16}
Krumholz, M.~R. \& Burkhart, B. 2016, \mnras, 458, 1671

\bibitem[{Krumholz {et~al.}(2018)Krumholz, Burkhart, Forbes, \&
  Crocker}]{Krumholz18}
Krumholz, M.~R., Burkhart, B., Forbes, J.~C., \& Crocker, R.~M. 2018, \mnras,
  477, 2716

\bibitem[{Law {et~al.}(2009)Law, Steidel, Erb, Larkin, Pettini, Shapley, \&
  Wright}]{Law09}
Law, D.~R., Steidel, C.~C., Erb, D.~K., {et~al.} 2009, \apj, 697, 2057

\bibitem[{Lehnert {et~al.}(2009)Lehnert, Nesvadba, Tiran, Matteo, Driel,
  Douglas, Chemin, \& Bournaud}]{Lehnert09}
Lehnert, M.~D., Nesvadba, N. P.~H., Tiran, L.~L., {et~al.} 2009, \apj, 699,
  1660

\bibitem[{Lehnert {et~al.}(2013)Lehnert, Tiran, Nesvadba, Driel, Boulanger, \&
  Matteo}]{Lehnert13}
Lehnert, M.~D., Tiran, L.~L., Nesvadba, N. P.~H., {et~al.} 2013, \aap, 555, A72

\bibitem[{Leitherer {et~al.}(1999)Leitherer, Schaerer, Goldader, Delgado,
  Robert, Kune, Mello, Devost, \& Heckman}]{Leitherer99}
Leitherer, C., Schaerer, D., Goldader, J.~D., {et~al.} 1999, \apjs, 123, 3

\bibitem[{Lequeux(1983)}]{Lequeux83}
Lequeux, J. 1983, \aap, 125, 394

\bibitem[{López {et~al.}(2016)López, Colina, Arribas, Pereira-Santaella, \&
  Alonso-Herrero}]{Lopez16}
López, J.~P., Colina, L., Arribas, S., Pereira-Santaella, M., \&
  Alonso-Herrero, A. 2016, \aap, 590, A67

\bibitem[{Maraston \& Strömbäck(2011)}]{Maraston11}
Maraston, C. \& Strömbäck, G. 2011, \mnras, 418, 2785

\bibitem[{Marquart {et~al.}(2007)Marquart, Fathi, Östlin, Bergvall, Cumming,
  \& Amram}]{Marquart07}
Marquart, T., Fathi, K., Östlin, G., {et~al.} 2007, \aap, 474, L9

\bibitem[{Martinez {et~al.}(2010)Martinez, Kolb, Sarazin, \&
  Tokovinin}]{Martinez10}
Martinez, P., Kolb, J., Sarazin, M., \& Tokovinin, A. 2010, The Messenger, 141,
  5

\bibitem[{Martinsson {et~al.}(2013)Martinsson, Verheijen, Westfall, Bershady,
  Schechtman-Rook, Andersen, \& Swaters}]{Martinsson13}
Martinsson, T. P.~K., Verheijen, M. A.~W., Westfall, K.~B., {et~al.} 2013,
  \aap, 557, A130

\bibitem[{Maschmann {et~al.}(2020)Maschmann, Melchior, Mamon, Chilingarian, \&
  Katkov}]{Maschmann20}
Maschmann, D., Melchior, A.-L., Mamon, G.~A., Chilingarian, I.~V., \& Katkov,
  I.~Y. 2020, \aap, 641, A171

\bibitem[{McGregor {et~al.}(2002)McGregor, Hart, Conroy, Pfitzner, Bloxham,
  Jones, Downing, Dawson, Young, Jarnyk, \& Harmelen}]{McGregor02}
McGregor, P.~J., Hart, J., Conroy, P.~G., {et~al.} 2002, Society of
  Photo-Optical Instrumentation Engineers (SPIE) Conference Series, 4841, 1581

\bibitem[{Menacho {et~al.}(2019)Menacho, Östlin, Bik, Bruna, Melinder, Adamo,
  Hayes, Herenz, \& Bergvall}]{Menacho19}
Menacho, V., Östlin, G., Bik, A., {et~al.} 2019, \mnras, 487, 3183

\bibitem[{Moiseev {et~al.}(2015)Moiseev, Tikhonov, \& Klypin}]{Moiseev15}
Moiseev, A.~V., Tikhonov, A.~V., \& Klypin, A. 2015, \mnras, 449, 3568

\bibitem[{Natta \& Panagia(1984)}]{Natta84}
Natta, A. \& Panagia, N. 1984, \apj

\bibitem[{Newman {et~al.}(2013)Newman, Genzel, Schreiber, Griffin, Mancini,
  Lilly, Renzini, Bouché, Burkert, Buschkamp, Carollo, Cresci, Davies,
  Eisenhauer, Genel, Hicks, Kurk, Lutz, Naab, Peng, Sternberg, Tacconi, Wuyts,
  Zamorani, \& Vergani}]{Newman13}
Newman, S.~F., Genzel, R., Schreiber, N. M.~F., {et~al.} 2013, \apj, 767, 104

\bibitem[{Oh {et~al.}(2020)Oh, Colless, Barsanti, Casura, Cortese, Sande,
  Owers, Scott, D'Eugenio, Bland-Hawthorn, Brough, Bryant, Croom, Foster,
  Groves, Lawrence, Richards, \& Sweet}]{Oh20}
Oh, S., Colless, M., Barsanti, S., {et~al.} 2020, \mnras, 495, 4638

\bibitem[{Oh {et~al.}(2022)Oh, Colless, D’Eugenio, Croom, Cortese, Groves,
  Kewley, van de Sande, Zovaro, Varidel, Barsanti, Bland-Hawthorn, Brough,
  Bryant, Casura, Lawrence, Lorente, Medling, Owers, \& Yi}]{Oh22}
Oh, S., Colless, M., D’Eugenio, F., {et~al.} 2022, \mnras, 512, 1765

\bibitem[{Osterbrock \& Ferland(2006)}]{Osterbrockbook}
Osterbrock, D.~E. \& Ferland, G.~J. 2006, {Astrophysics of gaseous nebulae and
  active galactic nuclei}, Astrophysics of gaseous nebulae and active galactic
  nuclei, 2nd. ed. by D.E. Osterbrock and G.J. Ferland. Sausalito, CA:
  University Science Books, 2006 (Astrophysics of gaseous nebulae and active
  galactic nuclei, 2nd. ed. by D.E. Osterbrock and G.J. Ferland. Sausalito, CA:
  University Science Books, 2006)

\bibitem[{{Planck Collaboration} {et~al.}(2016){Planck Collaboration}, Ade,
  Aghanim, Arnaud, Ashdown, Aumont, Baccigalupi, Banday, Barreiro, Bartlett,
  Bartolo, Battaner, Battye, Benabed, Benoît, Benoit-Lévy, Bernard,
  Bersanelli, Bielewicz, Bock, Bonaldi, Bonavera, Bond, Borrill, Bouchet,
  Boulanger, Bucher, Burigana, Butler, Calabrese, Cardoso, Catalano, Challinor,
  Chamballu, Chary, Chiang, Chluba, Christensen, Church, Clements, Colombi, L.,
  Combet, Coulais, Crill, Curto, Cuttaia, Danese, Davies, Davis, Bernardis,
  Rosa, Zotti, Delabrouille, Desert, Valentino, Dickinson, Diego, Dolag, Dole,
  Donzelli, Doré, Douspis, Ducout, Dunkley, Dupac, Efstathiou, Elsner,
  Enßlin, Eriksen, Farhang, Fergusson, Finelli, Forni, Frailis, Fraisse,
  Franceschi, Frejsel, Galeotta, Galli, Ganga, Gauthier, Gerbino, Ghosh, Giard,
  Giraud-Héraud, Giusarma, Gjerløw, González-Nuevo, Górski, Gratton,
  Gregorio, Gruppuso, Gudmundsson, Hamann, Hansen, Hanson, Harrison, Helou,
  Henrot-Versillé, Hernández-Monteagudo, Herranz, Hildebrandt, Hivon, Hobson,
  Holmes, Hornstrup, Hovest, Huang, Huffenberger, Hurier, Jaffe, Jaffe, Jones,
  Juvela, Keihänen, Keskitalo, Kisner, Kneissl, Knoche, Knox, Kunz,
  Kurki-Suonio, Lagache, Lähteenmäki, Lamarre, Lasenby, Lattanzi, Lawrence,
  Leahy, Leonardi, Lesgourgues, Levrier, Lewis, Liguori, Lilje, Linden-Vørnle,
  López-Caniego, Lubin, Macías-Pérez, Maggio, Maino, Mandolesi, Mangilli,
  Marchini, Maris, Martin, Martinelli, Martínez-González, Masi, Matarrese,
  McGehee, Meinhold, Melchiorri, Melin, Mendes, Mennella, Migliaccio, Millea,
  Mitra, Miville-Deschenes, Moneti, Montier, Morgante, Mortlock, Moss, Munshi,
  Murphy, Naselsky, Nati, Natoli, Netterfield, Norgaard-Nielsen, Noviello,
  Novikov, Novikov, Oxborrow, Paci, Pagano, Pajot, Paladini, Paoletti,
  Partridge, Pasian, Patanchon, Pearson, Perdereau, Perotto, Perrotta,
  Pettorino, Piacentini, Piat, Pierpaoli, Pietrobon, Plaszczynski,
  Pointecouteau, Polenta, Popa, Pratt, Prézeau, Prunet, Puget, Rachen, Reach,
  Rebolo, Reinecke, Remazeilles, Renault, Renzi, Ristorcelli, Rocha, Rosset,
  Rossetti, Roudier, d'Orfeuil, Rowan-Robinson, Rubiño-Martín, Rusholme,
  Said, Salvatelli, Salvati, Sandri, Santos, Savelainen, Savini, Scott,
  Seiffert, Serra, Shellard, Spencer, Spinelli, Stolyarov, Stompor, Sudiwala,
  Sunyaev, Sutton, Suur-Uski, Sygnet, Tauber, Terenzi, Toffolatti, Tomasi,
  Tristram, Trombetti, Tucci, Tuovinen, Türler, Umana, Valenziano, Valiviita,
  Tent, Vielva, Villa, Wade, Wandelt, Wehus, White, White, Wilkinson, Yvon,
  Zacchei, \& Zonca}]{Planck15}
{Planck Collaboration}, Ade, P. A.~R., Aghanim, N., {et~al.} 2016, \aap, 594,
  A13

\bibitem[{Rayner {et~al.}(2009)Rayner, Cushing, \& Vacca}]{Rayner09}
Rayner, J.~T., Cushing, M.~C., \& Vacca, W.~D. 2009, \apjs, 185, 289

\bibitem[{Riffel {et~al.}(2015)Riffel, Storchi-Bergmann, \& Riffel}]{Riffel15}
Riffel, R.~A., Storchi-Bergmann, T., \& Riffel, R. 2015, \mnras, 451, 3587

\bibitem[{Robitaille \& Bressert(2012)}]{Robitaille12}
Robitaille, T. \& Bressert, E. 2012, Astrophysical Source Code Library

\bibitem[{Rodrigues {et~al.}(2016)Rodrigues, Hammer, Flores, Puech, \&
  Athanassoula}]{Rodrigues16}
Rodrigues, M., Hammer, F., Flores, H., Puech, M., \& Athanassoula, E. 2016,
  \mnras, 465, 1157

\bibitem[{Rubin {et~al.}(1980)Rubin, Ford, \& Thonnard}]{Rubin80}
Rubin, V.~C., Ford, W. K.~J., \& Thonnard, N. 1980, \apj, 238, 471

\bibitem[{Schreiber {et~al.}(2006)Schreiber, Genzel, Lehnert, Bouché, Verma,
  Erb, Shapley, Steidel, Davies, Lutz, Nesvadba, Tacconi, Eisenhauer, Abuter,
  Gilbert, Gillessen, \& Sternberg}]{ForsterSchreiber06}
Schreiber, N. M.~F., Genzel, R., Lehnert, M.~D., {et~al.} 2006, \apj, 645, 1062

\bibitem[{Schreiber \& Wuyts(2020)}]{Schreiber20}
Schreiber, N. M.~F. \& Wuyts, S. 2020, \araa, 58, annurev

\bibitem[{Scott {et~al.}(2018)Scott, Sande, Croom, Groves, Owers, Poetrodjojo,
  D'Eugenio, Medling, Barat, Barone, Bland-Hawthorn, Brough, Bryant, Cortese,
  Foster, Green, Oh, Colless, Drinkwater, Driver, Goodwin, P., Federrath,
  Harischandra, Jin, Lawrence, Lorente, Mannering, O'Toole, Richards, Sanchez,
  Schaefer, Sealey, Sharp, Sweet, Taranu, \& Varidel}]{Scott18}
Scott, N., Sande, J. v.~d., Croom, S.~M., {et~al.} 2018, \mnras, 481, 2299

\bibitem[{Skrutskie {et~al.}(2006)Skrutskie, Cutri, Stiening, Weinberg,
  Schneider, Carpenter, Beichman, Capps, Chester, Elias, Huchra, Liebert,
  Lonsdale, Monet, Price, Seitzer, Jarrett, Kirkpatrick, Gizis, Howard, Evans,
  Fowler, Fullmer, Hurt, Light, Kopan, Marsh, McCallon, Tam, Dyk, \&
  Wheelock}]{Skrutskie06}
Skrutskie, M.~F., Cutri, R.~M., Stiening, R., {et~al.} 2006, \aj, 131, 1163

\bibitem[{Strickland \& Heckman(2009)}]{Strickland09}
Strickland, D.~K. \& Heckman, T.~M. 2009, \apj, 697, 2030

\bibitem[{Swinbank {et~al.}(2012)Swinbank, Smail, Sobral, Theuns, Best, \&
  Geach}]{Swinbank12}
Swinbank, A.~M., Smail, I., Sobral, D., {et~al.} 2012, \apj, 760, 130

\bibitem[{Sánchez {et~al.}(2012)Sánchez, Kennicutt, Paz, Ven, Vilchez,
  Wisotzki, Walcher, Mast, Aguerri, Albiol-Pérez, Alonso-Herrero, Alves,
  Bakos, Bartáková, Bland-Hawthorn, Boselli, Bomans, Castillo-Morales,
  Cortijo-Ferrero, Lorenzo-Cáceres, Olmo, Dettmar, Díaz, Ellis,
  Falcón-Barroso, Flores, Gallazzi, García-Lorenzo, Delgado, Gruel, Haines,
  Hao, Husemann, Iglesias-Páramo, Jahnke, Johnson, Jungwiert, Kalinova,
  Kehrig, Kupko, López-Sánchez, Lyubenova, Marino, Mármol-Queraltó,
  Márquez, Masegosa, Meidt, Méndez-Abreu, Monreal-Ibero, Montijo, Mourão,
  Palacios-Navarro, Papaderos, Pasquali, Peletier, Pérez, Pérez, Quirrenbach,
  Relaño, Rosales-Ortega, Roth, Ruiz-Lara, Sánchez-Blázquez, Sengupta,
  Singh, Stanishev, Trager, Vazdekis, Viironen, Wild, Zibetti, \&
  Ziegler}]{Sanchez12}
Sánchez, S.~F., Kennicutt, R.~C., Paz, A. G.~d., {et~al.} 2012, \aap, 538, A8

\bibitem[{Terlevich \& Melnick(1981)}]{Terlevich81}
Terlevich, R. \& Melnick, J. 1981, \mnras, 195, 839

\bibitem[{{The Astropy Collaboration} {et~al.}(2018){The Astropy
  Collaboration}, Price-Whelan, Sipőcz, Günther, Lim, Crawford, Conseil,
  Shupe, Craig, Dencheva, Ginsburg, VanderPlas, Bradley, Pérez-Suárez,
  Val-Borro, Aldcroft, Cruz, Robitaille, Tollerud, Ardelean, Babej, Bach,
  Bachetti, Bakanov, Bamford, Barentsen, Barmby, Baumbach, Berry, Biscani,
  Boquien, Bostroem, Bouma, Brammer, Bray, Breytenbach, Buddelmeijer, Burke,
  Calderone, Rodríguez, Cara, Cardoso, Cheedella, Copin, Corrales, Crichton,
  D'Avella, Deil, Depagne, Dietrich, Donath, Droettboom, Earl, Erben, Fabbro,
  Ferreira, Finethy, Fox, Garrison, Gibbons, Goldstein, Gommers, Greco,
  Greenfield, Groener, Grollier, Hagen, Hirst, Homeier, Horton, Hosseinzadeh,
  Hu, Hunkeler, Ivezić, Jain, Jenness, Kanarek, Kendrew, Kern, Kerzendorf,
  Khvalko, King, Kirkby, Kulkarni, Kumar, Lee, Lenz, Littlefair, Ma, Macleod,
  Mastropietro, McCully, Montagnac, Morris, Mueller, Mumford, Muna, Murphy,
  Nelson, Nguyen, Ninan, Nöthe, Ogaz, Oh, Parejko, Parley, Pascual, Patil,
  Patil, Plunkett, Prochaska, Rastogi, Janga, Sabater, Sakurikar, Seifert,
  Sherbert, Sherwood-Taylor, Shih, Sick, Silbiger, Singanamalla, P., Sladen,
  Sooley, Sornarajah, Streicher, Teuben, Thomas, Tremblay, Turner, Terrón,
  Kerkwijk, Vega, Watkins, Weaver, Whitmore, Woillez, Zabalza, \&
  Contributors}]{astropy18}
{The Astropy Collaboration}, Price-Whelan, A.~M., Sipőcz, B.~M., {et~al.}
  2018, \aj, 156, 123

\bibitem[{{The Astropy Collaboration} {et~al.}(2013){The Astropy
  Collaboration}, Robitaille, Tollerud, Greenfield, Droettboom, Bray, Aldcroft,
  Davis, Ginsburg, Price-Whelan, Kerzendorf, Conley, Crighton, Barbary, Muna,
  Ferguson, Grollier, Parikh, Nair, Günther, Deil, Woillez, Conseil, Kramer,
  Turner, Singer, Fox, Weaver, Zabalza, Edwards, Bostroem, Burke, Casey,
  Crawford, Dencheva, Ely, Jenness, Labrie, Lim, Pierfederici, Pontzen, Ptak,
  Refsdal, Servillat, \& Streicher}]{astropy13}
{The Astropy Collaboration}, Robitaille, T.~P., Tollerud, E.~J., {et~al.} 2013,
  \aap, 558, A33

\bibitem[{Varidel {et~al.}(2020)Varidel, Croom, Lewis, Fisher, Glazebrook,
  Catinella, Cortese, Krumholz, Bland-Hawthorn, Bryant, Groves, Brough,
  Federrath, Lawrence, Lorente, Owers, Richards, López-Sánchez, Sweet, Sande,
  \& Vaughan}]{Varidel20}
Varidel, M.~R., Croom, S.~M., Lewis, G.~F., {et~al.} 2020, \mnras, 495, 2265

\bibitem[{Wallace \& Hinkle(1997)}]{Wallace97}
Wallace, L. \& Hinkle, K. 1997, \apjs, 111, 445

\bibitem[{Winge {et~al.}(2009)Winge, Riffel, \& Storchi-Bergmann}]{Winge09}
Winge, C., Riffel, R.~A., \& Storchi-Bergmann, T. 2009, \apjs, 185, 186

\bibitem[{Wisnioski {et~al.}(2018)Wisnioski, Mendel, Schreiber, Genzel, Wilman,
  Wuyts, Belli, Beifiori, Bender, Brammer, Chan, Davies, Davies, Fabricius,
  Fossati, Galametz, Lang, Lutz, Nelson, Momcheva, Rosario, Saglia, Tacconi,
  Tadaki, Übler, \& Dokkum}]{Wisnioski18}
Wisnioski, E., Mendel, J.~T., Schreiber, N. M.~F., {et~al.} 2018, \apj
  [\eprint{1711.02111}]

\bibitem[{Wisnioski {et~al.}(2015)Wisnioski, Schreiber, Wuyts, Wuyts, Bandara,
  Wilman, Genzel, Bender, Davies, Fossati, Lang, Mendel, Beifiori, Brammer,
  Chan, Fabricius, Fudamoto, Kulkarni, Kurk, Lutz, Nelson, Momcheva, Rosario,
  Saglia, Seitz, Tacconi, \& Dokkum}]{Wisnioski15}
Wisnioski, E., Schreiber, N. M.~F., Wuyts, S., {et~al.} 2015, \apj, 799, 209

\bibitem[{Östlin {et~al.}(2001)Östlin, Amram, Bergvall, Masegosa, Boulesteix,
  \& Márquez}]{Ostlin01}
Östlin, G., Amram, P., Bergvall, N., {et~al.} 2001, \aap, 374, 800

\bibitem[{Östlin {et~al.}(1999)Östlin, Amram, Masegosa, Bergvall, \&
  Boulesteix}]{Ostlin99}
Östlin, G., Amram, P., Masegosa, J., Bergvall, N., \& Boulesteix, J. 1999,
  \aaps, 137, 419

\bibitem[{Östlin {et~al.}(2004)Östlin, Cumming, Amram, Bergvall, Kunth,
  Márquez, Masegosa, \& Zackrisson}]{Ostlin04}
Östlin, G., Cumming, R.~J., Amram, P., {et~al.} 2004, \aap, 419, L43

\bibitem[{Östlin {et~al.}(2014)Östlin, Hayes, Duval, Sandberg,
  Rivera-Thorsen, Marquart, Orlitová, Adamo, Melinder, Guaita, Atek, Cannon,
  Gruyters, Herenz, Kunth, Laursen, Mas-Hesse, Micheva, Otí-Floranes, Pardy,
  Roth, Schaerer, \& Verhamme}]{Ostlin14}
Östlin, G., Hayes, M., Duval, F., {et~al.} 2014, \apj, 797, 11

\bibitem[{Östlin {et~al.}(2015)Östlin, Marquart, Cumming, Fathi, Bergvall,
  Adamo, Amram, \& Hayes}]{Ostlin15}
Östlin, G., Marquart, T., Cumming, R.~J., {et~al.} 2015, \aap, 583, A55

\end{thebibliography}

\begin{appendix}
 \onecolumn

\section{Spatially resolved emission line maps and kinematics}\label{sec:appendixA}

\begin{figure*}[!h]
   \includegraphics[width=0.99\hsize]{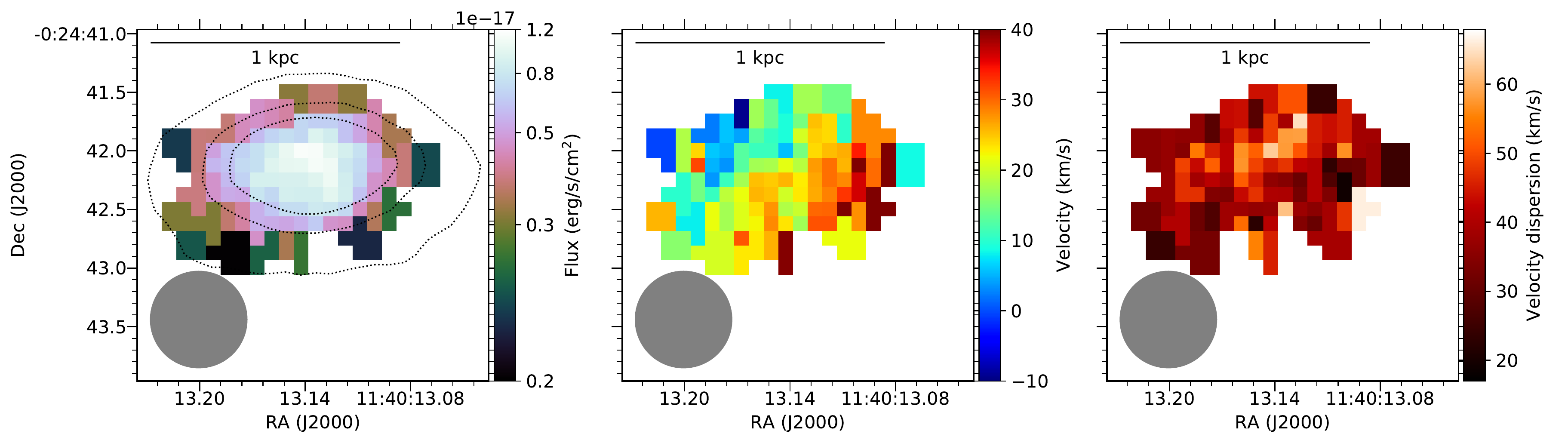}
      \caption{Spatially resolved maps of the \brg\ emission line of J1140-0024. \emph{Left:} \brg\ intensity map with as contours the \Ks\ band continuum flux created by convolving the datacube with the 2MASS \Ks\ response curve. The grey circle represents the FWHM of the observed PSF as derived in Sect. \ref{subsec:resolved}. The 1 kpc scale bar at the top is derived using the observed redshift and the cosmological parameters of \citet{Planck15}.
      \emph{Middle:} \brg\ velocity map constructed from a single Gaussian fit. \emph{Right:} Observed velocity dispersion map corrected for the instrumental resolution. This map is not corrected for beam smearing (Sect. \ref{subsec:resolved}). As the beam smearing correction only has a minor effect on the velocity dispersion maps we have chosen to show the observed maps before correction.
    \label{fig:mapJ1140}}
\end{figure*}

\begin{figure*}[!h]
   \includegraphics[width=0.99\hsize]{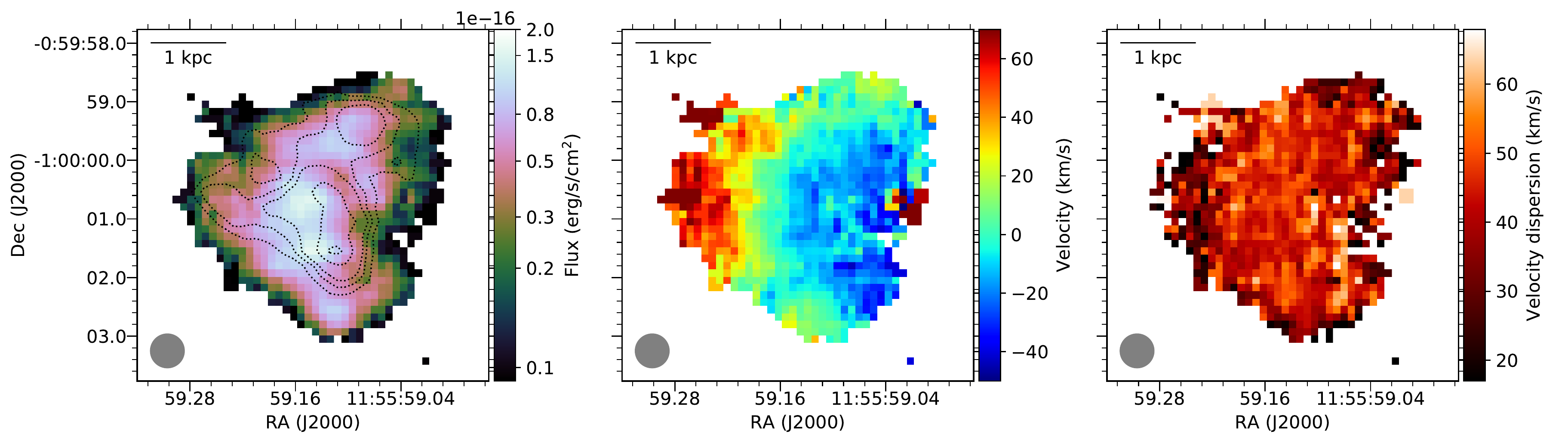}
      \caption{Spatially resolved maps of the \paa\ line of J1155-0100. See caption of Fig. \ref{fig:mapJ1140}.
    \label{fig:mapJ1155}}
\end{figure*}

\begin{figure*}[!h]
   \includegraphics[width=0.99\hsize]{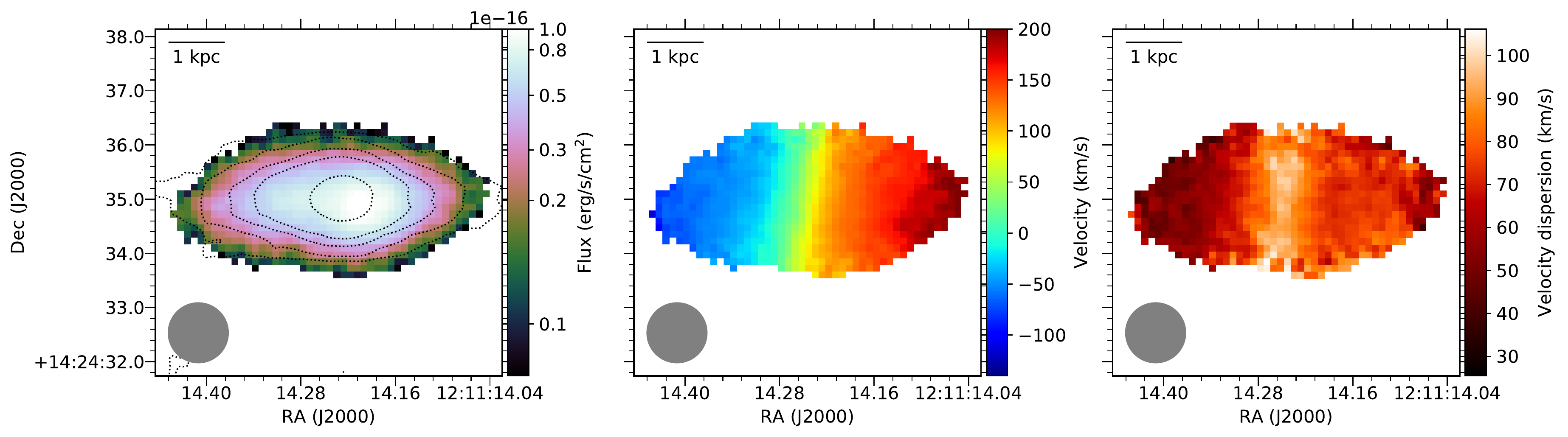}
      \caption{Spatially resolved maps of the \paa\ line of J1211+1424. See caption of Fig. \ref{fig:mapJ1140}.
    \label{fig:mapJ1211}}
\end{figure*}

\begin{figure*}
   \includegraphics[width=0.99\hsize]{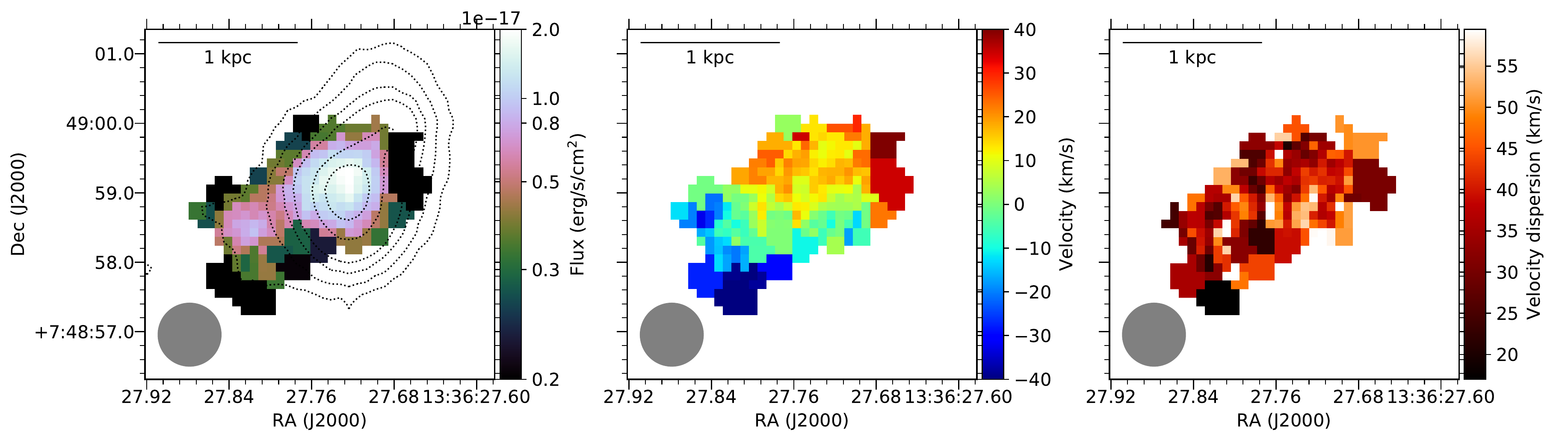}
      \caption{Spatially resolved maps of the \brg\ line of J1336+0748. See caption of Fig. \ref{fig:mapJ1140}.
    \label{fig:mapJ1336}}
\end{figure*}

\begin{figure*}
   \includegraphics[width=0.99\hsize]{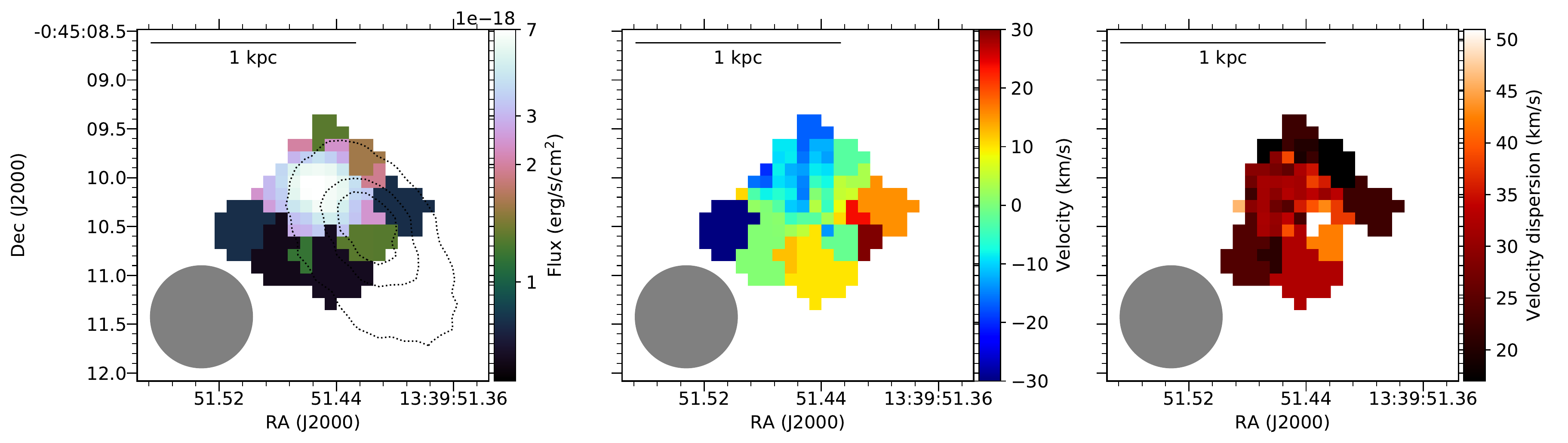}
      \caption{Spatially resolved maps of the \brg\ line of J1339-0045. See caption of Fig. \ref{fig:mapJ1140}.
    \label{fig:mapJ1339}}
\end{figure*}

\begin{figure*}
   \includegraphics[width=0.99\hsize]{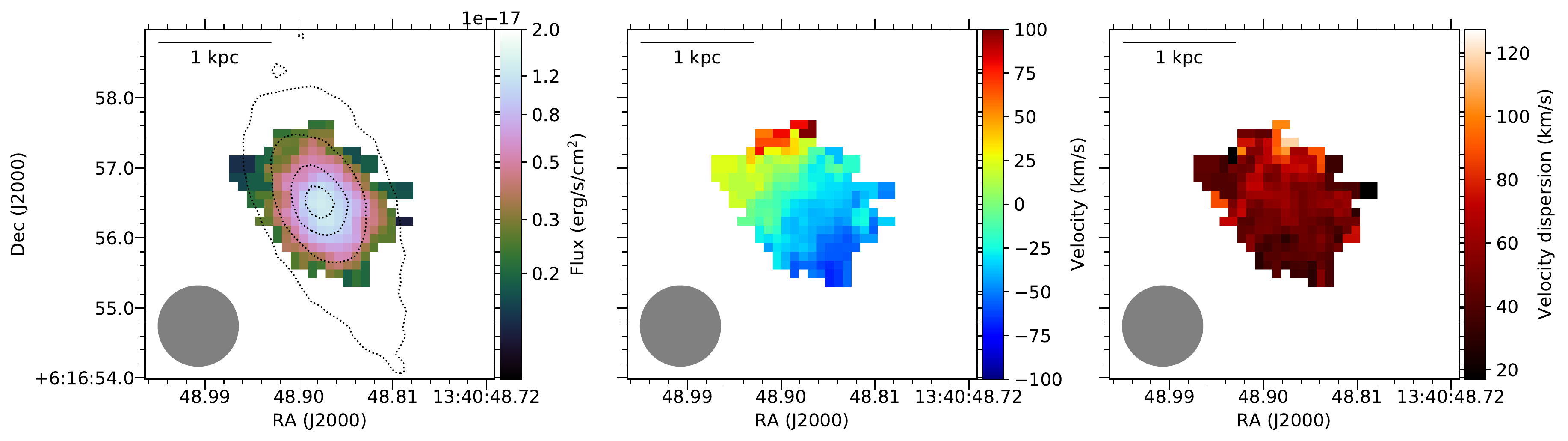}
      \caption{Spatially resolved maps of the \brg\ line of J1340+0616. See caption of Fig. \ref{fig:mapJ1140}.
    \label{fig:mapJ1340}}
\end{figure*}

\begin{figure*}
   \includegraphics[width=0.99\hsize]{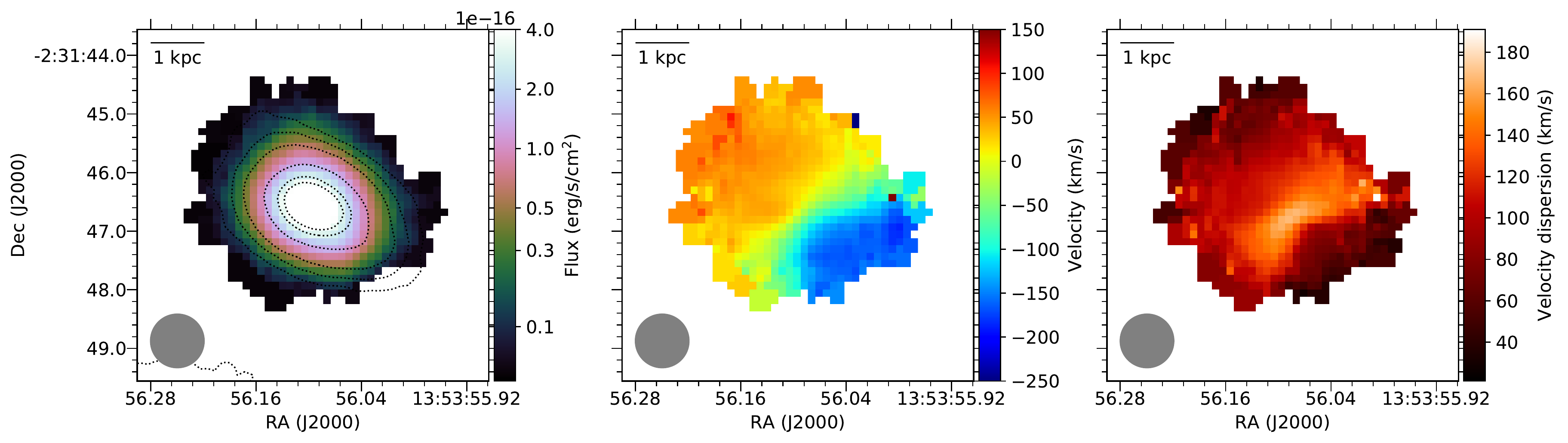}
      \caption{Spatially resolved maps of the \paa\ line of J1353-0231. See caption of Fig. \ref{fig:mapJ1140}.
    \label{fig:mapJ1353}}
\end{figure*}

\begin{figure*}
   \includegraphics[width=0.99\hsize]{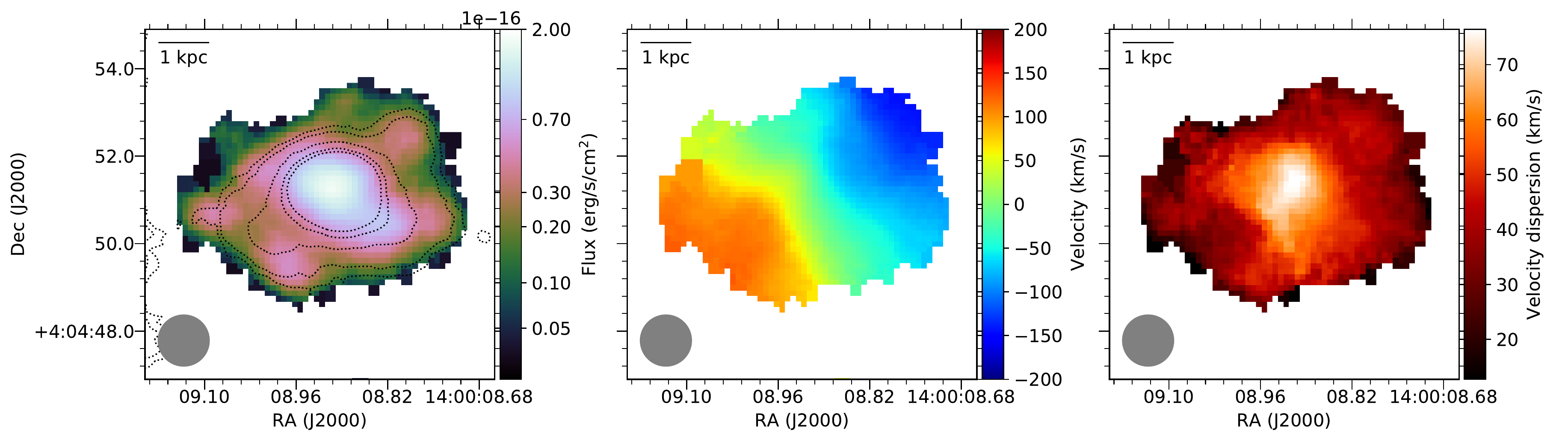}
      \caption{Spatially resolved maps of the \paa\ line of J1400+0404. See caption of Fig. \ref{fig:mapJ1140}.
    \label{fig:mapJ1400}}
\end{figure*}

\begin{figure*}
   \includegraphics[width=0.99\hsize]{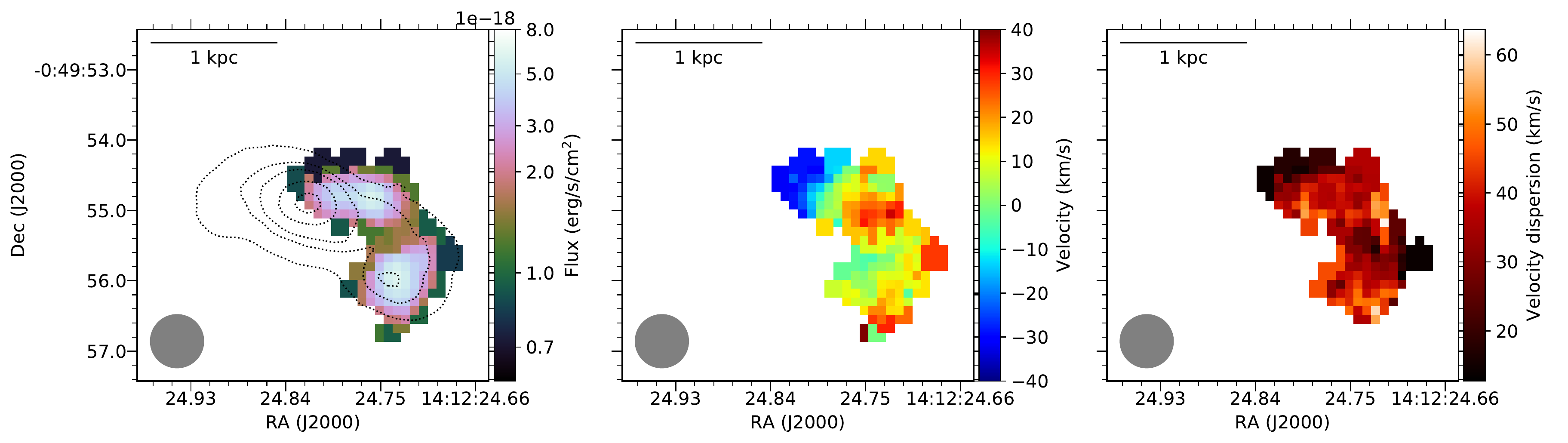}
      \caption{Spatially resolved maps of the \paa\ line of J1412-0049. See caption of Fig. \ref{fig:mapJ1140}.
    \label{fig:mapJ1412}}
\end{figure*}

\begin{figure*}
   \includegraphics[width=0.99\hsize]{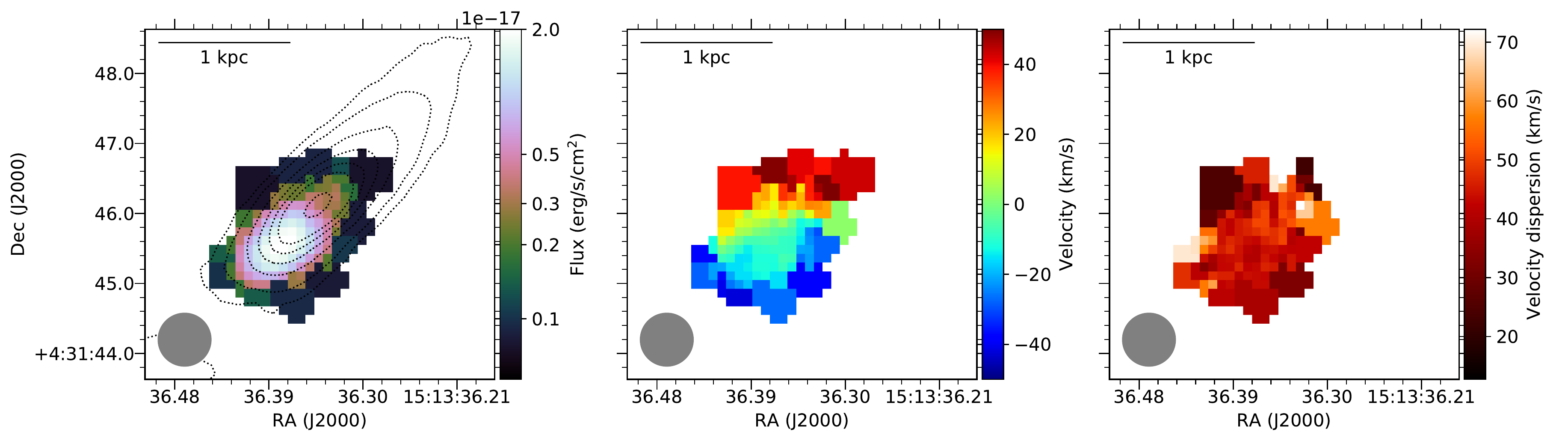}
      \caption{Spatially resolved maps of the \paa\ line of J1513+0431. See caption of Fig. \ref{fig:mapJ1140}.
    \label{fig:mapJ1513}}
\end{figure*}

\begin{figure*}
   \includegraphics[width=0.99\hsize]{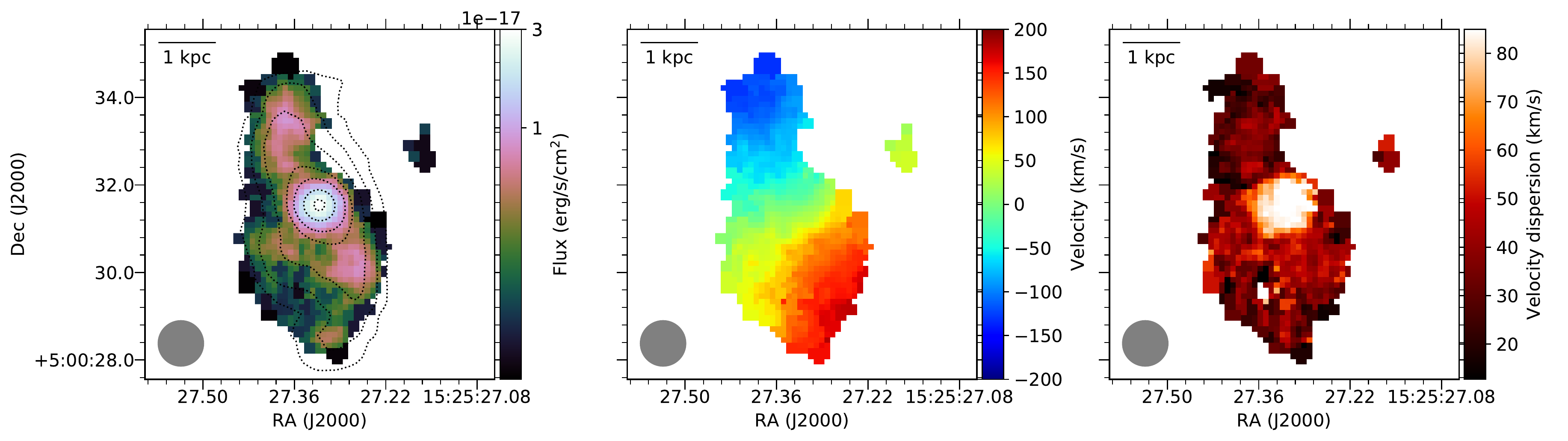}
      \caption{Spatially resolved maps of the \paa\ line of J1525+0500. See caption of Fig. \ref{fig:mapJ1140}.
    \label{fig:mapJ1525}}
\end{figure*}

\begin{figure*}
   \includegraphics[width=0.99\hsize]{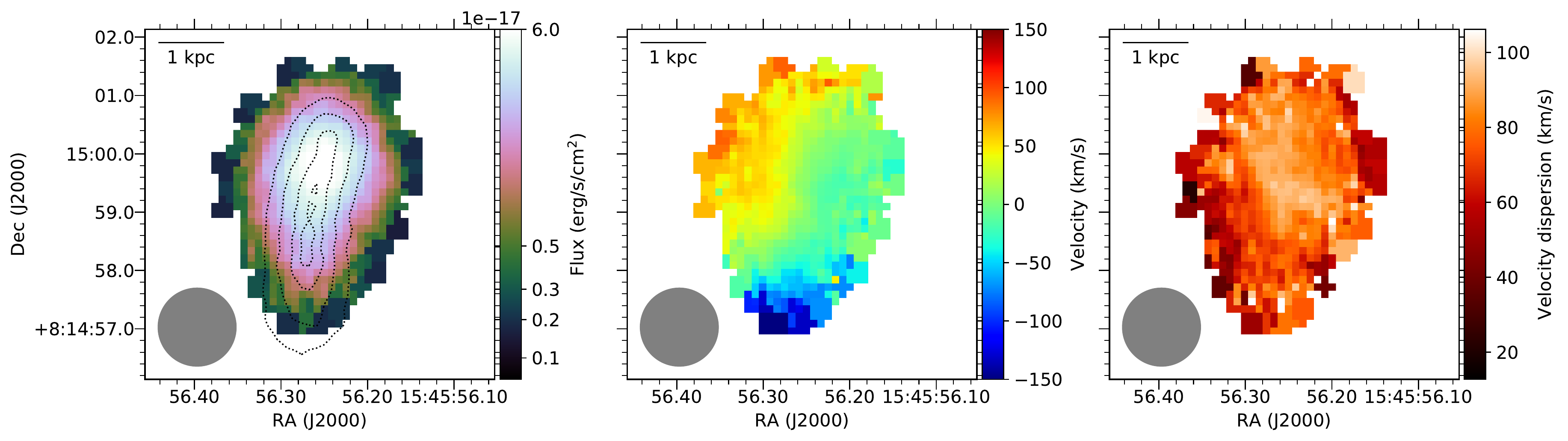}
      \caption{Spatially resolved maps of the \paa\ line of J1545+0814. See caption of Fig. \ref{fig:mapJ1140}.
    \label{fig:mapJ1545}}
\end{figure*}

\begin{figure*}[!ht]
   \includegraphics[width=0.99\hsize]{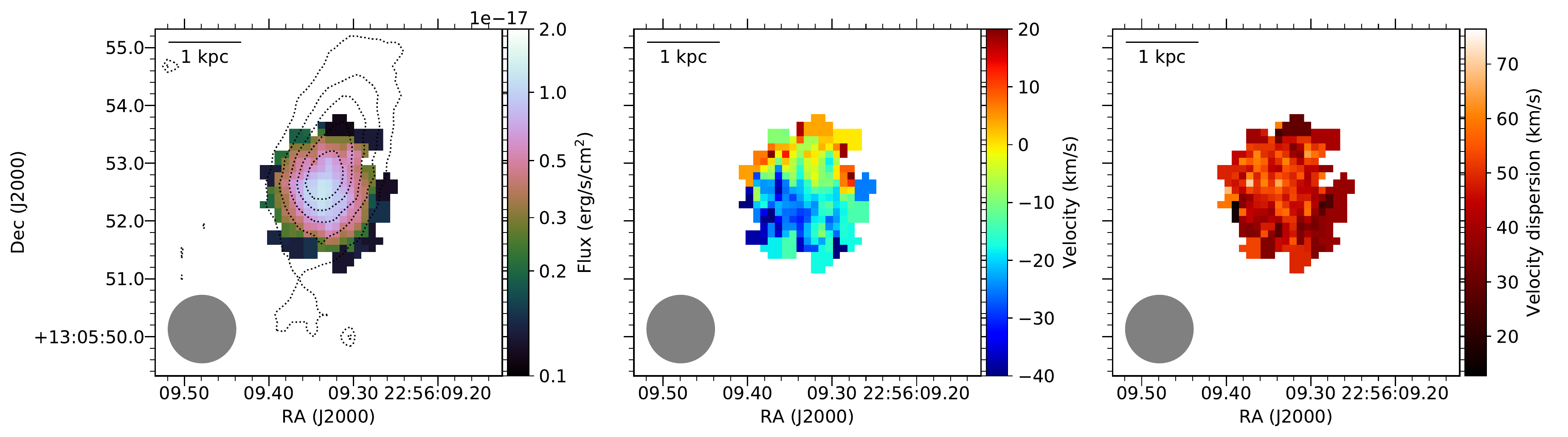}
      \caption{Spatially resolved maps of the \paa\ line of J2256+1305. See caption of Fig. \ref{fig:mapJ1140}.
    \label{fig:mapJ2256}}
\end{figure*}

\end{appendix}

%
%

\end{document}